\pdfoutput=1%

\documentclass[twocolumn, twocolappendix]{aastex631}
\usepackage[figuresright]{rotating}
\usepackage{physics}
\usepackage{enumitem}
\let\tablenum\relax
\usepackage{siunitx}

\allowdisplaybreaks[1]
\renewcommand\arcsec{\mbox{$^{\prime\prime}$}}
\newcommand\kmps{\mbox{$\si{\km\per\second}$}}
\newcommand\pcc{\mbox{$\si{\per\cm\cubed}$}}
\newcommand\psqc{\mbox{$\si{\per\cm\squared}$}}
\renewcommand\micron{\mbox{$\si{\micro\metre}$}}
\newcommand\ireight{IR08572~NW}
\newcommand\irone{IR01250}
\newcommand\ufive{U5101}
\newcommand\nfour{N4418}

\received{May 25, 2024}
\revised{October 6, 2024}
\accepted{October 7, 2024}
\submitjournal{ApJ}

\shorttitle{CO Rovib. Lines in Nearby U/LIRGs}
\shortauthors{S. Onishi et al.}

\begin{document}

\title{Systematic Study of the Inner Structure of Molecular Tori in Nearby U/LIRGs\\using Velocity Decomposition of CO Rovibrational Absorption Lines
\footnote{This research is based on data collected at the Subaru Telescope, which is operated by the National Astronomical Observatory of Japan. We are honored by and grateful for the opportunity to observe the universe from Maunakea, which has cultural, historical, and natural significance in Hawaii.}}

\AuthorCollaborationLimit=10
\correspondingauthor{Shusuke Onishi}
\email{s\_onishi@ir.isas.jaxa.jp}

\author[0000-0002-1765-7012]{Shusuke Onishi}
\affiliation{Institute of Space and Astronautical Science, Japan Aerospace Exploration Agency, 3-1-1 Yoshinodai, Chuo-ku, Sagamihara, Kanagawa 252-5210, Japan}


\author[0000-0002-6660-9375]{Takao Nakagawa}
\affiliation{Institute of Space and Astronautical Science, Japan Aerospace Exploration Agency, 3-1-1 Yoshinodai, Chuo-ku, Sagamihara, Kanagawa 252-5210, Japan}

\author[0000-0002-9850-6290]{Shunsuke Baba}
\affiliation{Institute of Space and Astronautical Science, Japan Aerospace Exploration Agency, 3-1-1 Yoshinodai, Chuo-ku, Sagamihara, Kanagawa 252-5210, Japan}
\affiliation{Gradiate School of Science and Engineering, Kagoshima University, 1-21-35 Korimoto, Kagoshima, Kagoshima 890-0065, Japan}


\author[0000-0002-5012-6707]{Kosei Matsumoto}
\affiliation{Institute of Space and Astronautical Science, Japan Aerospace Exploration Agency, 3-1-1 Yoshinodai, Chuo-ku, Sagamihara, Kanagawa 252-5210, Japan}
\affiliation{Department of Physics, Graduate School of Science, The University of Tokyo, 7-3-1 Hongo, Bunkyo-ku, Tokyo 113-0033, Japan}
\affiliation{Sterrenkundig Observatorium, Department of Physics and Astronomy, Universiteit Gent, Krijgslaan 281 S9, B-9000 Gent, Belgium}

\author{Naoki Isobe}
\affiliation{Institute of Space and Astronautical Science, Japan Aerospace Exploration Agency, 3-1-1 Yoshinodai, Chuo-ku, Sagamihara, Kanagawa 252-5210, Japan}

\author{Mai Shirahata}
\affiliation{Institute of Space and Astronautical Science, Japan Aerospace Exploration Agency, 3-1-1 Yoshinodai, Chuo-ku, Sagamihara, Kanagawa 252-5210, Japan}

\author[0000-0002-7914-6779]{Hiroshi Terada}
\affiliation{National Astronomical Observatory of Japan, 2-21-1 Osawa, Mitaka, Tokyo 181-8588, Japan}

\author[0000-0001-9855-0163]{Tomonori Usuda}
\affiliation{National Astronomical Observatory of Japan, 2-21-1 Osawa, Mitaka, Tokyo 181-8588, Japan}

\author[0000-0003-4842-565X]{Shinki Oyabu}
\affiliation{Institute of Liberal Arts and Sciences, Tokushima University, 1-1 Minami-josanjima-cho, Tokushima, Tokushima 770-8502, Japan}




\begin{abstract}
Determining the inner structure of the molecular torus around an active galactic nucleus is essential for understanding its formation mechanism.
However, spatially resolving the torus is difficult because of its small size.
To probe the clump conditions in the torus, we therefore perform the systematic velocity-decomposition analyses of the gaseous ${}^{12}\mathrm{CO}$ rovibrational absorption lines ($v=0\to 1, \Delta{J}=\pm{1}$) at $\lambda\sim 4.67\,\micron$ observed toward four (ultra)luminous infrared galaxies using the high-resolution ($R\sim 5000\text{--}10{,}000$) spectroscopy from the Subaru Telescope.
We find that each transition has two to five distinct velocity components with different line-of-sight (LOS) velocities ($V_\mathrm{LOS}\sim {-240}\text{--}{+100}\,\kmps$) and dispersions ($\sigma_V\sim 15\text{--}190\,\kmps$); i.e., the components (a), (b), $\cdots$, beginning with the broadest one in each target, indicating that the tori have clumpy structures.
By assuming a hydrostatic disk ($\sigma_V\propto R_\mathrm{rot}^{-0.5}$), we find that the tori have dynamic inner structures, with the innermost component (a) outflowing with velocity $|V_\mathrm{LOS}|\sim 160\text{--}240\,\kmps$, and the outer components (b) and (c) outflowing more slowly or infalling with $|V_\mathrm{LOS}|\lesssim 100\,\kmps$.
In addition, we find that the innermost component (a) can be attributed to collisionally excited hot ($\gtrsim 530\,\mathrm{K}$) and dense ($\log\qty(n_\mathrm{H_2}/\pcc)\gtrsim 6$) clumps, based on the level populations.
Conversely, the outer component (b) can be attributed to cold ($\sim 30\text{--}140\,\mathrm{K}$) clumps radiatively excited by a far-infrared-to-submillimeter background with a brightness temperature higher than $\sim 20\text{--}400\,\mathrm{K}$.
These observational results demonstrate the clumpy and dynamic structure of tori in the presence of background radiation.
\end{abstract}



\section{Introduction}\label{sec:intro}
Active galactic nuclei (AGNs) are the luminous central regions of active galaxies.
They have Eddington ratios exceeding $\sim 10^{-4}$ owing to mass accretion onto central supermassive black holes (SMBHs).
AGNs are classified into types 1 and 2 on the basis of their optical line widths \citep{Khachikian1974}.
The unified model of AGNs \citep[e.g.,][]{Miller1983,Antonucci1985a,Antonucci1993b} suggests that the two types of AGNs are intrinsically identical, with the observational differences caused predominantly by the inclination of a donut-like, geometrically thick structure surrounding the central SMBH.
The donut-like structure is called a ``molecular torus''.
To clarify the mechanism responsible for maintaining the geometrical thickness of the torus, it is essential to understand its inner structure.
However, spatially resolving the torus is difficult because it is expected to be only a few parsecs in size.
Although radio and near-infrared (NIR) interferometry and radio polarimetry with torus-scale beams are being achieved toward the nearest AGNs, such as NGC~1068 (\citealp[e.g.,][]{Garcia-Burillo2016,Imanishi2018,Imanishi2020,Lopez-Rodriguez2020,GRAVITYCollaboration2020a}; see also \citealp{Gamez-Rosas2022}), the structure inside the torus has not yet been resolved.
\par
For these reasons, the inner structure of the torus has mainly been discussed using theoretical models.
Those models consider the torus to consist of many dense molecular clouds (clumps) because a clump--clump collisional disk maintains its geometrical thickness better due to turbulence and to the outflowing motions of the clumps (clumpy torus models; \citealp[e.g.,][]{Beckert2004,Vollmer2004,Nenkova2008a}) than do continuous gas disks.
Clumpy torus models are also required to reproduce the observed optical depth of silicate dust at $\lambda\sim 9.7\,\micron$\footnote{Note that not all of the mid-infrared flux at the wavelength $\lambda\sim 8\text{--}13\,\micron$ originates in the torus. Some flux also comes from polar dust, according to recent mid-infrared interferometry \citep[e.g.,][]{Honig2013,Tristram2014,Asmus2019}.} \citep[e.g.,][]{Nenkova2002a, Dullemond2005a}.
\par
\begin{figure}
    \centering
    \includegraphics[width=\linewidth]{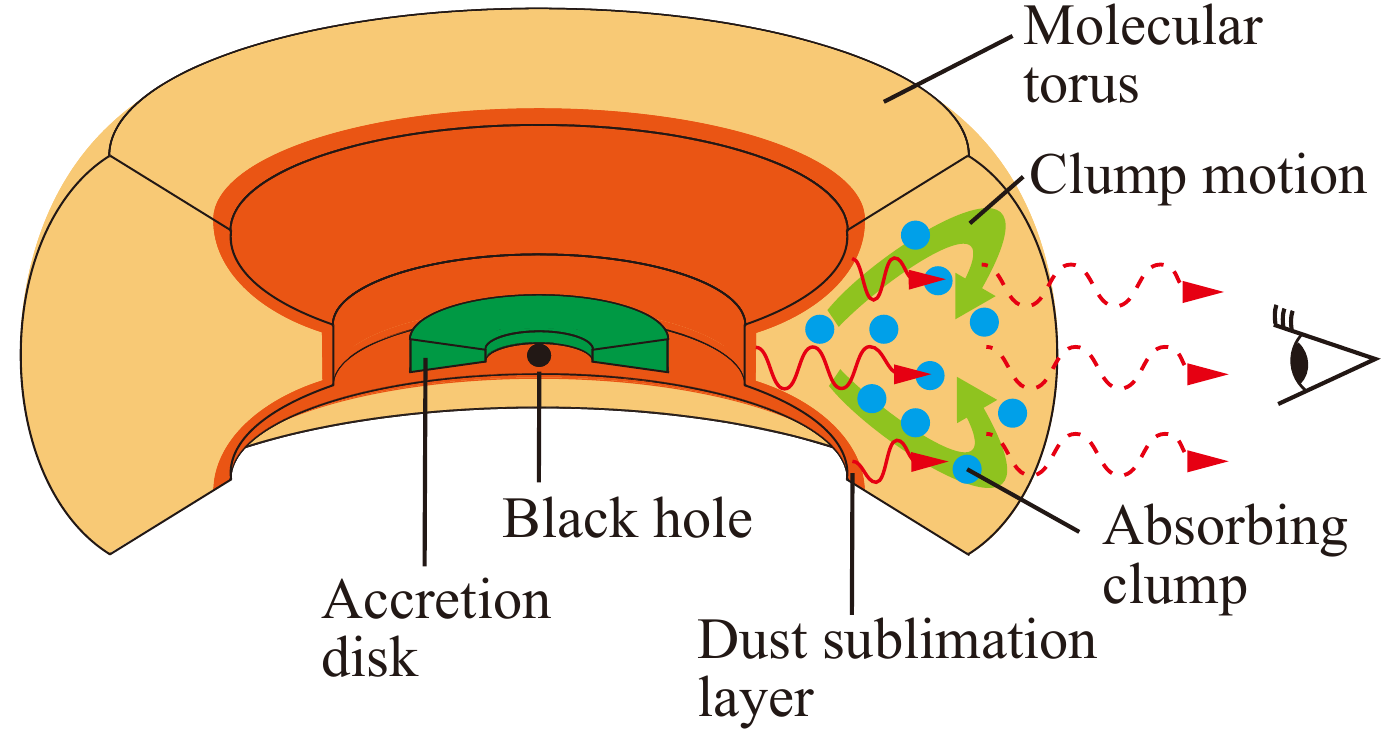}
    \caption{Schematic illustration of the geometry of the dust sublimation layer (the continuum source) and the absorbing clumps that cause CO rovibrational absorption lines (Reprinted from \citealp{Onishi2021}).}\label{fig:torus_clumpy}
\end{figure}

\citet{Wada2012a} and \citet{Wada2016} have proposed a radiation-hydrodynamic (RHD) model they call the ``radiation-driven fountain model,'' where the molecular torus is formed by outflowing and inflowing gas around the black hole and the accretion disk.
Other RHD and magnetohydrodynamic (MHD) simulations \citep[e.g.,][]{Namekata2016,Chan2017,Kudoh2020,Venanzi2020} also support this conclusion.
The molecular torus is then predicted to be not a static structure but instead a dynamic structure.
Thus, the clumps in the torus are expected to be inflowing or outflowing as shown in the simplified schematic image in Figure~\ref{fig:torus_clumpy}.
Therefore, observational studies of the dynamical and physical properties and of the spatial distributions of the clumps are required to clarify the inner structure of the torus.
\par
To determine the properties of the clumps, we have observed the absorption lines of the fundamental rovibrational transitions ($v=0\to 1,\ J\to J\pm{1}$) of gaseous carbon monoxide (CO) molecules at $\lambda\sim 4.67\,\micron$.
The CO rovibrational absorption lines have two advantages for this observational study.
First, we can preferentially observe CO absorption by the clumps in the torus, avoiding host-galaxy contamination because the main NIR continuum source at $\lambda\sim 4.67\,\micron$ is expected to be hot dust at the dust sublimation layer, the radius of which is $\lesssim 1\,\mathrm{pc}$ \citep[e.g.,][]{Rees1969a,Rieke1978,Barvainis1987,Landt2011,Mor2012,Ichikawa2014,Matsumoto2022}.
\citet{Matsumoto2022} performed a numerical simulation of the CO absorption in the central $32\,\mathrm{pc}$ region of the Circinus galaxy and indicated that more than 50\% of $4.7\,\micron$ continuum comes from the hot dust in the central $1.5\,\mathrm{pc}$ region when the inclination angle is $\le 70^\circ$ from the pole (see Figure~4 in \citealp{Matsumoto2022})\footnote{We should note that star clusters can also contribute to the nuclear activity in a luminous infrared galaxy NGC~4418 \citep{Varenius2014,Ohyama2019,Ohyama2023}, which is one of the targets in this paper; thus, a stellar contamination can exist in the $4.7\,\micron$ continuum of NGC~4418.}.
Figure~\ref{fig:torus_clumpy} illustrates the assumed geometry of the dust sublimation layer and the clumps inside the torus.
Second, a spectroscopic observation can simultaneously cover many absorption lines with multiple lower rotational levels $v=0,\ J$ owing to the crowdedness of the lines. 
This enables us to determine accurately the level populations at $v=0$ based on the optical depths of the lines and thus the physical properties (e.g., the excitation temperature and the CO column density) of absorbing clumps inside the torus.
\par
Some previous low-resolution spectroscopic observations of CO rovibrational absorption bands have supported the assumption that the absorption is caused predominantly by warm gas in the AGN, although individual transitions were not resolved.
For example, by comparing the Spitzer low-resolution ($R\sim 80$) spectrum of the CO rovibrational absorption band in the ultraluminous infrared galaxy (ULIRG) IRAS~00183$-$7111 with a local thermodynamic equilibrium (LTE) and isothermal slab model \citep{Cami2002}, \citet{Spoon2004} found that the absorbing CO gas was warm ($T\sim 720\,\mathrm{K}$).
By comparing low-resolution spectra from AKARI ($R\sim 150$) and Spitzer ($R\sim 80$) with the model of \citet{Cami2002}, \citet{Baba2018} also found that the absorbing CO gas had a warmer excitation temperature in 10 nearby ULIRGs ($T\sim 200\text{--}500\,\mathrm{K}$) than in a typical starburst galaxy ($T\lesssim 100\,\mathrm{K}$).
In addition, by combining radio interferometry from the Atacama large millimeter/submillimeter array (ALMA) and low-resolution NIR spectroscopy from AKARI ($R\sim 150$), \citet{Baba2022} found that the warm ($T\sim 1000\,\mathrm{K}$) CO gas from which the rovibrational absorption bands originate existed within the central $\sim 20\,\mathrm{pc}$ region of the AGN of the nearby ULIRG IRAS~17208$-$0014.
\par
Accordingly, if each rovibrational transition is resolved using high-resolution spectroscopy, by performing a velocity decomposition of each transition we can obtain dynamical properties, such as the line-of-sight (LOS) velocity and the velocity dispersion, of inflowing or outflowing clumps inside the torus.
Because the velocity dispersion increases near the central black hole, it enables determinations of the relative spatial distributions of the inflowing and outflowing clumps.
In addition, we can obtain physical properties of the clumps, such as the excitation temperature and the column density of CO molecules, based on the level populations derived from the optical depths of rovibrational transitions with multiple lower rotational levels.
For these reasons, high-resolution spectroscopy of the CO rovibrational absorption transitions provides a suitable probe into the inner structure of the molecular torus.
\par
In fact, some previous high-resolution spectroscopic observations have been successful in probing the dynamical and physical properties of clumps inside the tori of nearby AGNs.
In the northwest core of the ULIRG IRAS~08572+3915, each rovibrational transition of gaseous CO was resolved using high-resolution spectroscopy from the United Kingdom Infrared Telescope ($R\sim 7500$; \citealp{Geballe2006a}) and Subaru Telescope ($R\sim 5000, 10{,}000$; \citealp{Shirahata2013b} and \citealp{Onishi2021}, respectively).
\citet{Shirahata2013b} found three velocity components with the LOS velocity relative to the host galaxy of $V_\mathrm{LOS}\sim -160$ (outflowing), 0 (systemic), and $+100\,\mathrm{km\,s^{-1}}$ (inflowing) in each transition.
In addition, \citet{Onishi2021} performed a velocity decomposition of each rovibrational transition and found that the outflowing clumps existed in the innermost region of the torus, whereas the inflowing clumps existed in the outer regions based on the velocity dispersion of each velocity component.
They also found that the outer clumps had the low kinetic temperatures of $\sim 25\,\mathrm{K}$ and that they were radiatively excited by background radiation, whereas the innermost clumps had the high kinetic temperatures of $\sim 720\,\mathrm{K}$ and that they were collisionally excited, based on the level populations.
Based on high-resolution ($R\sim 5000$) spectroscopy from the Subaru Telescope, \citet{Ohyama2023} found that each transition in the luminous infrared galaxy (LIRG) NGC~4418 exhibited an extremely large CO column density ($N_\mathrm{CO}\sim 5\times 10^{23}\,\si{\per\cm\squared}$) and a warm excitation temperature ($T_\mathrm{ex}\sim 170\,\mathrm{K}$).
Recently, middle-resolution ($R\sim 3300$) spectroscopy of CO rovibrational transitions from the James Webb Space Telescope (JWST) are performed toward nearby AGNs in VV~114~E \citep{Gonzalez-Alfonso2024,Buiten2024} and NGC~3256 \citep{Pereira-Santaella2024}.
In particular, using the NIR spectrum of CO rovibrational transitions obtained with a 0\farcs{2} ($\sim 70\,\mathrm{pc}$ at the distance of VV~114~E) aperture, \citet{Gonzalez-Alfonso2024} performed velocity decomposition of each CO rovibrational absorption transition observed toward the ``s2'' core in the southwest nucleus of the merging galaxy VV~114~E.
They found three velocity components with different temperatures in the range $\sim 10\text{--}500\,\mathrm{K}$ and identified the s2 core as an AGN based on the extreme CO excitation\footnote{
    Note that \citet{Evans2022} and \citet{Rich2023} identified the adjacent core, which is the ``s1'' core in the notation of \citet{Gonzalez-Alfonso2024}, as an AGN based on mid-infrared colors and the extremely low equivalent width of polycyclic aromatic hydrocarbon emission, respectively.
    In addition, \citet{Buiten2024} found that the s1 core lacked the $2.3\,\micron$ CO bandhead ($\Delta{v}=\pm 2$) whereas the s2 core showed it.
    Because the $2.3\,\micron$ CO bandhead is the characteristic of cool stellar atmospheres, they identified the s1 core as an AGN and the s2 core as a star-forming region and attributed the extreme CO excitation to an intense local radiation field inside a dusty young massive star cluster.
    This interpretation of the extreme CO excitation is contrary to that of \citet{Gonzalez-Alfonso2024}.
    On the other hand, \citet{Buiten2024} identified the s1 core as an AGN, attributing the lack of highly excited CO molecules to geometric dilution of the intense radiation from a bright point source.
}.
As in \ireight{}, the hottest component was found to be outflowing.
\par
Despite the usefulness of the velocity-decomposition technique, it has been applied only to a limited number of galaxies (IRAS~08572+3915~NW and VV~114~E), and it is unclear whether the inner structures of tori are similar in other AGNs.
Hence, performing velocity decompositions toward more AGNs is essential for systematically investigating the inner structure of the molecular torus.
In this paper, we study the dynamical and physical properties of clumps inside tori and compare the inner structures of some AGNs.
To achieve this aim, we perform velocity decomposition of the CO rovibrational absorption transitions in the following four nearby LIRGs and ULIRGs: IRAS~01250+2832, UGC~5101, NGC~4418, and IRAS~08572+3915~NW, using high-resolution NIR spectroscopy with the spectral resolutions $R\sim 5000\text{--}10{,}000$, or, equivalently, the velocity resolutions $\Delta{V}\sim 30\text{--}60\,\kmps$.
\par
This is the first systematic velocity-decomposition study of the CO rovibrational absorption transitions in AGNs.
Section~\ref{sec:targets} describes the characteristics of the targets in this paper.
Then, Section~\ref{sec:obs_data} provides the information about the observational conditions and data reduction.
Section~\ref{sec:analyses} presents the velocity-decomposition results for the observed CO rovibrational absorption transitions.
Subsequently, Section~\ref{sec:discussion} estimates the spatial distributions of clumps attributed to each velocity component inside the tori of the targets.
Next, we compare the level populations of each velocity component with a non-LTE model to determine the physical properties of clumps, and we compare the derived properties with RHD and MHD models.
Finally, Section~\ref{sec:conclusion} presents the conclusions of this paper.
Throughout this paper, we adopt a flat cosmology with $\Omega_\mathrm{M}=0.307$, $\Omega_\mathrm{\Lambda}=1-\Omega_\mathrm{M}$, and $H_0=67.8\,\kmps\,\mathrm{Mpc^{-1}}$ of \citet{PlanckCollaboration2014}.

\section{Targets}\label{sec:targets}
As mentioned in Section~\ref{sec:intro}, we analyze the CO spectra observed for the following four LIRGs and ULIRGs with AGNs: IRAS~01250+2832, UGC~5101, NGC~4418, and IRAS~08572+3915~NW.
In this section, we first describe the characteristics of the four targets.
Table~\ref{tab:targ_info} summarizes the general parameters related to this paper.

\subsection{IRAS~01250+2832}\label{subsec:ir01}
IRAS~01250+2832 (hereafter referred to as \irone{}) is a LIRG\footnote{Strictly speaking, \irone{} does not satisfy the definition of a LIRG: $\log\qty(L_\mathrm{IR}/L_\odot)\ge 11$. However, we herein regard this source as a LIRG, in accordance with \citet{Oyabu2011} under a different cosmology from this paper.} whose infrared luminosity ($L_\mathrm{IR}\equiv L(8\text{--}1000\,\micron)$) is $\log\qty(L_\mathrm{IR}/L_\odot)=10.97$ \citep{Oyabu2011}.
\citet{Oyabu2011} found this object with a compact hot dust component first based on AKARI MIR All-Sky Survey.
Although they made an estimate of its redshift using the optical spectrum obtained with the Shane $3\,\mathrm{m}$ telescope at the Lick observatory, the uncertainty of the redshift was not evaluated.
As explained in Appendix~\ref{app:rs_ir01}, we thus reanalyzed the optical spectrum to determine the redshift of \irone{} and its corresponding uncertainty.
As a result, we found the redshift of \irone{} to be $z=0.04254\pm 0.00010\,(\mathrm{stat.})\pm 0.00002\,(\mathrm{sys.})$, corresponding to the luminosity distance of $D_\mathrm{L}=194\,\mathrm{Mpc}$ under the assumed cosmology.
The statistical error corresponds to the velocity uncertainty $c\Delta{z}=30\,\kmps$.
\par
\irone{} is likely to have an AGN at its center because its spectral energy distribution (SED) at $\lambda\sim 2\text{--}10\,\micron$ is well reproduced by a single blackbody with the high temperature $T\sim 500\,\mathrm{K}$, indicating an AGN-dominated NIR spectrum \citep{Oyabu2011}.
This conclusion is also supported by the lack of $3.3\,\micron$ polycyclic aromatic hydrocarbon (PAH) emission feature \citep{Oyabu2011,Yamada2013a}.
A $3.1\,\micron$ $\mathrm{H_2O}$-ice absorption with an optical depth $\tau_{3.1}\sim 0.18$ \citep{Yamada2013a} indicates that the AGN is obscured to some extent.

\subsection{UGC~5101}\label{subsec:u51}
UGC~5101 (hereafter referred to as \ufive{}) is a ULIRG with an infrared luminosity $\log\qty(L_\mathrm{IR}/L_\odot)=12.02$ \citep{Sanders1996,Sanders2003}.
The redshift determined from the optical lines is $z=0.03940$ \citep{Abazajian2004}, which corresponds to the luminosity distance $D_\mathrm{L}=179\,\mathrm{Mpc}$ under the assumed cosmology.
The uncertainty in the redshift is $\Delta{z}=4\times 10^{-5}$, or $c\Delta{z}=12\,\kmps$.
\par
\ufive{} is likely to have an AGN at its center based on a large hardness ratio $-0.29\pm 0.05$ between X-ray emission above and below $2\,\mathrm{keV}$ \citep{Iwasawa2011} and the detections of the AGN transmitted component in the hard X-ray band above $10\,\mathrm{keV}$ \citep{Oda2017}.
In contrast to \ireight{} and \irone{}, a $3.3\,\micron$ PAH emission feature is detected in this galaxy with a rest equivalent width of $\sim 33\,\mathrm{nm}$ \citep{Imanishi2008}, indicating that it is undergoing star-formation activity.
Thus, this object is probably an AGN-starburst composite.
The PAH emission from this source is likely to originate in the foreground of the attenuating dust, and the emission from behind it is unlikely to show PAH emission \citep{Imanishi2001}.
Similarly, the apparent $9.7\,\micron$ silicate dust absorption feature exhibits a smaller optical depth, $\tau_{9.7}\sim 1.9$ \citep{Dartois2007}, than that in \ireight{}, likely because of unabsorbed mid-infrared (MIR) emission \citep{Dietrich2018}, as indicated in other AGNs \citep{Aalto2015,Gonzalez-Alfonso2015}.
The $3.1\,\micron$ $\mathrm{H_2O}$-ice and the $3.4\,\micron$ carbonaceous dust absorption features, with $\tau_{3.1}\sim 1.0$ and $\tau_{3.4}\sim 0.6$, respectively, indicate a more heavily obscured AGN, with a dust extinction $A_V=102\text{--}166$ \citep{Imanishi2008}.
The dust extinction corresponds to $N_\mathrm{H}\sim (2\text{--}3)\times 10^{23}\,\mathrm{cm^{-2}}$ if we assume $N_\mathrm{H}/A_V=1.8\times 10^{21}\,\mathrm{cm^{-2}}$ \citep{Bohlin1978a}.
Model fitting to the $\sim 0.3\text{--}100\,\mathrm{keV}$ X-ray spectrum also indicates a large LOS hydrogen column density: $N_\mathrm{H,X}=(9.5\text{--}16)\times 10^{23}\,\mathrm{cm^{-2}}$ \citep{Oda2017,Yamada2021}.
In addition, model fitting to the infrared SED yields an AGN luminosity $\sim 70\%$ of the infrared luminosity, and the fraction obtained at $\lambda\sim 5\,\micron$ is similar to that, according to the best-fit model \citep{Dietrich2018}.

\subsection{NGC~4418}\label{subsec:n44}
NGC~4418 (hereafter referred to as \nfour{}) is a LIRG with the infrared luminosity $\log\qty(L_\mathrm{IR}/L_\odot)=11.06$ \citep{Sanders1996,Sanders2003}.
The redshift determined from the optical lines is $z=0.007085$ \citep{Albareti2017}, which corresponds to the luminosity distance $D_\mathrm{L}=31.5\,\mathrm{Mpc}$ under the assumed cosmology.
The uncertainty in the redshift is $\Delta{z}=9\times 10^{-6}$, or $c\Delta{z}=3\,\kmps$.
\par
\nfour{} is likely to have an AGN at its center based on the warm MIR-to-far-infrared (FIR) color $S(25\,\micron)/S(60\,\micron)=0.22\,(>0.1)$ \citep{DeGrijp1985,Sanders1988a,Sanders2003} and the large ratio of the $\mathrm{H_2}\ v=1\to 0\ S(1)$ line to the $\mathrm{H_I\ Br\gamma}$ line, $L_\mathrm{S(1)}/L_\mathrm{Br\gamma}>8$ \citep{Mouri1992,Imanishi2004}.
Similar to \ufive{} and in contrast to \ireight{} and \irone{}, the $3.3\,\micron$ PAH emission feature is detected in this source with a rest equivalent width of $\sim 11.5\text{--}16\,\mathrm{nm}$ \citep{Imanishi2010b,Yamada2013a}, indicating the presence of star-formation activity.
Thus, this object also is probably an AGN-starburst composite.
In addition, a radio observation exhibits signs of a nuclear starburst, which is the detection of possible super star clusters within a $\sim 41\,\mathrm{pc}$ region in the nucleus \citep{Varenius2014}, in contrast to \ufive{}, for which star formation occurs in the outer regions.
The AGN in this LIRG is the most heavily obscured of the targets in this work, with a $9.7\,\micron$ silicate dust absorption feature having the optical depth $\tau_{9.7}\sim 5.6\text{--}7.8$ \citep{Roche1986,Dudley1997,Spoon2001,Stierwalt2013,Roche2015}; this corresponds to $N_\mathrm{H,9.7}\sim (0.9\text{--}3)\times 10^{23}\,\mathrm{cm^{-2}}$, as shown in Table~\ref{tab:targ_info}.
In addition, the non-detection of hard X-ray photons in \nfour{} indicates a Compton-thick AGN with $N_\mathrm{H,X}\gg 10^{24}\,\mathrm{cm^{-2}}$ \citep{Maiolino2003,Yamada2021}.
\par
This object has been well studied using radio interferometry with ALMA, and the compact core of the optically thick warm-dust continuum is observed to have a FWHM $\sim 20\,\mathrm{pc}$, a brightness temperature $\sim 400\,\mathrm{K}$, and an $\mathrm{H_2}$ column density $\sim 10^{25.7}\,\mathrm{cm^{-2}}$ \citep{Sakamoto2021a}.
In addition, bright vibrationally excited HCN emission is detected at (sub)millimeter wavelengths in the compact nuclear region of this source, with a diameter of $\sim 30\,\mathrm{pc}$ \citep{Imanishi2018a,Falstad2021}, which also indicates that the core is heavily obscured and that trapped MIR photons vibrationally excite the HCN molecules \citep{Gonzalez-Alfonso2019}.

\subsection{IRAS 08572+3915 NW}\label{subsec:target_ir08}
Observations of CO rovibrational lines toward IRAS~08572+3915 (hereafter referred to as IR08572) were published by \citet{Geballe2006a,Shirahata2013b,Onishi2021}.
In this paper, we reanalyze this target to reduce the spectra in a common manner with that for the other targets.\par
We here summarize the characteristics of IR08572 briefly; see Section~2 of \citet{Onishi2021} for details.
IR08572 is a ULIRG whose infrared luminosity is $\log\qty(L_\mathrm{IR}/L_\odot)=12.18$ \citep{Sanders1996,Sanders2003}.
Its redshift is $z=0.0583$ \citep{Evans2002}, which corresponds to the luminosity distance $D_\mathrm{L}=269\,\mathrm{Mpc}$ under the assumed cosmology.
Although the uncertainty in this redshift is not evaluated in \citet{Evans2002}, we conservatively assume it to be $\Delta{z}=5\times 10^{-5}$, or $c\Delta{z}=15\,\kmps$ based on the number of significant digits.
IR08572 has two NIR cores separated by $5\farcs4=5.6\,\mathrm{kpc}$ \citep{Scoville2000} toward the northwest (NW) and southeast (SE) directions, and approximately $80\%$ of the MIR luminosity comes from the NW core \citep{Soifer2000}.
\par
In this paper, we focus on the NW core.
\ireight{} is classified as an AGN based on the lack of a $3.3\,\micron$ PAH emission feature \citep{Imanishi2008} and on the detection of hard X-ray photons, despite their small number \citep{Iwasawa2011,Yamada2021}.
The AGN is likely to be heavily dust-obscured based on the ice and carbonaceous dust absorption features at $3.1$ and $3.4\,\micron$, respectively \citep{Dudley1997,Dartois2007,Imanishi2007,Stierwalt2013,Imanishi2008,Doi2019}.
In addition, fitting the XCLUMPY \citep{Tanimoto2019} model to the $0.3\text{--}20\,\mathrm{keV}$ X-ray spectrum indicates the large LOS hydrogen column density $N_\mathrm{H,X}=(8.5^{+12.9}_{-2.8})\times 10^{23}\,\mathrm{cm^{-2}}$ \citep{Yamada2021}.
Further, model fitting to the infrared SED predicts the AGN luminosity to be $\sim 90\%$ of the total infrared luminosity, and other contributions to the NIR continuum at $\lambda\sim 5\,\micron$ are likely to be negligible, according to the best-fit model \citep{Dietrich2018}.

\begin{splitdeluxetable*}{cccccccBccccc}
\tablecaption{General Parameters of the Targets\label{tab:targ_info}}
\centering
\tablehead{
\colhead{Target} & \colhead{$z$} & \colhead{$c\Delta{z}$} & \colhead{$D_\mathrm{L}$} & \colhead{$\log(L_\mathrm{IR}/L_\odot)$} & \colhead{$f_\mathrm{AGN}$} &\colhead{Rest $\mathrm{EW_{3.3}}$} & \colhead{$\tau_{3.1}$} & \colhead{$\tau_\mathrm{3.4}$} & \colhead{$\tau_{9.7}$} & \colhead{$N_\mathrm{H,9.7}$} & \colhead{$N_\mathrm{H,X}$}\\
\colhead{} & \colhead{} & \colhead{($\kmps$)} & \colhead{(Mpc)} & \colhead{} & \colhead{} & \colhead{(nm)} & \colhead{} & \colhead{} & \colhead{} & \colhead{($10^{23}\,\mathrm{cm^{-2}}$)} & \colhead{($10^{23}\,\mathrm{cm^{-2}}$)}
}
\decimalcolnumbers%
\startdata
{\irone{}} & {0.04254} & {30} & {194} & {10.97} & {\ldots} & {$<21$\tablenotemark{b}} & {0.18\tablenotemark{b}} & {\ldots} & {\ldots} & {\ldots} & {\ldots}\\
{\ufive{}} & {0.03940} & {12} & {179} & {12.02} & {$0.76\pm 0.04$} & {33\tablenotemark{a}} & {1.0\tablenotemark{a}} & {0.6\tablenotemark{a}} & {1.9\tablenotemark{f}} & {$0.3\text{--}0.7$} & {$9.5\text{--}16$\tablenotemark{k,l}}\\
{\nfour{}} & {0.007085} & {3} & {31.5} & {11.06} & {\ldots} & {$11.5\text{--}16$\tablenotemark{b,c}} & {\ldots} & {\ldots} & {$5.6\text{--}7.8$\tablenotemark{d,e,g,h,i}} & {$0.9\text{--}3$} & {$>10$\tablenotemark{m}}\\
{IR08572} & {0.0583} & {15} & {269} & {12.18} & {$0.91\pm 0.05$} & {$<5$\tablenotemark{a}} & {0.3\tablenotemark{a}} & {0.8\tablenotemark{a}} & {$3.8\text{--}5.2$\tablenotemark{d,e,f,n}} & {$0.6\text{--}2$} & {$>10$, $5.7\text{--}21.4$ \tablenotemark{j,k}}\\
\enddata%
\tablecomments{
Column (1): target name. Columns (2) and (3): redshift and corresponding velocity uncertainty. References are \citet{Abazajian2004}, \citet{Albareti2017}, and \citet{Evans2002} for \ufive{,} \nfour{,} and IR08572, respectively. The redshift of \irone{} is estimated in this work (Appendix~\ref{app:rs_ir01}). Column (4): luminosity distance based on the redshift. Column (5): infrared ($8\text{--}1000\,\micron$) luminosity based on the IRAS flux \citep{Sanders2003} and the formula in Table 1 of \citet{Sanders1996}, except for \irone{}. (See text for the details.) Column (6): AGN fraction to the infrared luminosity based on the model fitting to SED \citep{Dietrich2018}. Column (7): rest equivalent width of the $3.3\,\micron$ PAH emission feature. Column (8): optical depth of the $3.1\,\micron$ $\mathrm{H_2O}$ ice absorption feature. Column (9): optical depth of the $3.4\,\micron$ carbonaceous dust absorption feature. Column (10): optical depth of the $9.7\,\micron$ silicate dust absorption feature. Column (11): LOS hydrogen column density based on the optical depth of silicate dust with $A_V/\tau_\mathrm{9.7}= 9\text{--}21$ \citep{Draine2003} and $N_\mathrm{H}/A_V=1.8\times 10^{21}\,\mathrm{cm^{-2}}$ \citep{Bohlin1978a} assumed. Column (12): LOS hydrogen column density based on the X-ray spectra.}
\tablerefs{${}^{a}$\citet{Imanishi2008}, ${}^{b}${\citet{Yamada2013a}}, ${}^{c}${\citet{Imanishi2010b}}, ${}^{d}${\citet{Dudley1997}}, ${}^{e}${\citet{Stierwalt2013}}, ${}^{f}${\citet{Dartois2007}}, ${}^{g}${\citet{Roche1986}}, ${}^{h}${\citet{Spoon2001}}, ${}^{i}${\citet{Roche2015}}, ${}^{j}${\citet{Iwasawa2011}}, ${}^{k}${\citet{Yamada2021}}, ${}^{l}${\citet{Oda2017}}, ${}^{m}${\citet{Maiolino2003}}, ${}^{n}${\citet{Imanishi2007}}}
\end{splitdeluxetable*}

\section{Observations and Data Reduction}\label{sec:obs_data}
This section explains the observational configurations and the methodology used in reducing the CO spectra observed toward the targets.

\subsection{Observational Configurations}
Using the Infrared Camera and Spectrograph (IRCS) on the Subaru Telescope, we performed $M$-band echelle spectroscopy toward \irone{}, \ufive{}, \nfour{}, and \ireight{}.
Table~\ref{tab:obsinfo} summarizes the observational information about these spectral data.
In observations toward \irone{} and \ireight{}, we used the $0\farcs27\times 9\farcs37$ aperture, which corresponds to the spectral resolution $R\sim 10{,}000$, or $\Delta{V}\sim 30\,\mathrm{km\,s^{-1}}$.
The slit size corresponds to the physical scales of $0.25\,\mathrm{kpc}\times 8.8\,\mathrm{kpc}$ and $0.35\,\mathrm{kpc}\times 12\,\mathrm{kpc}$ in \irone{} and \ireight{}, respectively.
Alternatively, in observations toward \ufive{} and \nfour{}, we used the $0\farcs54\times 9\farcs37$ aperture, which corresponds to the spectral resolution $R\sim 5000$, or $\Delta{V}\sim 60\,\mathrm{km\,s^{-1}}$.
The slit size corresponds to the physical scales of $0.47\,\mathrm{kpc}\times 8.1\,\mathrm{kpc}$ and $0.041\,\mathrm{kpc}\times 1.4\,\mathrm{kpc}$ in \ufive{} and \nfour{}, respectively.
Because the slit width is larger than a typical AGN scale of a few parsecs, NIR emission from star forming regions around the AGN core can not be completely excluded.
However, we assume that the NIR spectra from the central region of an AGN are preferentially observed because the dominant NIR continnum source at $\sim 4.67\,\micron$ is expected to be hot dust at the dust sublimation layer \citep{Matsumoto2022} as mentioned in Section~\ref{sec:intro}.
In the observations toward \ireight{}, we positioned the slit at the position angle $\text{PA}=55^\circ$ east of north to avoid the SE core, whereas we positioned it to $\text{PA}=0^\circ$ in observations toward the other targets.
\par
In addition, for some observations, we used adaptive optics (AO) with natural guide stars for \irone{} and \ufive{} and with laser guide stars for \ireight{} to improve the signal-to-noise ratio ($S/N$).
After wavefront correction by the AO system, the seeing sizes were reduced to approximately half or less of the natural seeing, as shown in Table~\ref{tab:obsinfo}.
\par
For sky subtraction, we performed all observations in the ABBA nodding mode, in which the telescope was nodded along the slit for 3\farcs0 (2004 and 2011), 3\farcs7 (2005), or 4\farcs0 (2009, 2010, and 2019) to place the target on positions A and B of the slit in turn.
For wavelength calibration and correcting for telluric transmission, we also observed standard stars with the early spectral types B or A so that the spectral shapes can be approximated well by a blackbody.
Table~\ref{tab:std_stars} summarizes the standard-star parameters.
The details of sky subtraction, wavelength calibration, and correction for telluric transmission are described in Section~\ref{subsec:extract_1dspec}.
\par
To determine the gas temperature accurately (Section~\ref{subsec:excite}) based on the level populations of $\mathrm{CO}$ molecules in warm ($T\sim 200\text{--}500\,\mathrm{K}$; \citealp{Baba2018}) gas, the observed wavelength ranges cover the CO rovibrational transitions for the high rotational levels $J\ge 18$ in all targets.
The included rotational levels are $0\text{--}58$, $0\text{--}58$, $0\text{--}18$, and $J=0\text{--}26$ for \irone{,} \ufive{,} \nfour{,} and \ireight{,} respectively, as shown in Table~\ref{tab:obsinfo}.

\subsection{Data Reduction}
\subsubsection{Extraction of one-dimensional raw spectra}\label{subsec:extract_1dspec}
We extracted one-dimensional raw spectra from the slit images of the objects and their standard stars using IRAF v2.16.1 \citep{Tody1986,Tody1993} via PyRAF v2.1.15 \citep{Pyraf} in the standard manner, as follows:
\begin{enumerate}
\item
    Slit images for which the target was in the A (B) position of the slit were stacked to reduce the noise.
    Then, we subtracted the stacked A and B images from each other to remove background sky emission.
    We thus obtained two subtracted spectral images, with the target in the positions A and B of the slit.
\item
    We corrected differences between the responses of pixels in the infrared detector array on the basis of the slit images illuminated by the calibration lamp (flat fielding).
    Subsequently, the signals for bad pixels, where the response was too low, too high, or deviating, or where the dark current was too high or deviating, were interpolated from signals of the surrounding pixels.
\item
    We extracted a one-dimensional raw spectrum from the two-dimensional slit image, which was stacked and bad-pixel cleaned using the \verb|apall| task of IRAF.
    These procedures yield two one-dimensional spectra without wavelength calibration for the A and B positions.
\end{enumerate}

\subsubsection{Wavelength calibration and correction of telluric features}
To minimize the systematic error caused by wavelength calibration, we fit the telluric absorption lines imprinted in the spectra of the standard stars with the telluric line model using the Molecfit v1.5.9 package \citep{Kausch2015,Smette2015}.
Table~\ref{tab:std_stars} summarizes the standard-star parameters.
The resulting accuracy is approximately one pixel width or less ($\lesssim 8\,\kmps$).
\par
After wavelength calibration, the object spectra are divided by the standard-star spectra and multiplied by the blackbody spectra with the corresponding effective temperature ($T_\mathrm{eff}$) to correct for transmission through the telluric atmosphere and for throughput bias.
To keep the differences small between the spectral shapes of the telluric-transmission lines in the spectra from the objects and from the standard stars, we chose the standard stars such that the air-mass differences between them and the target objects were $\lesssim 0.2$.
After that, we converted the wavelengths to local standard-of-rest (LSR) values using the \verb|rvcorrect| task of IRAF.

\subsubsection{Estimation of Flux Errors}\label{subsec:flux_error}
We evaluated flux errors as follows:
First, we separated the observed slit images for each configuration into four independent groups based on whether the slit was in the A or B position of the ABBA nodding sequence and whether the data were derived in the former (f) or latter (l) half of each observational sequence.
Second, we reduced each group (A-f, A-l, B-f, and B-l) to four one-dimensional spectra, as described in the preceding sections.
Standard stars for each group are shown in Table~\ref{tab:std_stars}.
Finally, we adopt the mean of these four fluxes as the flux value and take the uncertainty of the mean as the flux error, using Student's $t$-distribution with three degrees of freedom.

\begin{splitdeluxetable*}{ccccccccBccccc}
\tablecaption{Observation Log of Data Used in This Paper\label{tab:obsinfo}}
\tablehead{
\colhead{No.} & \colhead{Date (UT)} & \colhead{Target} & \colhead{Slit Size} & \colhead{$R$} & \colhead{(ECH, XDS)} & \colhead{$\lambda_{11}\,(\micron)$} & \colhead{$\lambda_{12}\,(\micron$)} & \colhead{No.} & \colhead{$J$} & \colhead{IT (min.)} & \colhead{AO} & \colhead{$\Delta_{\mathrm{s},K}$ ($\arcsec$)}
}
\decimalcolnumbers%
\startdata
1 & 2010/3/1 & \ireight{} & $0\farcs27\times 9\farcs37$ & {10000} & $(-3200, -5500)$ & \ldots & 4.73--4.84 & 1 & {$R$: 11--26} & 72 & No & 0.8\\
2 & 2019/1/19 & \ireight{} & $0\farcs27\times 9\farcs37$ & {10000} & $(6500, -6100)$ & 5.00--5.13 & \ldots & 2 & {$P$: 7--19} & 160 & Yes & 0.3\\
3 & 2019/1/20 & \ireight{} & $0\farcs27\times 9\farcs37$ & {10000} & $(12000, -5650)$ & 4.90--5.04 & \ldots & 3 & {$R$: 0--3, $P$: 1--10} & 128 & Yes & 0.4\\
4 & 2019/1/20 & \ireight{} & $0\farcs27\times 9\farcs37$ & {10000} & $(-10000, -6100)$ & \ldots & 4.82--4.92 & 4 & {$R$: 2--13} & 100.8 & Yes & 0.4\\
5 & 2009/8/26 & \irone{} & $0\farcs27\times 9\farcs37$ & {10000} & $(-10000, -5000)$ & 5.26--5.37 & 4.82--4.92 & 5 & {$R$: 0--4, $P$: 1--6, 37--44} & {68} & {Yes} & {0.3}\\
6 & 2009/8/28 & \irone{} & $0\farcs27\times 9\farcs37$ & {10000} & $(6500, -6100)$ & 5.00--5.13 & 4.59--4.71 & 6 & {$R$: 20--41, $P$: 15--26} & {60} & {Yes} & {0.2}\\
7 & 2009/12/14 & \irone{} & $0\farcs27\times 9\farcs37$ & {10000} & $(-3200, -5500)$ & 5.16--5.28 & 4.73--4.84 & 7 & {$R$: 2--16, $P$: 29--37} & {60} & {Yes} & {0.2}\\
8 & 2011/8/24 & \irone{} & $0\farcs27\times 9\farcs37$ & {10000} & $(3200, -5500)$ & 5.06--5.19 & 4.64--4.75 & 8 & {$R$: 14--30, $P$: 20--30} & {46.2} & {Yes} & {0.1}\\
9 & 2011/10/24 & \irone{} & $0\farcs27\times 9\farcs37$ & {10000} & $(12000, -5650)$ & 4.90--5.04 & 4.50--4.62 & 9 & {$R$: 35--58, $P$: 5--18} & {100.8} & {Yes} & {0.1}\\
10 & 2004/10/23 & \ufive{} & $0\farcs54\times 9\farcs37$ & {5000} & $(-8500, -5500)$ & \ldots & 4.80--4.91 & 10 & {$R$: 0--4, $P$: 1--6} & {28} & {No} & {0.8\tablenotemark{a}}\\
11 & 2004/10/24 & \ufive{} & $0\farcs54\times 9\farcs37$ & {5000} & $(-8500, -5500)$ & \ldots & 4.80--4.91 & 11 & {$R$: 0--4, $P$: 1--6} & {36} & {No} & {0.8\tablenotemark{a}}\\
12 & 2005/5/27 & \ufive{} & $0\farcs54\times 9\farcs37$ & {5000} & $(-10000, -5500)$ & \ldots & 4.82--4.92 & 12 & {$R$: 0--2, $P$: 1--7} & {32} & {No} & {0.8}\\
13 & 2009/12/14 & \ufive{} & $0\farcs54\times 9\farcs37$ & {5000} & $(6500, -6100)$ & 5.00--5.13 & 4.59--4.71 & 13 & {$R$: 18--38, $P$: 16--27} & {80} & {Yes} & {0.3}\\
14 & 2009/12/14 & \ufive{} & $0\farcs54\times 9\farcs37$ & {5000} & $(12000, -6100)$ & 4.90--5.04 & 4.50--4.62 & 14 & {$R$: 33--58, $P$: 6--19} & {56} & {Yes} & {0.3}\\
15 & 2010/2/28 & \nfour{} & $0\farcs54\times 9\farcs37$ & {5000} & $(1000, -5800)$ & \ldots & 4.67--4.78 & 15 & {$R$: 0--2, $P$: 1--9} & {68} & {No} & {0.8}\\
16 & 2010/3/1 & \nfour{} & $0\farcs54\times 9\farcs37$ & {5000} & $(-6000, -5400)$ & \ldots & 4.77--4.88 & 16 & {$P$: 9--18} & {68} & {No} & {0.8}\\
\enddata%

\tablecomments{
Columns (1) and (9): data number in this paper. Column (2): observation date. Column (3): target name. Column (4): the width $\times$ the length of the slit used in the observation. Column (5): spectral resolution. The velocity resolutions corresponding to the spectral resolutions of $R\sim 5000$ and 10,000 are $\Delta{V}\sim 60$ and $30\,\kmps$, respectively. Column (6): unique configuration number for the angles of the echelle grating (ECH) and the cross disperser (XDS) of Subaru IRCS. Columns (7), (8): observed wavelength ranges whose echelle orders are $m=11$ and $12$, respectively. Unused ranges are denoted with ellipsis dots. Column (10): The lower rotational levels of $R$ and $P$ branches of ${}^{12}$CO rovibrational absorption lines, which are included in the observed wavelength ranges. Column (11): on-source integration time. Column (12): with or without AO. Column (13): FWHM seeing size in $K$ band.
}

\tablenotetext{a}{Because the seeing size in $K$ ($\lambda\sim 2.2\,\mathrm{\mu{m}}$) band is not measured, it is inferred from the auto-guider (AG) seeing ($\lambda\sim 0.73\,\mathrm{\mu{m}}$) based on the assumption that the seeing size varies as $\propto\lambda^{-0.2}$ \citep{Fried1974, Fried1975, Fried1977}.}

\end{splitdeluxetable*}

\begin{deluxetable*}{ccccccccc}
\tablecaption{Data Groups and Their Standard Stars\label{tab:std_stars}}
\tablehead{
\colhead{No.} & \colhead{Target} & \colhead{Group} & \colhead{$m_\mathrm{air,obj}$} & \colhead{STD Star} & \colhead{$M_V$} & \colhead{Type} & \colhead{$T_\mathrm{eff}$ (K)} & \colhead{$m_\mathrm{air,std}$}
}
\decimalcolnumbers
\startdata
1 & \ireight{} & A-f, B-f & {$1.10^{+0.11}_{-0.03}$} & HR 2088 & 1.90 & A2 & 8840 & $1.17\pm 0.01$\\
1 & \ireight{} & A-l, B-l & {$1.21^{+0.11}_{-0.07}$} & HR 2088 & 1.90 & A2 & 8840 & $1.17\pm 0.01$\\
2 & \ireight{} & A-f, B-f & {$1.07^{+0.04}_{-0.01}$} & HR 4534 & 2.14 & A3 & 8550 & $1.11^{+0.02}_{-0.01}$\\
2 & \ireight{} & A-l, B-l & {$1.25^{+0.30}_{-0.14}$} & HR 3982 & 1.35 & B7 & 14000 & $1.37\pm 0.06$\\
3 & \ireight{} & A-f, B-f & {$1.17^{+0.17}_{-0.08}$} & HR 0936 & 2.12 & B8 & 12500 & $1.26\pm 0.02$\\
3 & \ireight{} & A-l, B-l & {$1.07^{+0.02}_{-0.01}$} & HR 4534 & 2.14 & A3 & 8550 & $1.13\pm 0.02$\\
4 & \ireight{} & A-f, B-f & {$1.15^{+0.08}_{-0.05}$} & HR 4534 & 2.14 & A3 & 8550 & $1.09\pm 0.01$\\
4 & \ireight{} & A-l, B-l & {$1.36^{+0.18}_{-0.12}$} & HR 3982 & 1.35 & B7 & 14000 & $1.33^{+0.06}_{-0.05}$\\
5 & \irone{} & A-f, B-f & {$1.02^{+0.07}_{-0.01}$} & HR 0936 & 2.12 & B8 & 12500 & $1.07\pm 0.01$\\
5 & \irone{} & A-l, B-l & {$1.03^{+0.03}_{-0.01}$} & HR 0936 & 2.12 & B8 & 12500 & $1.07\pm 0.01$\\
6 & \irone{} & A-f, B-f & {$1.02^{+0.02}_{-0.01}$} & HR 7557 & 0.77 & A7 & 7800 & $1.09\pm 0.01$\\
6 & \irone{} & A-l, B-l & {$1.05^{+0.05}_{-0.03}$} & HR 7557 & 0.77 & A7 & 7800 & $1.09\pm 0.01$\\
7 & \irone{} & A-f, B-f & {$1.02\pm 0.01$} & HR 8781 & 2.49 & B9 & 10700 & $1.04\pm 0.01$\\
7 & \irone{} & A-l, B-l & {$1.09^{+0.13}_{-0.05}$} & HR 8781 & 2.49 & B9 & 10700 & $1.04\pm 0.01$\\
8 & \irone{} & A-f, B-f & {$1.02^{+0.03}_{-0.01}$} & HR 0936 & 2.12 & B8 & 12500 & $1.07\pm 0.01$\\
8 & \irone{} & A-l, B-l & {$1.02^{+0.05}_{-0.01}$} & HR 0936 & 2.12 & B8 & 12500 & $1.07\pm 0.01$\\
9 & \irone{} & A-f, B-f & {$1.02^{+0.03}_{-0.01}$} & HR 0936 & 2.12 & B8 & 12500 & $1.08\pm 0.01$\\
9 & \irone{} & A-l, B-l & {$1.05^{+0.07}_{-0.03}$} & HR 0936 & 2.12 & B8 & 12500 & $1.08\pm 0.01$\\
10 & \ufive{} & A-f, B-f & {$1.66^{+0.07}_{-0.05}$} & HR 2618 & 1.50 & B2 & 20600 & $1.80^{+0.09}_{-0.06}$\\
10 & \ufive{} & A-l, B-l & {$1.55\pm 0.04$} & HR 2618 & 1.50 & B2 & 20600 & $1.80^{+0.09}_{-0.06}$\\
11 & \ufive{} & A-f, B-f & {$1.69\pm 0.09$} & HR 2653 & 3.02 & B3 & 17000 & $1.66\pm 0.04$\\
11 & \ufive{} & A-l, B-l & {$1.52^{+0.06}_{-0.05}$} & HR 2653 & 3.02 & B3 & 17000 & $1.66\pm 0.04$\\
12 & \ufive{} & A-f, B-f & {$1.73^{+0.07}_{-0.06}$} & HR 4295 & 2.37 & A1 & 9200 & $1.55\pm 0.03$\\
12 & \ufive{} & A-l, B-l & {$1.90^{+0.09}_{-0.08}$} & HR 5028 & 2.75 & A2 & 8840 & $1.95\pm 0.03$\\
13 & \ufive{} & A-f, B-f & {$1.56\pm 0.01$} & HR 4905 & 1.77 & A0 & 9700 & $1.31\pm 0.01$\\
13 & \ufive{} & A-l, B-l & {$1.39^{+0.06}_{-0.04}$} & HR 4905 & 1.77 & A0 & 9700 & $1.31\pm 0.01$\\
14 & \ufive{} & A-f, B-f & {$1.34\pm 0.01$} & HR 4905 & 1.77 & A0 & 9700 & $1.29\pm 0.01$\\
14 & \ufive{} & A-l, B-l & {$1.35^{+0.03}_{-0.01}$} & HR 4905 & 1.77 & A0 & 9700 & $1.29\pm 0.01$\\
15 & \nfour{} & A-f, B-f & {$1.09^{+0.06}_{-0.02}$} & HR 5685 & 2.61 & B8 & 12500 & $1.15^{+0.06}_{-0.01}$\\
15 & \nfour{} & A-l, B-l & {$1.28^{+0.19}_{-0.12}$} & HR 5685 & 2.61 & B8 & 12500 & $1.15^{+0.06}_{-0.01}$\\
16 & \nfour{} & A-f, B-f & {$1.08^{+0.05}_{-0.01}$} & HR 5685 & 2.61 & B8 & 12500 & $1.16\pm 0.01$\\
16 & \nfour{} & A-l, B-l & {$1.24^{+0.15}_{-0.09}$} & HR 5685 & 2.61 & B8 & 12500 & $1.16\pm 0.01$\\
\enddata
\tablecomments{Column (1): data number in this paper, corresponding to that in Table~\ref{tab:obsinfo}. Column (2): corresponding object name. Column (3): data group where the spectrum of the standard star is used. Column (4): median and range of air mass of objects. Column (5): names of standard stars. Columns (6) and (7): $V$-band magnitude and spectral types of the standard stars, respectively \citep{Hoffleit1995}. Column (8): effective temperatures of the standard stars \citep{Pecaut2013}. The effective temperatures of the main sequence stars of corresponding types are used in this paper. Column (9): median and range of air mass of standard stars.}
\end{deluxetable*}

\section{Analyses of the Spectra}\label{sec:analyses}

This section explains the methods used to determine the continuum for each CO spectrum and to decompose the velocity components for each transition.

\subsection{Continuum determination and derived spectra}\label{subsec:continuum}
\begin{figure}
    \centering
    \includegraphics[width=\linewidth]{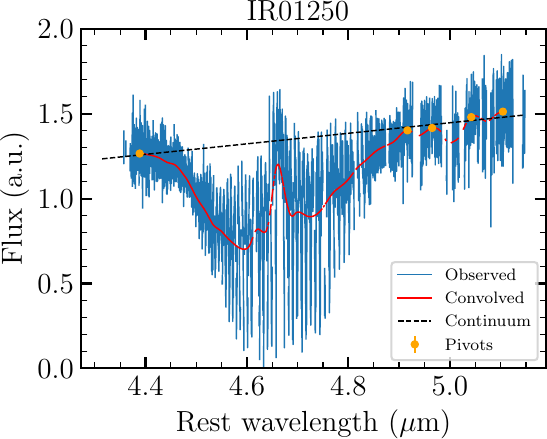}
    \caption{
        Adopted continuum of \irone{.}
        The observed high-dispersion spectrum and that convolved to $R\sim 180$ are colored blue and red, respectively.
        The dashed black lines and orange dots represent the adopted continuum and the pivots used to estimate it, respectively.
    }\label{fig:contfit_ir01}
\end{figure}

We determined the continuum for \irone{} using Subaru spectra because the Subaru observations cover wavelength ranges wide enough for this.
For the other targets, it is difficult to determine the continuum for each CO rovibrational absorption band reliably because crowded lines are distributed across the observed wavelength ranges.
Thus, we determined the continuum levels for \ufive{,} \nfour{}, and \ireight{} using low-resolution spectra obtained by \citet{Baba2018} and \citet{Baba2018a} with wider wavelength ranges obtained using AKARI ($R\sim 165$) and Spitzer ($R\sim 80$).
The continuum for each target is determined as follows:

\paragraph{IRAS~01250+2832}
We determined the continuum for this source from the flux levels of the pivots in the Subaru spectra as illustrated in Figure~\ref{fig:contfit_ir01}.
The continuum is assumed to be linear over the range $\lambda\sim 4.4\text{--}5.1\,\micron$ based on a previous analysis of its AKARI SED \citep{Oyabu2011}, which indicates that this continuum can be approximated by a blackbody spectrum.

\begin{figure}
    \centering
    \includegraphics[width=0.8\linewidth]{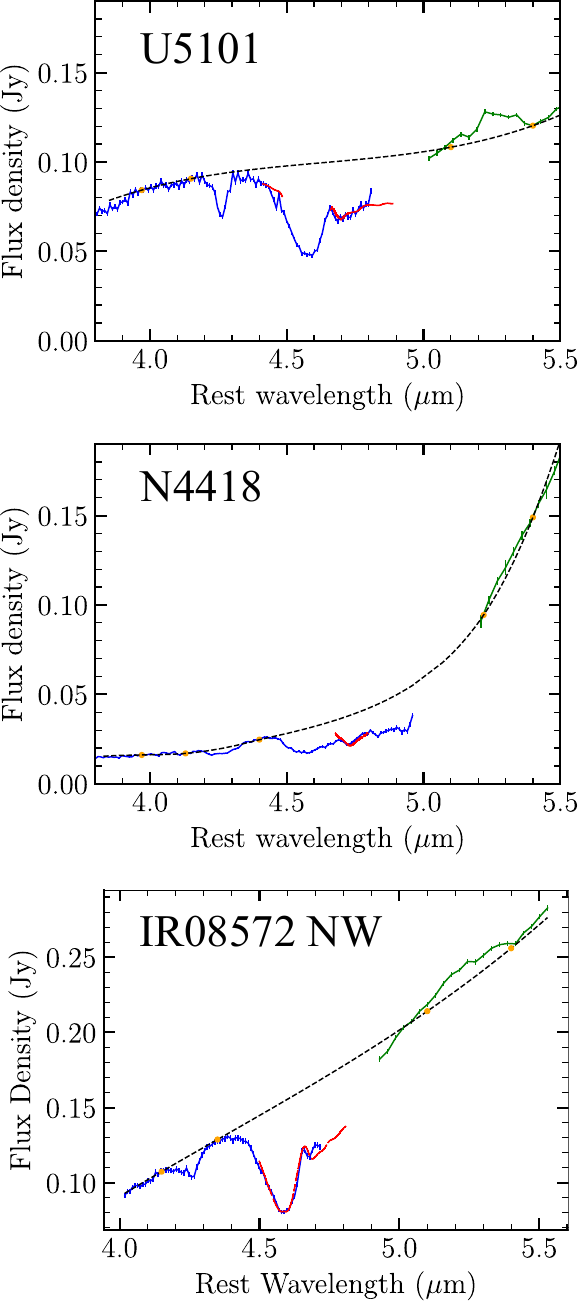}
    \caption{
        Subaru spectra (this work) scaled to the AKARI and Spitzer spectra \citep{Baba2018,Baba2018a} of \ufive{,} \nfour{,} and \ireight{}.
        The spectra of \ireight{} are reprinted from \citet{Onishi2021}.
        The low-resolution spectra of AKARI ($R\sim 160$) and Spitzer ($R\sim 80$) and the Subaru spectra, which are convolved to $R\sim 160$ and flux scaled, are colored blue, green, and red, respectively.
        The dashed black lines and orange dots are, respectively, the continuum determined from the AKARI and Spitzer spectra and the pivots used to estimate it.
        The adopted continuum values for \ireight{} and \nfour{} are the same as those of \citet{Baba2018} and \citet{Baba2018a}, respectively, whereas the continuum of \ufive{} is not.
        After this flux-scaling process, we rescaled the flux levels of \nfour{}.
        (See the text for the details.)
    }\label{fig:scaletoAK}
\end{figure}

\paragraph{UGC~5101}
To apply the AKARI/Spitzer continuum to the Subaru spectra for \ufive{}, we first scaled the flux levels of the Subaru spectra after matching the wavelength resolution to that of the AKARI/Spitzer spectra.
The top panel in Figure~\ref{fig:scaletoAK} shows the scaled Subaru and AKARI/Spitzer spectra together with the continuum.
We adopted a continuum different from that of \citet{Baba2018} by moving the location of the $\sim 4.4\,\micron$ pivot to $\sim 4.2\,\micron$ because the flux level at $\sim 4.4\,\micron$ appears to be reduced by the broad ice-absorption feature of $\mathrm{H_2O}$ combination mode at $\lambda\sim 4.5\,\micron$ \citep{Boogert2015}.

\paragraph{NGC~4418}
The flux levels of the \nfour{} spectra are scaled after matching the wavelength resolution as for \ufive{}.
The middle panel of Figure~\ref{fig:scaletoAK} shows the scaled Subaru and AKARI/Spitzer spectra with the continuum.
Subsequently, we found that the flux levels of \nfour{} around the band center were larger than unity, indicating that the intrinsic absorption in the Subaru spectrum is deeper than in the AKARI spectrum owing to galactic emission in the larger aperture of AKARI.
We therefore rescaled the Subaru spectrum of \nfour{} by the factor $1.30\pm 0.07$ so that the flux levels around the band center of the CO rovibrational absorption at $\sim 4.66\text{--}4.67\,\micron$ became unity.

\paragraph{IRAS~08572+3915~NW}
We scaled the flux levels of the \ireight{} spectra after matching the wavelength resolution as for \ufive{} and \nfour{}.
The continuum was determined in the same manner as in \citet{Onishi2021}, although the emission lines of $\mathrm{H_I\ Pf\beta}$ and $\mathrm{H_2}\ v=0\to 0\ S(9)$ were not excluded.
The bottom panel of Figure~\ref{fig:scaletoAK} shows the scaled Subaru and AKARI/Spitzer spectra together with the continuum.
\par
\begin{figure*}
    \centering
    \begin{tabular}{c}
        \includegraphics[width=0.95\linewidth]{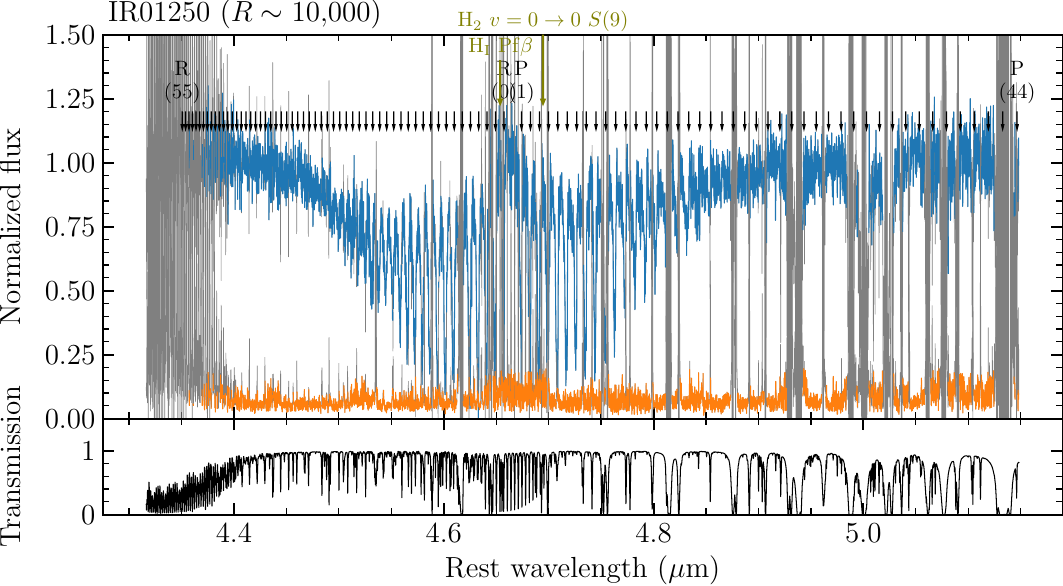}\\[3em]
        \includegraphics[width=0.95\linewidth]{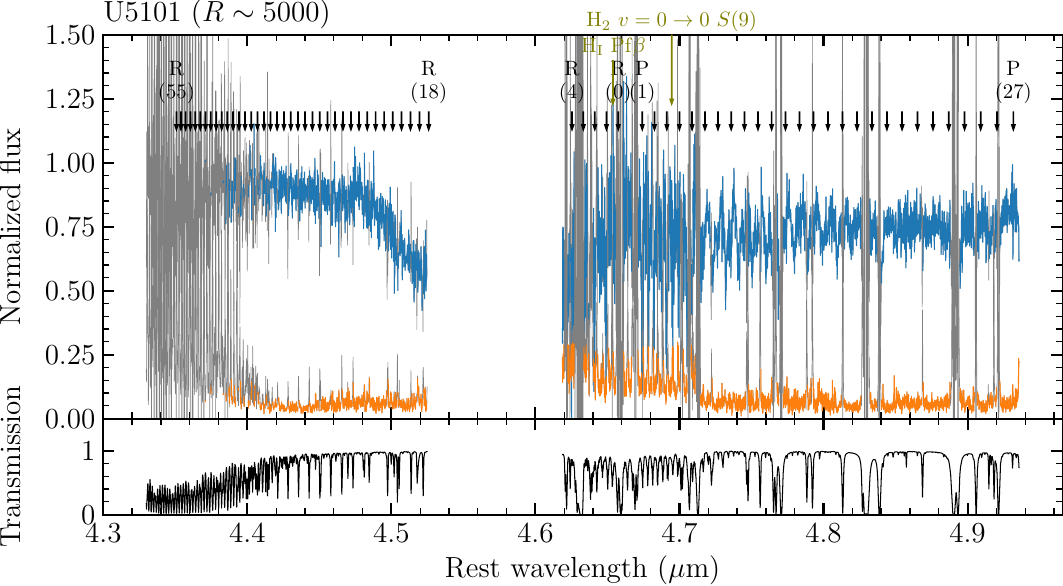}
    \end{tabular}
    \caption{
        Derived spectra of \irone{} and \ufive{} normalized with the continuum (top) and the telluric transmission (bottom).
        Top: Normalized flux values of each target are shown as a solid blue line, and flux errors are shown as a solid orange line.
        The continuum level is unity.
        The rest wavelength of the $\mathrm{H_I}$ Pf$\beta$ and $\mathrm{H_2}$ $v=0\to 0\ S(9)$ lines is denoted by the vertical olive-green arrows.
        The excluded bad data points with large flux error or small telluric transmission are colored gray.
        (See text for the thresholds.)
        The rest wavelength of the CO rovibrational absorption lines is denoted by the vertical arrows.
        Note that the wavelength range of $\sim 4.53\text{--}4.62\,\micron$ has not been observed and has no data points in \ufive{}.
        Bottom: Telluric transmission at the corresponding wavelength ranges corrected by the redshift.
    }\label{fig:reducted_spec_ir01_u05}
\end{figure*}
\begin{figure*}
    \centering
    \begin{tabular}{c}
        \includegraphics[width=0.95\linewidth]{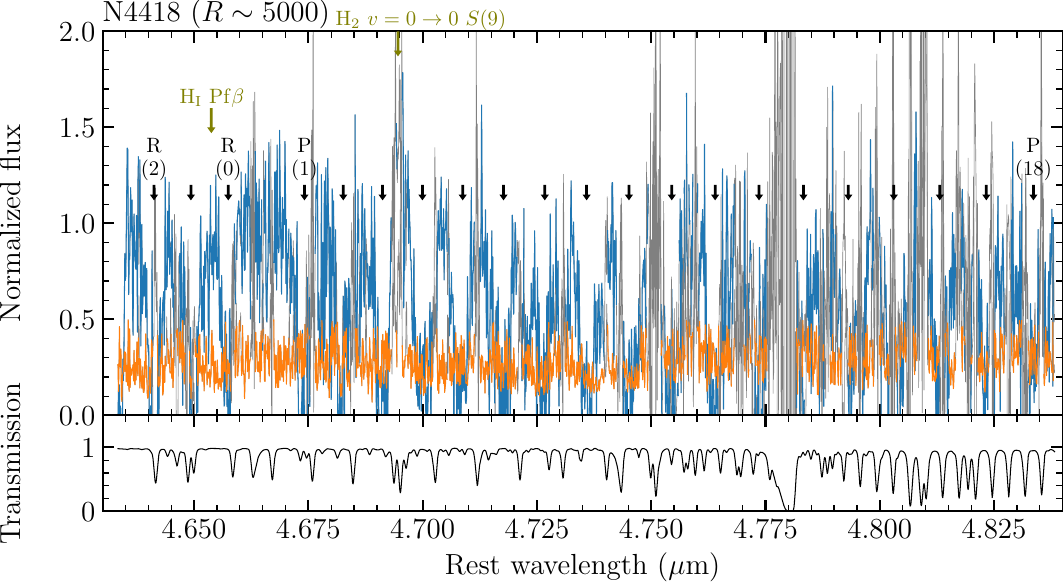}\\[3em]
        \includegraphics[width=0.95\linewidth]{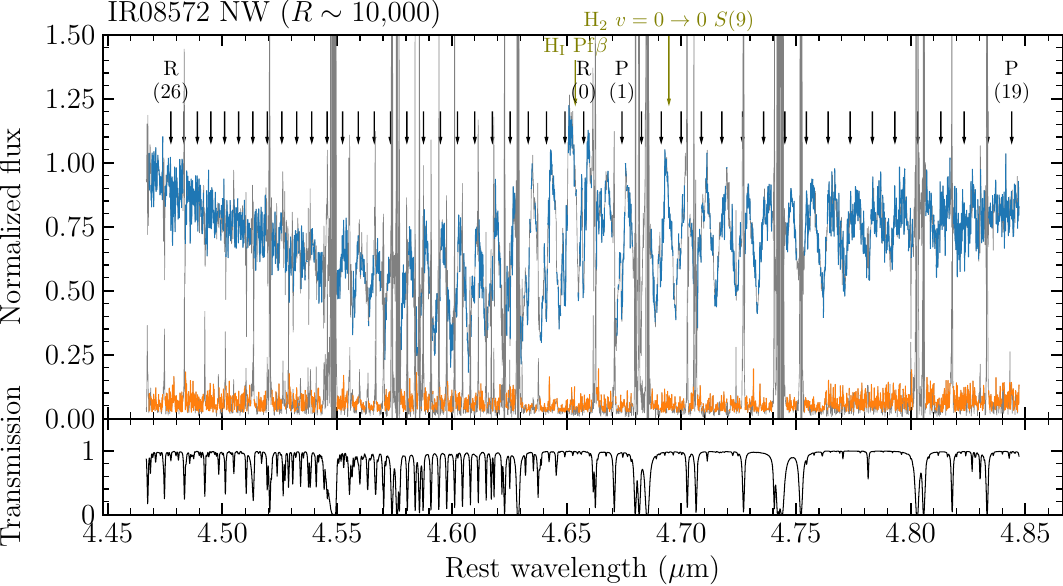}
    \end{tabular}
    \caption{
        Derived spectra of \nfour{} and \ireight{} normalized by the continuum (top) and telluric transmission (bottom).
        Notations are the same as those in Figure~\ref{fig:reducted_spec_ir01_u05}.
    }\label{fig:reducted_spec_n44_ir08}
\end{figure*}

After normalizing the continuum for each target, we derived the final spectra for the CO rovibrational absorption lines.
Figures~\ref{fig:reducted_spec_ir01_u05} and \ref{fig:reducted_spec_n44_ir08} show the continuum-normalized spectra of \irone{,} \ufive{,} \nfour{}, and \ireight{}.
Noisy data points were excluded based on telluric transmission and flux errors.
Points with telluric transmission less than 0.6 in \irone{,} \ufive{,} and \nfour{} and less than 0.7 in \ireight{} were excluded.
Points with flux errors larger than 0.2 in \irone{} and \ireight{}, 0.3 in \ufive{}, and 0.5 in \nfour{} were also excluded.
As a result, the derived $S/N$ values are $\sim 15$, $\sim 15$, $\sim 3\text{--}4$, and $\sim 20$ against the continuum levels for \irone{,} \ufive{,} \nfour{,} and \ireight{}, respectively.

\clearpage

\subsection{Velocity Decomposition}\label{subsec:vel_decomp}
This section describes the velocity-decomposition methods and the results for the velocity components detected in the CO rovibrational absorption lines.
\par
\begin{figure*}
    \centering
    \includegraphics[width=0.95\linewidth]{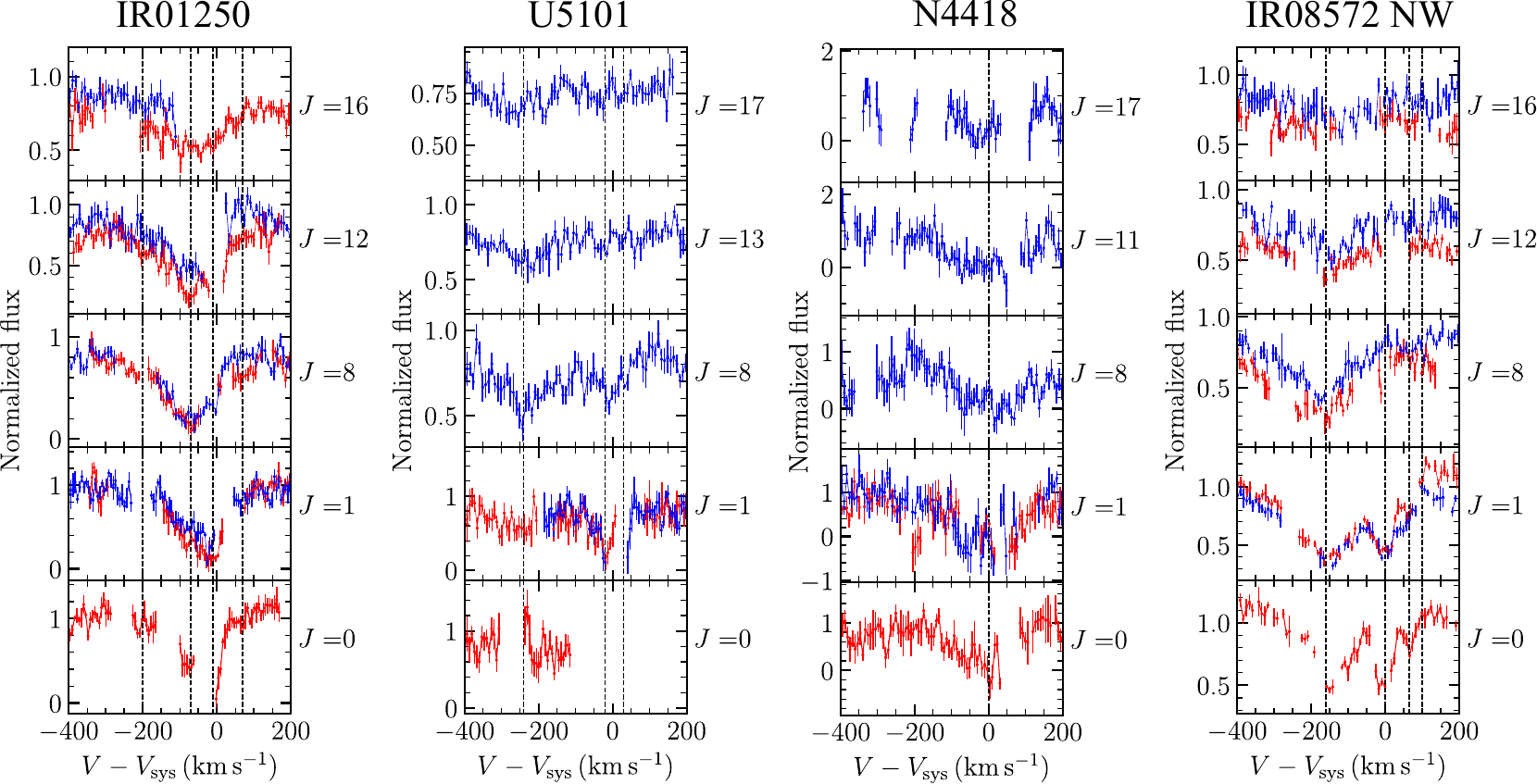}
    \caption{
        Spectra of the $R$-branch (red) and $P$-branch (blue) in each rotational level.
        In principle, we show spectra in rotational levels $J=0,1,8,12$, and 16, so that they cover the entire range of $J$ observed toward all targets in common. (See Table~\ref{tab:obsinfo} for observed $J$ ranges for each target.)
        When the rotational level is unobservable from the ground due to the telluric transmission, its adjacent levels are shown here.
        The abscissa is the LOS velocity relative to the system.
        The ordinate is the normalized flux.
        The vertical dashed lines denote the visually inspected valley position of each velocity component: $V-V_\mathrm{sys}\sim(-200,-70,-10,0)$, $(-240,-10,+30)$, $(0)$, $(-160,0,+65,+100)\,\kmps{}$ for \irone{}, \ufive{}, \nfour{}, and \ireight{}, respectively.
        Exact positions are estimated with the velocity decomposition in Section~\ref{subsec:vel_decomp}.
    }\label{fig:vel_spec_merged}
\end{figure*}

Figure~\ref{fig:vel_spec_merged} shows the velocity profiles of some gaseous CO transitions in the observed spectra of \irone{}, \ufive{}, \nfour{}, and \ireight{}.
Some velocity components with diverse velocity centroids and dispersions appear in each target, as indicated by previous studies \citep{Shirahata2006,Shirahata2012,Shirahata2013b,Shirahata2017,Onishi2021}.
In \irone{,} \ufive{,} and \ireight{}, outflowing velocity components with $\qty|V_\mathrm{LOS}|\sim 150\text{--}200\,\kmps{}$ appear at high rotational levels ($J\gtrsim 8$), and systemic components with $\qty|V_\mathrm{LOS}|\sim 0\,\kmps{}$ appear at low rotational levels ($J\lesssim 1$).
However, no outflowing velocity components appear in \nfour{}.

To decompose the velocity components for these targets, we fitted Gaussian profiles to the optical depths of the CO absorption lines using Lmfit v1.0.0 package \citep{Lmfit}, as in \citet{Onishi2021}.
Line features considered in the velocity decomposition process are summarized in Table~\ref{tab:consider_infit}.
We assume the absorption transitions from $v=0$ to $v=1$ to be dominant and the emission transitions to be negligible.
These assumptions are reasonable because the temperature of the molecular gas ($T_\mathrm{gas}$) inside the torus is expected to be less than the temperature $T_\mathrm{sub}\sim 1500\,\mathrm{K}$ of the dust-sublimation layer \citep{Netzer1993}, while the energy difference between the CO vibrational levels with $v=0$ and $v=1$ is typically $T_{v01}\sim 3000\,\mathrm{K}$, or $T_\mathrm{gas}(<T_\mathrm{sub})<T_{v01}$.
\par
We herein assume the NIR source to be fully covered by the absorbing gas for the simplicity; thus, the optical depth $\tau(\lambda)$ can be expressed as $\tau(\lambda)=-\ln(F_\lambda/F_\mathrm{c})$, using the normalized flux $F_\lambda/F_\mathrm{c}$, where $F_\mathrm{c}$ is the continuum flux.
The effect of the area-covering factor of the absorbing gas is taken into account later in the level population analyses described in Section~\ref{subsec:excite} because traditional $\chi^2$ model fitting cannot constrain the area-covering factor ($C_f$) well.
There are two reasons for this: (1) Only a lower limit for $C_f$ can be imposed based on the absorption depth, and (2) both $C_f$ and the CO column density for each rotational level ($N_J$) of a saturated absorption line are strongly degenerate, as mentioned in some studies of optical absorption lines in quasars \citep[e.g.,][]{Liang2018,Ishita2021}.
Thus, we constrain $C_f$ using Markov chain Monte Carlo (MCMC)-based fitting after the number of free parameters is reduced.
\par
The optical depth $\tau_{JJ'}(\lambda)$ for each absorption line $R(J)$ ($v=0\to 1,\ J\to J'=J+1$) and $P(J)$ ($v=0\to 1,\ J\to J'=J-1$) can be written as the sum of the optical depth $\tau_{JJ',i}(\lambda)$ of the $i$th velocity component in each absorption transition as follows:
\begin{equation}
    \tau_{JJ'}(\lambda)=\sum_i\tau_{JJ',i}(\lambda).\label{eq:sum_tau}
\end{equation}
The number of velocity components is estimated on the basis of the Akaike information criterion (AIC) so that the AIC difference between the second-best model relative to the best model satisfies $\Delta\mathrm{AIC}>14$, which indicates that the second-best model is rejected with a significance level of $>99.9\%$.
See Appendix~\ref{sec:aic} for the details.
The AIC difference $\Delta\mathrm{AIC}$ for each target is summarized in Column (3) of Table~\ref{tab:velocities_each_comp}.
\par
The optical depth of each velocity component is expressed in terms of the column density $N_{J,i}$ of the CO molecules at $v=0,\ J$ as
\begin{eqnarray}
    \tau_{JJ',i}(\lambda)&=&\frac{\pi e^2}{m_\mathrm{e} c^2}N_{J,i}f_{JJ'}\lambda_{JJ'}^2\phi_{\lambda JJ',i},\label{eq:opt_depth}\\
    \phi_{\lambda JJ',i}&=&\frac{1}{\sqrt{2\pi}\sigma_{\lambda JJ',i}}\exp\left[-\frac{(\lambda-\lambda_{0JJ',i})^2}{2\sigma_{\lambda JJ',i}^2}\right],\label{eq:line_prof}\\
    \lambda_{0JJ',i}&=&\left(1+\frac{V_{\mathrm{LOS},i}}{c}\right)\lambda_{JJ'},\\
    \sigma_{\lambda JJ',i}&=&\frac{\sigma_{V,i}}{c}\lambda_{JJ'},\label{eq:vel2lam}
\end{eqnarray}
where $\pi e^2/m_\mathrm{e}c^2=8.8523\times 10^{-13}\,\mathrm{cm}$; $f_{JJ'}$ is the oscillator strength of the transition ($v=0\to 1,\ J\to J'$); $\lambda_{0JJ',i}$ and $\sigma_{\lambda JJ',i}$ are, respectively, the central wavelength and the standard deviation of the $i$th component of the absorption transition; $V_{\mathrm{LOS},i}$ and $\sigma_{V,i}$ are, respectively, the LOS velocity centroid and dispersion of the $i$th component; and $\lambda_{JJ'}$ is the rest wavelength of the transition.
Herein, we assume the line profile function $\phi_{\lambda}$ to be a Gaussian because the Einstein $A$-coefficients for these transitions are at most $\sim 35\,\mathrm{s^{-1}}$; thus, the FWHM due to the natural broadening is less than $\sim 2\times 10^{-12}\,\micron$, which is negligible compared to the observed widths of the absorption lines.
We obtain the oscillator strength $f_{JJ'}$ for $v=0\to 1,\ J\to J'$ from the Einstein $A$-coefficients $A_{J'J}$ for $v=1\to 0,\ J'\to J$, as follows:
\begin{equation}
    f_{JJ'}=\frac{2J'+1}{2J+1}\frac{m_\mathrm{e}c}{8\pi^2 e^2}\lambda_{JJ'}^2A_{J'J}\label{eq:osc_str}
\end{equation}
\citep{Goorvitch1994}.
We also consider the ${}^{13}\mathrm{CO}$ rovibrational absorption lines with the common velocity centroid and dispersion for each velocity component.
Table~\ref{tab:line_info} summarizes the rest wavelengths and the oscillator strengths of the ${}^{12}\mathrm{CO}$ and ${}^{13}\mathrm{CO}$ rovibrational transitions used in this paper.
In the fitting procedure, we determine $\lambda_{0JJ',i}$ and $\sigma_{\lambda JJ',i}$ from $V_{\mathrm{LOS},i}$ and $\sigma_{V,i}$, respectively.
In short, the free parameters are $V_{0,i}$ and $\sigma_{V,i}$ for the $i$th velocity component, together with $N_{J,i}$ for each rotational level $J$ of the $i$th velocity component.
\par
We also consider some other emission and absorption features in the observed wavelength ranges in order to determine the parameters of gaseous CO accurately.
Table~\ref{tab:consider_infit} shows the features we considered for each target.
The emission features are $\mathrm{H_I\ Pf\beta}\ (\lambda=4.6538\,\micron)$ and $\mathrm{H_2}\ v=0\to 0\ S(9)\ (\lambda=4.6946\,\micron)$.
The line widths of $\mathrm{H_I\ Pf\beta}$ are fixed at $\sigma_V=228$\footnote{Based on the velocity dispersion of the $\mathrm{H_I}\ 21\,\si{\cm}$ absorption line observed toward \ufive{} \citep{Gereb2015}.}, 500\footnote{A conservative value because no previous studies have measured the line widths of atomic-hydrogen lines in \nfour, and the line width of $\mathrm{H_I\ Pf\beta}$ in our spectra cannot be determined by fitting due to their low $S/N$ values.}, and $241$\footnote{Based on the velocity dispersion of the $\mathrm{H_I\ Br\gamma}$ emission line observed toward \ireight{} \citep{Goldader1995}.}\,\kmps{} in \ufive, \nfour, and \ireight, respectively, while the width is a free parameter in \irone.
The line widths of $\mathrm{H_2}\ S(9)$ are fixed at $\sigma_V=500,500$\footnote{A conservative value because no previous studies have measured the line widths of molecular-hydrogen lines in \irone{} and \ufive{}, and the line width of $\mathrm{H_2}\ v=0\to 0\ S(9)$ in our spectra cannot be determined by fitting due to their low $S/N$ values.}, $149$\footnote{Based on the velocity dispersion of the $\mathrm{H_2}\ v=1\to 0\ S(1)$ emission line observed toward \nfour{} \citep{Imanishi2004}.}, and $430$\footnote{Based on the velocity dispersion of the $\mathrm{H_2}\ v=1\to 0\ S(1)$ emission line observed toward \ireight{} \citep{Goldader1995}.}\,\kmps{} in \irone, \ufive, \nfour, and \ireight, respectively.
The absorption features include ices of $\mathrm{H_2O}$ (combination mode; $\lambda=4.5\,\micron$, $\Delta\lambda=0.55\,\micron$), $\mathrm{OCN^-}$ (apolar and polar; $\lambda=4.598,4.617\,\micron$, $\Delta\lambda=0.032,0.055\,\micron$, respectively), CO ($\mathrm{CO_2}$-dominant apolar, pure apolar, and polar; $\lambda=4.665,4.673,4.681\,\micron,\ \Delta\lambda=0.0065,0.0076,0.0232\,\micron$, respectively), and OCS ($\lambda=4.90\,\micron,\ \Delta\lambda=0.053\,\micron$), where $\lambda$ is the band center and $\Delta\lambda$ is the FWHM bandwidth.
The band centers and widths of the ice absorption features are based on \citet{Boogert2015}.

\begin{deluxetable*}{lcccccccccc}
\tablecaption{Wavelength, Oscillator Strength, and Lower-state Energy of the ${}^{12}\mathrm{CO}$ and the ${}^{13}\mathrm{CO}$ Rovibrational $(v=0\to 1,\ J\to J'=J\pm 1)$ Lines Considered in This Paper\label{tab:line_info}}
\tablehead{
{} & \multicolumn5c{$\mathrm{{}^{12}CO}$} & \multicolumn5c{$\mathrm{{}^{13}CO}$}\\
{} & {} & \multicolumn2c{$R$-branch} & \multicolumn2c{$P$-branch} & {} & \multicolumn2c{$R$-branch} & \multicolumn2c{$P$-branch}\\
{} & {} & \multicolumn2c{($J'=J+1$)} & \multicolumn2c{($J'=J-1$)}& {} & \multicolumn2c{($J'=J+1$)} & \multicolumn2c{($J'=J-1$)}\\
{$J$} & {$E_J$} & {$\lambda_{JJ'}$} & {$f_{JJ'}$} & {$\lambda_{JJ'}$} & {$f_{JJ'}$}& {$E_J$} & {$\lambda_{JJ'}$} & {$f_{JJ'}$} & {$\lambda_{JJ'}$} & {$f_{JJ'}$}\\
{} & {(K)} & {$\mathrm{(\mu m)}$} & {($10^{-6}$)} & {$\mathrm{(\mu m)}$} & {($10^{-6}$)}& {(K)} & {$\mathrm{(\mu m)}$} & {($10^{-6}$)} & {$\mathrm{(\mu m)}$} & {($10^{-6}$)}}
\startdata
0 & 0.0 & 4.6575 & 11.6587 & $\ldots$ & $\ldots$ & 0.0 & 4.7626 & 11.1501 & $\ldots$ & $\ldots$\\
1 & 5.5 & 4.6493 & 7.7884 & 4.6742 & 3.8715 & 5.3 & 4.7544 & 7.4497 & 4.7792 & 3.7017\\
2 & 16.6 & 4.6412 & 7.0260 & 4.6826 & 4.6371 & 15.9 & 4.7463 & 6.7141 & 4.7877 & 4.4331\\
3 & 33.2 & 4.6333 & 6.7034 & 4.6912 & 4.9585 & 31.7 & 4.7383 & 6.4093 & 4.7963 & 4.7421\\
4 & 55.3 & 4.6254 & 6.5271 & 4.6999 & 5.1308 & 52.9 & 4.7305 & 6.2407 & 4.8050 & 4.9078\\
5 & 83.0 & 4.6177 & 6.4224 & 4.7088 & 5.2382 & 79.3 & 4.7227 & 6.1410 & 4.8138 & 5.0083\\
6 & 116.2 & 4.6100 & 6.3526 & 4.7177 & 5.3079 & 111.1 & 4.7150 & 6.0724 & 4.8227 & 5.0748\\
7 & 154.9 & 4.6024 & 6.3056 & 4.7267 & 5.3558 & 148.1 & 4.7075 & 6.0281 & 4.8317 & 5.1232\\
8 & 199.1 & 4.5950 & 6.2690 & 4.7359 & 5.3908 & 190.4 & 4.7000 & 5.9924 & 4.8408 & 5.1551\\
9 & 248.9 & 4.5876 & 6.2459 & 4.7451 & 5.4154 & 237.9 & 4.6926 & 5.9695 & 4.8501 & 5.1780\\
10 & 304.2 & 4.5804 & 6.2283 & 4.7545 & 5.4303 & 290.8 & 4.6853 & 5.9511 & 4.8594 & 5.1953\\
11 & 365.0 & 4.5732 & 6.2164 & 4.7640 & 5.4428 & 348.9 & 4.6782 & 5.9415 & 4.8689 & 5.2081\\
12 & 431.3 & 4.5662 & 6.2049 & 4.7736 & 5.4530 & 412.3 & 4.6711 & 5.9316 & 4.8784 & 5.2159\\
13 & 503.1 & 4.5592 & 6.1989 & 4.7833 & 5.4596 & 481.0 & 4.6641 & 5.9234 & 4.8881 & 5.2239\\
14 & 580.5 & 4.5524 & 6.1973 & 4.7931 & 5.4610 & 555.0 & 4.6572 & 5.9230 & 4.8979 & 5.2269\\
15 & 663.4 & 4.5456 & 6.1961 & 4.8031 & 5.4646 & 634.2 & 4.6504 & 5.9192 & 4.9078 & 5.2259\\
16 & 751.7 & 4.5389 & 6.1946 & 4.8131 & 5.4616 & 718.7 & 4.6437 & 5.9182 & 4.9178 & 5.2283\\
17 & 845.6 & 4.5324 & 6.1956 & 4.8233 & 5.4622 & 808.4 & 4.6371 & 5.9196 & 4.9280 & 5.2280\\
18 & 945.0 & 4.5259 & 6.1987 & 4.8336 & 5.4604 & 903.5 & 4.6306 & 5.9230 & 4.9382 & 5.2255\\
19 & 1049.8 & 4.5195 & 6.2004 & 4.8440 & 5.4567 & 1003.7 & 4.6242 & 5.9247 & 4.9486 & 5.2210\\
20 & 1160.2 & 4.5132 & 6.2037 & 4.8546 & 5.4512 & 1109.2 & 4.6179 & 5.9280 & 4.9591 & 5.2184\\
21 & 1276.1 & 4.5071 & 6.2083 & 4.8652 & 5.4476 & 1220.0 & 4.6116 & 5.9326 & 4.9697 & 5.2144\\
22 & 1397.4 & 4.5010 & 6.2142 & 4.8760 & 5.4394 & 1336.0 & 4.6055 & 5.9384 & 4.9804 & 5.2092\\
23 & 1524.2 & 4.4950 & 6.2212 & 4.8869 & 5.4334 & 1457.3 & 4.5995 & 5.9418 & 4.9912 & 5.2030\\
24 & 1656.5 & 4.4891 & 6.2260 & 4.8980 & 5.4265 & 1583.8 & 4.5935 & 5.9462 & 5.0022 & 5.1957\\
25 & 1794.2 & 4.4832 & 6.2316 & 4.9091 & 5.4222 & 1715.5 & 4.5877 & 5.9547 & 5.0133 & 5.1913\\
26 & 1937.5 & 4.4775 & 6.2381 & 4.9204 & 5.4136 & 1852.4 & 4.5819 & 5.9607 & 5.0245 & 5.1825\\
27 & 2086.1 & 4.4719 & 6.2453 & 4.9318 & 5.4078 & 1994.6 & 4.5762 & 5.9673 & 5.0358 & 5.1767\\
28 & 2240.3 & 4.4664 & 6.2531 & 4.9434 & 5.3980 & 2142.0 & 4.5706 & 5.9714 & 5.0473 & 5.1703\\
29 & 2399.8 & 4.4609 & 6.2616 & 4.9550 & 5.3911 & 2294.6 & 4.5651 & 5.9792 & 5.0589 & 5.1597\\
30 & 2564.9 & 4.4556 & 6.2675 & 4.9668 & 5.3837 & 2452.4 & 4.5597 & 5.9876 & 5.0706 & 5.1523\\
31 & 2735.3 & 4.4503 & 6.2740 & 4.9788 & 5.3722 & 2615.4 & 4.5544 & 5.9933 & 5.0824 & 5.1445\\
32 & 2911.2 & 4.4452 & 6.2840 & 4.9908 & 5.3639 & 2783.6 & 4.5491 & 5.9994 & 5.0944 & 5.1363\\
33 & 3092.5 & 4.4401 & 6.2915 & 5.0031 & 5.3552 & 2956.9 & 4.5440 & 6.0092 & 5.1065 & 5.1315\\
34 & 3279.2 & 4.4351 & 6.2994 & $\ldots$ & $\ldots$ & 3135.5 & 4.5389 & 6.0163 & $\ldots$ & $\ldots$\\
35 & 3471.3 & 4.4302 & 6.3077 & $\ldots$ & $\ldots$ & 3319.2 & 4.5340 & 6.0237 & $\ldots$ & $\ldots$\\
36 & 3668.8 & 4.4254 & 6.3165 & $\ldots$ & $\ldots$ & 3508.1 & 4.5291 & 6.0315 & $\ldots$ & $\ldots$\\
37 & 3871.7 & 4.4207 & 6.3226 & $\ldots$ & $\ldots$ & 3702.2 & 4.5243 & 6.0397 & $\ldots$ & $\ldots$\\
38 & 4080.0 & 4.4160 & 6.3321 & $\ldots$ & $\ldots$ & 3901.4 & 4.5196 & 6.0482 & $\ldots$ & $\ldots$\\
39 & 4293.7 & 4.4115 & 6.3419 & $\ldots$ & $\ldots$ & 4105.8 & 4.5150 & 6.0539 & $\ldots$ & $\ldots$\\
40 & 4512.7 & 4.4070 & 6.3492 & $\ldots$ & $\ldots$ & 4315.3 & 4.5104 & 6.0631 & $\ldots$ & $\ldots$\\
\enddata
\tablecomments{The oscillator strength was calculated from the Einstein $A$-coefficient. See the text for the details. $E_J$, $\lambda$, and Einstein $A$-coefficients were derived from the high-resolution transmission molecular absorption database \citep{Coxon2004,Li2015,Gordon2017}.}
\end{deluxetable*}

\begin{splitdeluxetable}{cccccBccccc}
\tablecaption{Considered emission and absorption features in the velocity decomposition\label{tab:consider_infit}}
\tablehead{
\colhead{} & \multicolumn{4}{c}{Gas} & \colhead{} & \multicolumn{4}{c}{Ice}\\
\colhead{} & \multicolumn{2}{c}{Absorption} & \multicolumn{2}{c}{Emission} & \colhead{} & \multicolumn{4}{c}{Absorption}\\
\colhead{Target} & \colhead{$\mathrm{{}^{12}CO}$} & \colhead{$\mathrm{{}^{13}CO}$} & \colhead{$\mathrm{H_I}$} & \colhead{$\mathrm{H_2}$} & \colhead{Target} & \colhead{$\mathrm{H_2O}$} & \colhead{$\mathrm{CO}$} & \colhead{$\mathrm{OCN^-}$} & \colhead{$\mathrm{OCS}$}}
\startdata
\irone{} & {\checkmark} & {\checkmark} & {\checkmark} & {\checkmark} & \irone{} & {\ldots} & {\ldots} & {\ldots} & {\ldots}\\
\ufive{} & {\checkmark} & {\checkmark}\tablenotemark{b} & {\checkmark} & {\checkmark} & \ufive{} & {\checkmark} & {\checkmark} & {\checkmark} & {\checkmark}\\
\nfour{} & {\checkmark} & {\checkmark} & {\checkmark} & {\checkmark} & \nfour{} & {\ldots} & {\checkmark} & {\ldots} & {\ldots}\\
\ireight{} & {\checkmark} & {\checkmark} & {\checkmark} & {\checkmark} & \ireight{} & {\ldots} & {\checkmark}\tablenotemark{a} & {\checkmark} & {\ldots}\\
\enddata
\tablecomments{Whether each feature was considered in the velocity decomposition in each object. Check marks denote considered features, and ellipsis dots denote not considered ones.}
\tablenotetext{a}{Polar CO ice was not considered in \ireight{.}}
\tablenotetext{b}{${}^{12}\mathrm{CO}/{}^{13}\mathrm{CO}$ ratio was fixed to 80 as a typical value in \ufive{}, because the velocity-decomposition fitting did not converge when the ratio was free and the absorption lines of ${}^{13}\mathrm{CO}$ were not visually inspected.}
\end{splitdeluxetable}

\begin{deluxetable*}{ccccccccc}
\tablecaption{Results of Velocity Decomposition and Derived Parameters\label{tab:velocities_each_comp}}
\tablehead{
{Target}  &  {$\chi^2_\nu$}  &  {$\Delta\mathrm{AIC}$}  &  {Component}  &  {$V_\mathrm{LOS}$}  &  {$\sigma_V$}  &  {$J$} & {$\mathrm{{}^{12}CO}/\mathrm{{}^{13}CO}$} & {Resolved}\\
{}  &  {}  & {}  &  {}  &  {$(\kmps)$}  &  {$(\kmps)$}  &  {} & {}
}
\decimalcolnumbers
\startdata
{\irone{}}  &  {$7989/4851$}  &  {61}  &  {(a)}  &  {$-210\pm 10$}  &  {$186\pm 4$}  &  {$\le 31$} & {$\ge 19.1$} & {\checkmark}\\
{}  &  {}  &  {}  &  {(b)}  &  {$-70\pm 1$}  &  {$51.9\pm 0.8$}  &  {$\le 28$} & {$\ge 52.7$} & {\checkmark}\\
{}  &  {}  &  {}  &  {(c)}  &  {$74\pm 1$}  &  {$15\pm 1$}  &  {$\le 8$} & {$\ge 6.7$} & {\ldots}\\
{}  &  {}  &  {}  &  {(d)}  &  {$-11.1\pm 0.9$}  &  {$14.5\pm 0.8$}  &  {$\le 19$} & {$8.0\pm 1.2$} & {\ldots}\\
{\ufive{}}  &  {$3813/2907$}  &  {168}  &  {(a)}  &  {$-241\pm 4$}  &  {$130\pm 5$}  &  {$\le 38$} & {80 (fix)} & {\checkmark}\\
{}  &  {}  &  {}  &  {(b)}  &  {$+29\pm 6$}  &  {$86\pm 6$}  &  {$\le 23$} & {80 (fix)} & {\checkmark}\\
{}  &  {}  &  {}  &  {(c)}  &  {$-15\pm 4$}  &  {$29\pm 4$}  &  {$\le 6$} & {80 (fix)} & {\ldots}\\
{\nfour{}}  &  {$1449/1367$}  &  {$48$}  &  {(a)}  &  {$+3\pm 10$}  &  {$142\pm 13$}  &  {$\le 18$} & {$\ge 3.1$} & {\checkmark}\\
{}  &  {}  &  {}  &  {(b)}  &  {$-13\pm 2$}  &  {$64\pm 3$}  &  {$\le 16$} & {$4.9\pm 0.7$} & {\checkmark}\\
{\ireight{}}  &  {$3783/2034$}  &  {963}  &  {(a)}  &  {$-161\pm 5$}  &  {$173\pm 4$}  &  {$\le 26$} & {$\ge 7.1$} & {\checkmark}\\
{}  &  {}  &  {}  &  {(b)}  &  {$-165\pm 2$}  &  {$80\pm 3$}  &  {$\le 12$} & {$\ge 19.3$} & {\checkmark}\\
{}  &  {}  &  {}  &  {(c)}  &  {$+100$ (fix)}  &  {42 (fix)}  &  {$\le 26$} & {$\ge 1.6$} & {\checkmark}\\
{}  &  {}  &  {}  &  {(d)}  &  {$-4\pm 1$}  &  {$25\pm 1$}  &  {$\le 6$} & {$\ge 12.5$} & {\checkmark}\\
{}  &  {}  &  {}  &  {(e)}  &  {$65\pm 1$}  &  {$13\pm 1$}  &  {$\le 9$} & {$\ge 3.5$} & {\ldots}\\
\enddata
\tablecomments{Column (1): target name. Column (2): chi-square value over the degree of freedom of the best fit model. Column (3): difference of AIC between the best fit model and the second-best fit model with the different number of velocity components. (See Appendix~\ref{sec:aic} for the details.) Column (4): velocity component name. Columns (5), (6), (7), and (8): LOS velocity, velocity dispersion, detected $J$ range (more than 1-$\sigma$ significance), and abundance ratio of $\mathrm{{}^{12}CO}/\mathrm{{}^{13}CO}$ for each velocity component, which are estimated in Section~\ref{subsec:vel_decomp}. Negative LOS velocities represent outflowing motion, while positive ones represent inflowing motion. Column (9): whether the velocity component is resolved with the spectral resolution, which is $\sim 10{,}000$ for \irone{} and \ireight{} and $\sim 5000$ for \ufive{} and \nfour{}. Checkmarks indicate that the velocity component is resolved spectrally, and ellipsis dots indicate that the component is unresolved.}
\end{deluxetable*}

\begin{sidewaysfigure*}
    \centering
    \includegraphics[width=0.9\linewidth]{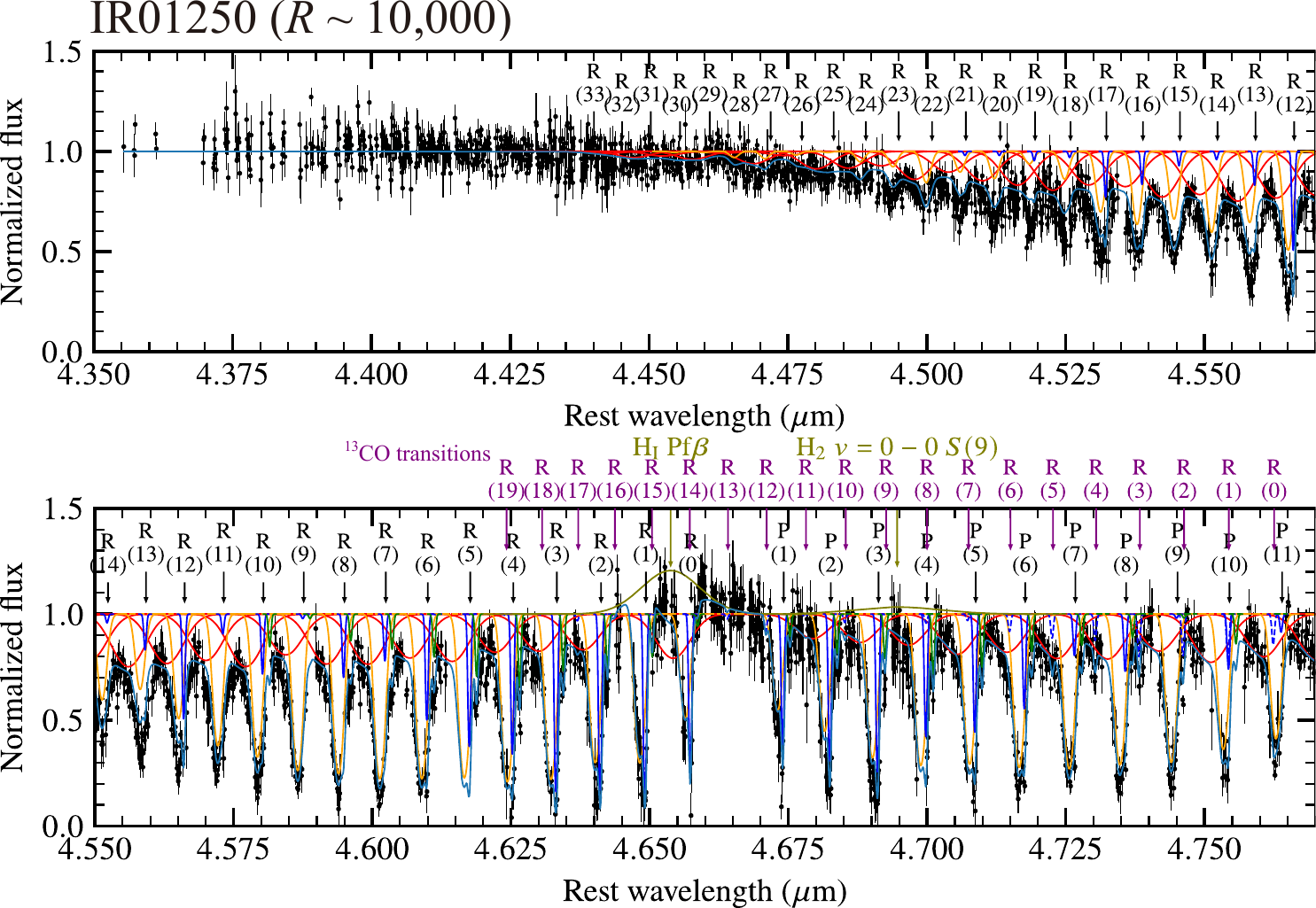}
    \caption{
        Decomposed spectrum of the gaseous CO rovibrational absorption lines in \irone{}.
        The abscissa is the rest wavelength.
        The ordinate is the normalized flux.
        Components (a)--(d) and their sum are denoted with red, orange, green, dark blue, and sky-blue lines, respectively.
        In addition, component (d) of $\mathrm{{}^{13}CO}$ absorption lines are shown with dark-blue dashed lines.
        The black and purple arrows denote the rest wavelength of each ${}^{12}\mathrm{CO}$ and ${}^{13}\mathrm{CO}$ rovibrational transition, respectively.
        Because the ${}^{13}\mathrm{CO}$ transitions are detected only in component (d) with $J\le 19$, the purple arrows denote the transitions in the $J$ range.
        The $\mathrm{H_I\ Pf\beta}$ and $\mathrm{H_2}\ v=0\to 0\ S(9)$ emission lines are colored olive-green.
    }\label{fig:fitres_colines_ir01}
\end{sidewaysfigure*}

As a result, we have found a number of velocity components, which we denote as components (a), (b), $\cdots$, (e), from the broadest component to the narrowest one in each target.
As a representative example, we show in Figure~\ref{fig:fitres_colines_ir01} the decomposed spectrum of \irone{}, where four velocity components are detected; i.e., $i=\text{(a)--(d)}$.
The decomposed spectra of the other targets are presented in Appendix~\ref{sec:decomp_spec_others}.
In addition, the decomposed velocity spectra of some CO lines in \irone{,} \ufive{,} \nfour{,} and \ireight{} are shown in Figure~\ref{fig:decmp_vel_specs}.
The best-fit reduced-$\chi^2$ value, the $\Delta{\mathrm{AIC}}$ value between the most likely model and the second candidate, the LOS velocity ($V_\mathrm{LOS}$) and dispersion ($\sigma_V$), the detected $J$ range (more than 1-$\sigma$ significance), and the $\mathrm{{}^{12}CO}/\mathrm{{}^{13}CO}$ ratio for each velocity component are listed in Table~\ref{tab:velocities_each_comp}.
\par
Each velocity component has different motions and excitation states in each target.
The broadest component (a) is outflowing with a LOS velocity $\qty|V_\mathrm{LOS}|\sim 150\text{--}200\,\kmps$ in \irone{,} \ufive{,} and \ireight{}, while it is systemic in \nfour{}.
This component is highly excited and is detected in nearly all the observed rotational levels up to $J\sim 18\text{--}38$.
The second-broadest component (b) also exhibits outflow with a LOS velocity $\qty|V_\mathrm{LOS}|\sim 70\text{--}150\,\kmps$ in \irone{} and \ireight{}, while instead it exhibits a slow inflow with $\qty|V_\mathrm{LOS}|\sim 30\,\kmps$ in \ufive{} and is nearly systemic in \nfour{}.
This component is moderately or highly excited up to $J\sim 12\text{--}28$.
Components (c) and (d) are resolved only in \ireight{}, where component (c) is inflowing with $\qty|V_\mathrm{LOS}|\sim 100\,\kmps$ and is highly excited up to $J\sim 26$, while component (d) is systemic and is detected only up to low energy levels with $J\sim 6$.
Multiple velocity components with different motions and excitation states have also been detected in the s2 core of VV~114~E, which is likely to be an AGN \citep{Gonzalez-Alfonso2024}\footnote{
    As mentioned in the footnote in Section~\ref{sec:intro}, it should be noted that \citet{Evans2022}, \citet{Rich2023}, and \citet{Buiten2024} identified the adjacent s1 core as an AGN based on the MIR colors, the low equivalent width of PAH emission, and the lack of the $2.3\,\micron$ CO bandhead, respectively, while \citet{Gonzalez-Alfonso2024} identified the s2 core as an AGN based on the extreme CO excitation.
}.
\par
In addition, the derived $\mathrm{{}^{12}CO}/\mathrm{{}^{13}CO}$ ratios shown in Table~\ref{tab:velocities_each_comp} are consistent with selective dissociation \citep[e.g.,][]{vanDishoeck1988}, which gives typical values of $\sim 40\text{--}100$ in the Milky Way \citep{Wilson1992,Romano2017}, in all velocity components except component (d) of \irone{} and component (b) of \nfour{}.
These two components do not contradict the assumption of selective dissociation because the absorption lines may be saturated and the column density of $\mathrm{{}^{12}CO}$ may be underestimated as also suggested in \citet{Ohyama2023}.
In fact, component (b) of \nfour{} has a small area-covering factor and a large CO column density, indicating line saturation.
We do not discuss the $\mathrm{{}^{12}CO}/\mathrm{{}^{13}CO}$ ratio further here because it is beyond the scope of this paper.

\onecolumngrid{}

\addtocounter{figure}{-1}
\begin{sidewaysfigure*}
    \centering
    \includegraphics[width=0.9\linewidth]{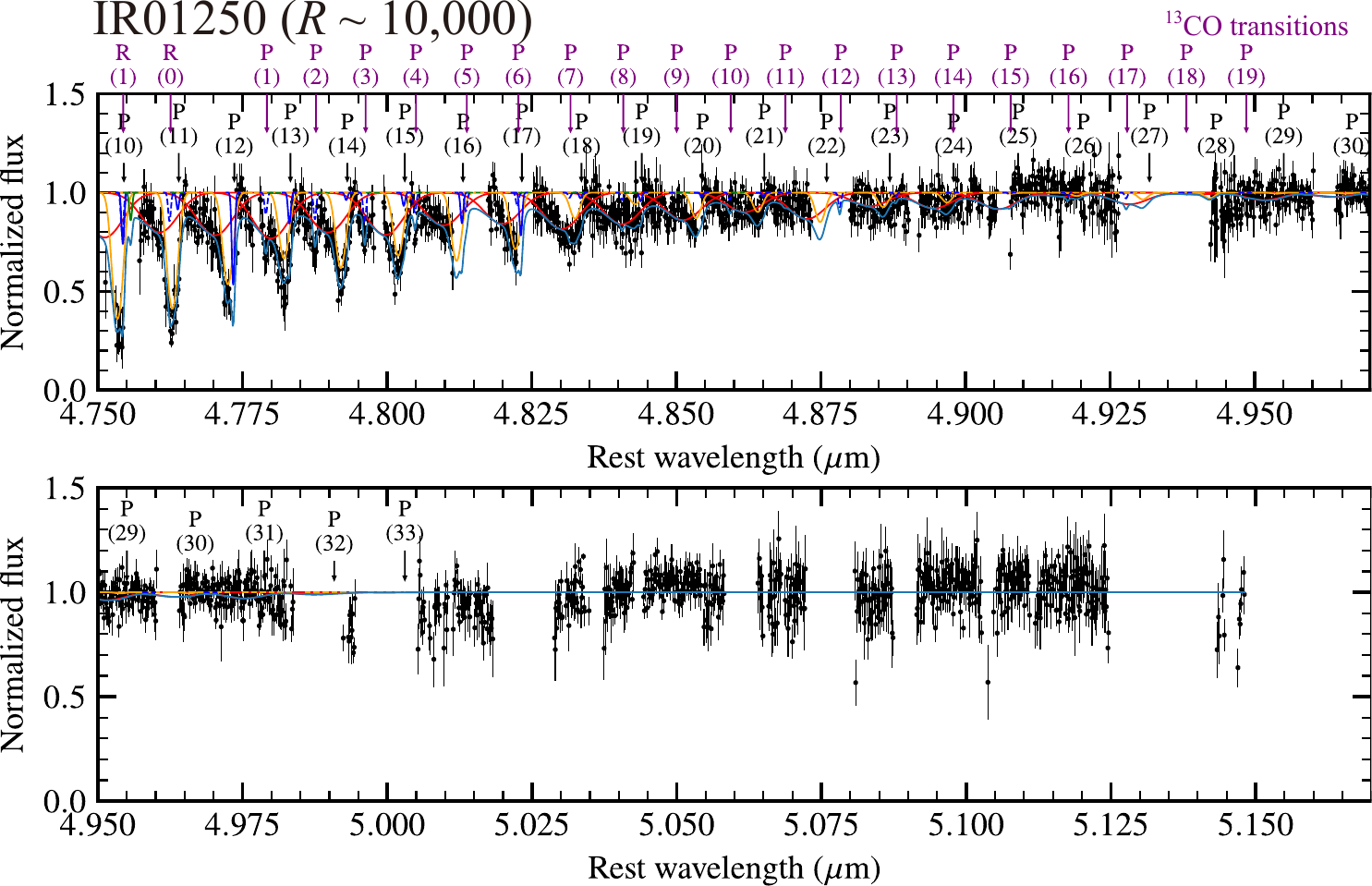}
    \caption{
        (Continued) Decomposed spectrum of the gaseous CO rovibrational absorption lines in \irone{}.
        The abscissa is the rest wavelength.
        The ordinate is the normalized flux.
        Components (a)--(d) and their sum are denoted with red, orange, green, dark blue, and sky-blue lines, respectively.
        In addition, component (d) of $\mathrm{{}^{13}CO}$ absorption lines are shown with dark-blue dashed lines.
        The black and purple arrows denote the rest wavelength of each ${}^{12}\mathrm{CO}$ and ${}^{13}\mathrm{CO}$ rovibrational transition, respectively.
        Because the ${}^{13}\mathrm{CO}$ transitions are detected only in component (d) with $J\le 19$, the purple arrows denote the transitions in the $J$ range.
    }
\end{sidewaysfigure*}

\twocolumngrid{}

\begin{figure*}
    \centering
    \includegraphics[width=0.9\linewidth]{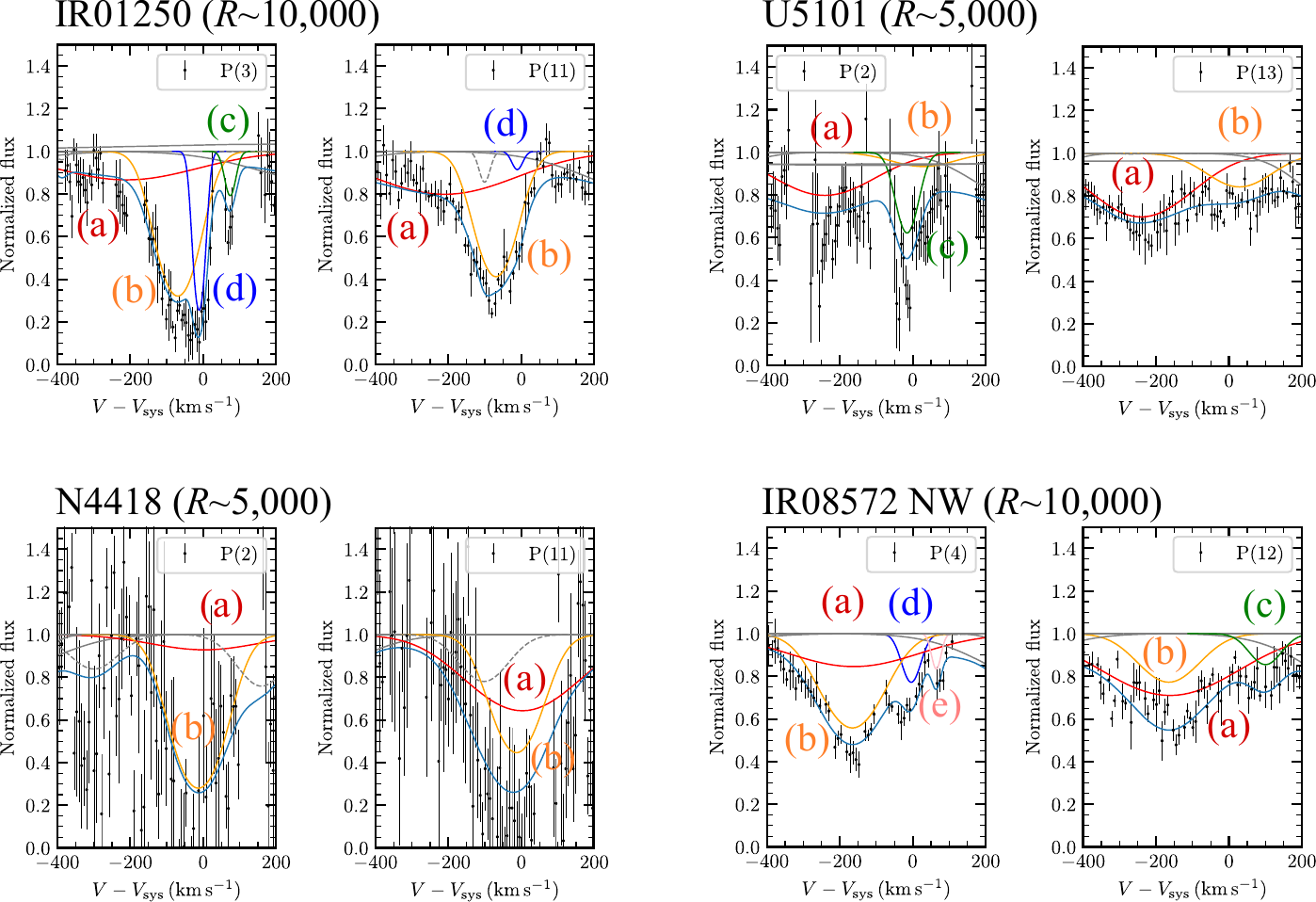}
    \caption{
        Velocity spectra and decomposed velocity components of some gaseous ${}^{12}\mathrm{CO}$ rovibrational transitions as representative examples.
        The transitions shown here are selected so that low and high rotational levels are covered, and all detected velocity components are visible in the figures for each target.
        The abscissa is the LOS velocity relative to the systemic velocity ($V-V_\mathrm{sys}$).
        The ordinate is the normalized flux.
        Component (a), (b), (c), (d), and (e) are shown with red, orange, green, dark blue, and pink lines, respectively.
        The sum of the detected velocity components are shown with sky-blue lines.
        The features other than gaseous ${}^{12}\mathrm{CO}$ are colored gray.
        In particular, the absorption transitions of ${}^{13}\mathrm{CO}$ are shown with dashed lines, and the emission transition of $\mathrm{H_2}\ v=0\to 0\ S(9)$ (visible in $P(3)$ panel of \irone{}) and the absorption transitions of ice are shown with solid lines.
    }\label{fig:decmp_vel_specs}
\end{figure*}

\clearpage

\section{Discussion}\label{sec:discussion}

\subsection{Location of Each Component}\label{subsec:location}
Section~\ref{subsec:vel_decomp} has shown that we detected discrete components with different LOS velocities or velocity dispersions in each CO rovibrational absorption line.
The detection of multiple discrete velocity components indicates that the torus medium is not a uniform medium but instead is highly clumpy, with the clumps forming clusters in separated spatial regions.
\par
In this section, we discuss the dynamical structure inside the torus based on the estimated distance from the central black hole for each velocity component.
We also compare the indicated structure with RHD and MHD torus models.

\subsubsection{Assumptions}\label{subsubsec:assump_estim}
To estimate the distance from the central black hole, we make two assumptions in this paper:

\paragraph{Assumption~(i): The Clumps are in Keplerian Rotation}
If the motions of the clumps in the innermost regions of the torus are governed by the central black hole, and if the motions of clumps can be described by Keplerian rotation together with turbulence and outflowing or inflowing motions, then the velocity of rotation of a clump is related to its radius of rotation by $V_\mathrm{rot}\propto R_\mathrm{rot}^{-0.5}$ [assumption~(i)].
It must therefore be checked whether the sphere of influence around the black hole covers the central regions of the torus in each target.
However, the black-hole masses of the target galaxies are difficult to estimate because the X-ray-to-optical radiation from the vicinity of their black holes is heavily obscured.
In this paper, we therefore estimate the radius of the sphere of influence as follows:
\begin{enumerate}
    \item
    We derive the black hole mass ($M_\mathrm{BH}$) from the $K$-band luminosity ($L_K$) using the following relation:
    \begin{equation}
        \log\qty(\frac{M_\mathrm{BH}}{M_\odot})\sim 1.199\log\qty(\frac{L_K}{L_{K,\odot}})-5.184
    \end{equation}
    \citep{Peng2006,Costagliola2013}.
    Herein, we use the Two Micron All Sky Survey (2MASS) $K$-band magnitudes \citep{Skrutskie2006} and the luminosity distances shown in Table~\ref{tab:targ_info} for the targets.
    In addition, in this paper we adopt the value $L_{K,\odot}=6.75\times 10^{31}\,\mathrm{erg\,s^{-1}}$ \citep{Willmer2018} for the $K$-band solar luminosity.
    \item
    We convert the derived mass into a stellar velocity dispersion ($\sigma_\mathrm{*}$) based on the $M_\mathrm{BH}$-$\sigma_\mathrm{*}$ relation \citep{Tremaine2002}
    \begin{equation}
        \log\qty(\frac{\sigma_*}{200\,\kmps})\sim 0.249\qty[\log\qty(\frac{M_\mathrm{BH}}{M_\odot})-8.13].
    \end{equation}
    \item
    The radius of the sphere of influence of the black hole is thus estimated to be $r_\mathrm{infl}=GM_\mathrm{BH}/\sigma_*^2$.
\end{enumerate}
Table~\ref{tab:radii_infl} summarizes the $K$-band luminosity, black hole mass, stellar velocity dispersion, and radius of the sphere of influence for each target.
The dynamics in the molecular torus, the size of which is expected to be a few parsecs and is comparable or smaller than the radii of the spheres of influence of the targets, is thus driven by the central black hole.
This assumption enables us to estimate the location of each component on the basis of its velocity dispersion.
\begin{deluxetable}{ccccc}
\tablecaption{Radius of the Sphere of Influence and Relevant Parameters of Each Target\label{tab:radii_infl}}
\tablehead{
{Target} & {$L_K$} & {$M_\mathrm{BH}$} & {$\sigma_*$} & {$r_\mathrm{infl}$}\\
{} & {$(10^{43}\,\mathrm{erg\,s^{-1}})$} & {$(10^8\,M_\odot)$} & {$(\kmps)$} & {(pc)}
}
\decimalcolnumbers%
\startdata
\irone{} & 0.21 & 0.25 & 130 & 6\\
\ufive{} & 2.2 & 4.1 & 260 & 30\\
\nfour{} & 0.12 & 0.13 & 110 & 4\\
\ireight{} & 0.58 & 0.84 & 180 & 10\\
\enddata
\tablecomments{Column (1): target name. Column (2): $K$-band luminosity based on the 2MASS $K$-band magnitude \citep{Skrutskie2006} and the luminosity distance in Table~\ref{tab:targ_info}. Column (3): black hole mass estimated from $L_K$ as in \citet{Peng2006}. Column (4): stellar velocity dispersion estimated from $M_\mathrm{BH}$ as in \citet{Tremaine2002}. Column (5): radius of the sphere of influence.}
\end{deluxetable}

\paragraph{Assumption~(ii): Hydrostatic and Triangular Torus Disks}
If the molecular torus is assumed to be a hydrostatic disk, as in many previous theoretical studies \citep[e.g.,][]{Beckert2004,Vollmer2004,Hopkins2012}, the ratio of the rotation velocity ($V_\mathrm{rot}$) to the velocity dispersion ($\sigma_V$) is roughly equal to that of the radius of rotation ($R_\mathrm{rot}$) to the disk height ($H$); i.e., $\sigma_V/V_\mathrm{rot}\sim H/R_\mathrm{rot}$.
In addition, assuming that the molecular torus is a triangular disk with $H/R_\mathrm{rot}\sim \text{constant}$ for simplicity, we can show that the ratio of the velocity dispersion to the rotation velocity also is constant; i.e., $\sigma_V/V_\mathrm{rot}\sim \text{constant}$ [assumption~(ii)].
Although a hydrostatic triangular disk with a constant $\sigma_V/V_\mathrm{rot}$ ratio is assumed for simplicity, it is also applicable to a hydrodynamic radiation-driven fountain model \citep{Wada2016}, which predicts that the ratio $V_\mathrm{rot}/\sigma_V$ is nearly constant in the inner region of the torus.
In fact, theoretical models that assume a triangular disk, such as the CLUMPY \citep{Nenkova2008a} and XCLUMPY \citep{Tanimoto2019} models, reproduce well the SEDs of AGNs in Seyfert galaxies.

Assumptions~(i) and (ii) lead to the following relationship between the radius of rotation and the velocity dispersion:
\begin{equation}
    R_\mathrm{rot}\propto\sigma_V^{-2}.\label{eq:r_rot_vsigm}
\end{equation}
For these reasons, in this section we estimate the ratios of the radius of rotation of each velocity component to that of component (a) based on Equation~(\ref{eq:r_rot_vsigm}).

\subsubsection{Estimates of the Radius of Rotation}\label{subsubsec:est_rot_radius}
The estimated ratios of the velocity components are shown in Column~(5) of Table~\ref{tab:res_mcmc}.
The ratio of the radius of rotation of component (b) to that of the innermost component (a) is $\sim 2\text{--}13$ in each target, and the ratio of the radius of rotation of component (c) to that of the innermost component (a) is $\sim 17$ in \ireight{}.
\par
The innermost component (a) is outflowing ($|V_\mathrm{LOS}|\sim 160\text{--}240\,\kmps$) in three of the four targets.
In addition, the second-innermost component (b), for which the ratio of the radii of rotation is $R_\mathrm{rot,b}/R_\mathrm{rot,a}\sim 5\text{--}13$, is also outflowing ($|V_\mathrm{LOS}|\sim 70\text{--}160\,\kmps$) in \irone{} and \ireight{}.
In contrast, component (b) of \ufive{} and component (c) of \ireight{}, are infalling ($|V_\mathrm{LOS}|\sim 30\text{--}100\,\kmps$).
\par
\begin{figure*}
    \centering
    \includegraphics[width=0.95\linewidth]{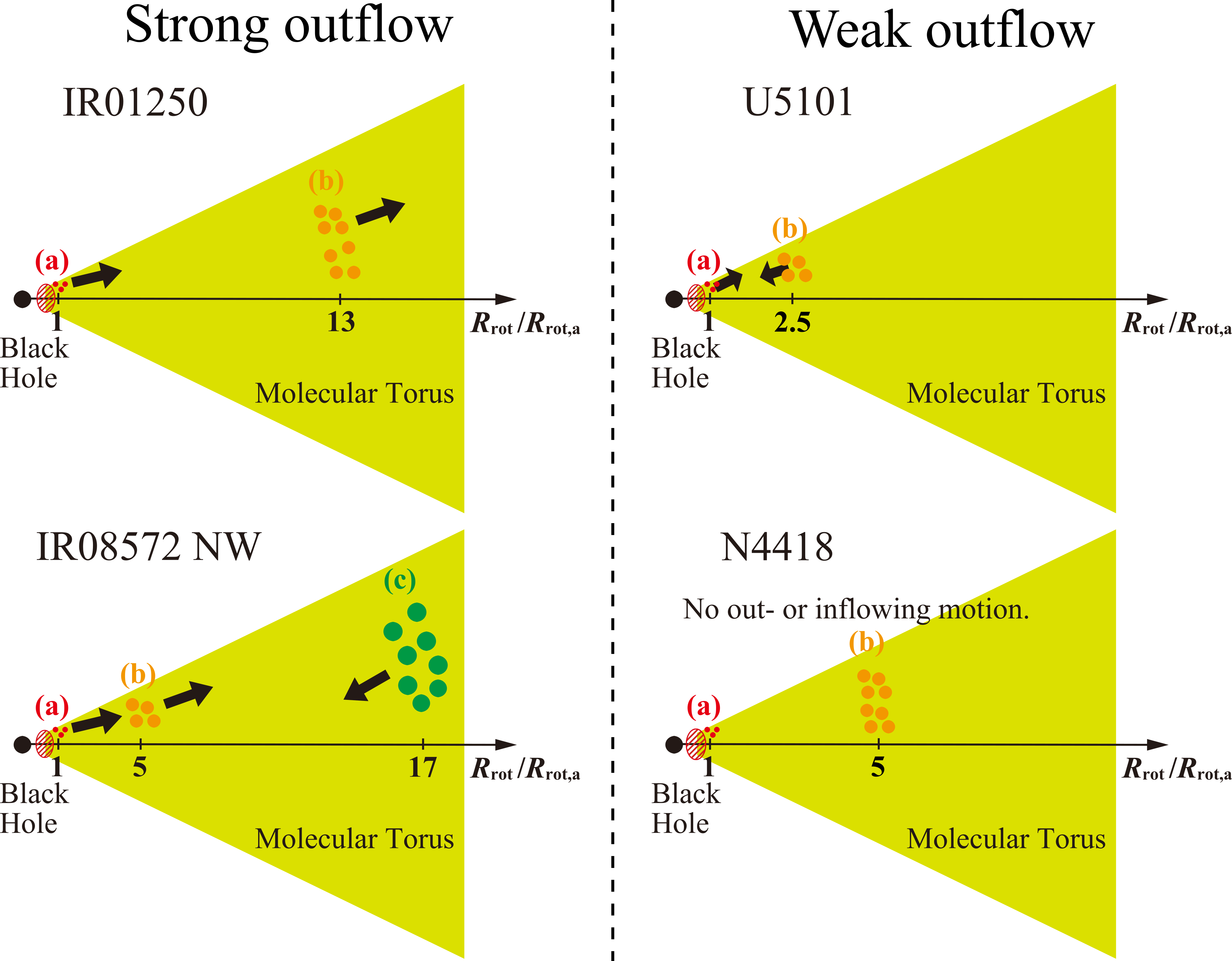}
    \caption{
        Schematic image of the geometry of the velocity components in the molecular torus.
        Components (c) and (d) are excluded here because they are unlikely to be related to the molecular torus according to their small velocity dispersions of $\sigma_V\sim 15\,\kmps$ as mentioned in Section~\ref{subsubsec:est_rot_radius}.
        The image of \ireight{} is adapted from \citet{Onishi2021} because changes of analyses do not make difference in the estimate of the radii of rotation.
    }\label{fig:torus_geom}
\end{figure*}

From these results, we can infer the geometry of each velocity component inside the torus, as shown in Figure~\ref{fig:torus_geom}, on the basis of the LOS velocity and the radius of rotation normalized by that of component (a).

In \irone{} and \ireight{,} clumps in the torus undergo outflow in large regions up to $R_\mathrm{rot}/R_\mathrm{rot,a}\sim 13$.
On the other hand, in \ufive{} and \nfour{,} either the outflowing region is small ($R_\mathrm{rot}/R_\mathrm{rot,a}\lesssim 2.5$) or else no outflowing region is found.
\par
We can make a rough estimate of the absolute values of the radii of rotation of the velocity components based on the dust-sublimation radius, assuming that the infrared luminosity ($L_\mathrm{IR}$) in Table~\ref{tab:targ_info} is similar to the bolometric luminosity of the AGN.
The dust sublimation radius is given by $R_\mathrm{sub}\sim 1\,\mathrm{pc}\times (L_\mathrm{AGN}/10^{46}\,\mathrm{erg\,s^{-1}})^{0.5}$ \citep[e.g.,][]{Barvainis1987,Netzer2015}, if radiation anisotropy can be ignored\footnote{
    If we consider radiation anisotropy \citep[e.g.,][]{Netzer1987}, the dust-sublimation radius can be smaller; thus, the radius shown here is an upper limit. In addition, the dust species makes a difference, with the radius ranging from $R_\mathrm{sub}\sim 0.5\,\mathrm{pc}$ (for graphite dust) to $\sim 1.3\,\mathrm{pc}$ (for silicate dust) when $L_\mathrm{AGN}\sim 10^{46}\,\mathrm{erg\,s^{-1}}$ \citep[e.g.,][]{Netzer2015}. In this paper, we adopt $R_\mathrm{sub}\sim 1\,\mathrm{pc}\times (L_\mathrm{AGN}/10^{46}\,\mathrm{erg\,s^{-1}})^{0.5}$ as a typical value.
}.
On this basis, the dust-sublimation radii of \irone{} and \nfour{} are $R_\mathrm{sub}\sim 0.1\,\mathrm{pc}$ and those of \ufive{} and \ireight{} are $R_\mathrm{sub}\sim 0.3\,\mathrm{pc}$.
If we assume that the innermost component (a) is located near the dust-sublimation layer, the radii of rotation of the velocity components in the torus are $R_\mathrm{rot}\sim 0.1\text{--}1\,\mathrm{pc}$ in \irone{} and \nfour{}, whereas they are $R_\mathrm{rot}\sim 0.3\text{--}5\,\mathrm{pc}$ in \ufive{} and \ireight{}.
These radii satisfy the condition that the velocity components are inside the sphere of influence of the black hole, the radii of which are shown in Table~\ref{tab:radii_infl}.
\par
Here, components (c) and (d) in \irone{} and components (d) and (e) in \ireight{} shown in Table~\ref{tab:velocities_each_comp} are omitted from Table~\ref{tab:res_mcmc} and not to be discussed in this paper anymore, because their corresponding radii of rotation are as large as $R_\mathrm{rot}/R_\mathrm{rot,a}\sim 50\text{--}200$\footnote{
    Note that the estimated ratios of the radii of rotation of components outside the sphere of influence may be different from their true values.
} or $R_\mathrm{rot}\gtrsim 15\,\mathrm{pc}$ in \irone{} and $R_\mathrm{rot}\gtrsim 50\,\mathrm{pc}$ in \ireight{}; i.e., these clumps are outside the sphere of influence shown in Table~\ref{tab:radii_infl}.
In addition, component (c) in \ufive{} is neither discussed because its velocity dispersion is too small to be resolved, and the true dispersion is unclear.
We thus focus on the velocity components inside the sphere of influence; i.e., components (a) and (b) in \irone{,} \ufive{,} and \nfour{} and components (a)--(c) in \ireight{}, in this paper.
\par
In conclusion, the velocity centroid and dispersion of each velocity component indicate the dynamical structures of the molecular tori, where clumps undergo outflow in the inner region and inflow in the outer region (Figure~\ref{fig:torus_geom}).
Moreover, the relative sizes of the outflowing regions differ between the tori of \irone{} and \ireight{} and those of \ufive{} and \nfour{.}
\clearpage

\subsection{Excitation Mechanisms}\label{subsec:excite}
Section~\ref{subsec:location} has shown the inferred dynamical structure inside the molecular torus of each object based on the velocity centroid and velocity dispersion of each velocity component.
In this section, we discuss the excitation mechanisms for each velocity component in \irone{,} \ufive{,} \nfour{,} and \ireight{} based on the level populations of CO molecules.
First, we show the level populations derived for each velocity component in each target and explain the configurations of RADEX models to estimate the gas temperature, column density, and other physical properties of each component.
Next, we discuss the excitation mechanism for each velocity component based on the estimated physical properties.

\subsubsection{Population Diagrams and RADEX Models}\label{subsubsec:radex_model}
\begin{figure*}
    \begin{tabular}{ccc}
        \includegraphics[width=0.3\linewidth]{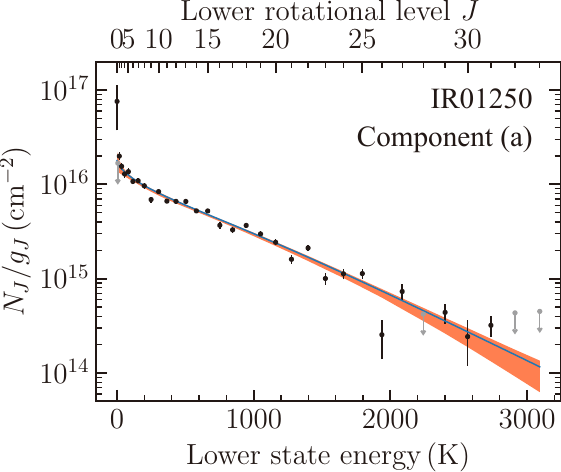}
        &
        \includegraphics[width=0.3\linewidth]{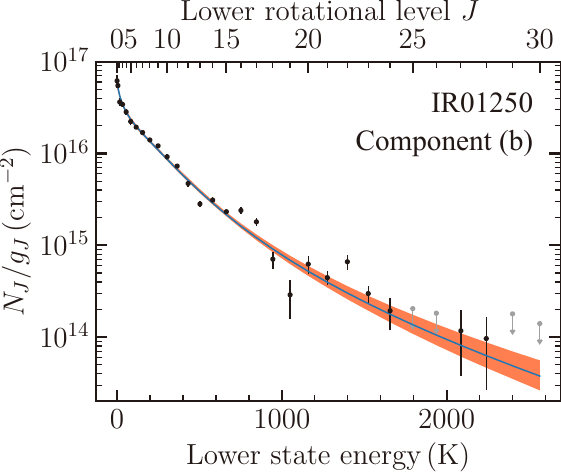}
        &
        {}\\[2em]
        \includegraphics[width=0.31\linewidth]{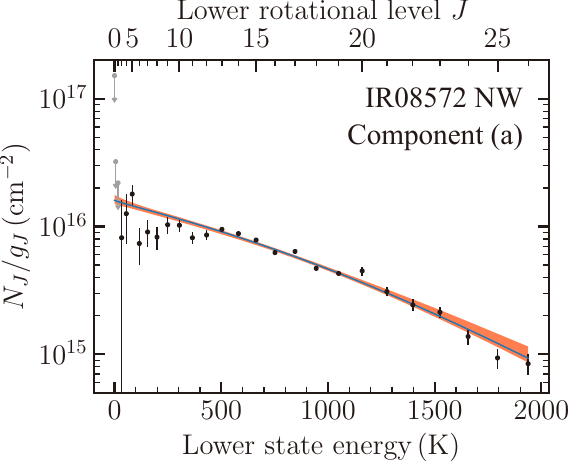}
        &
        \includegraphics[width=0.3\linewidth]{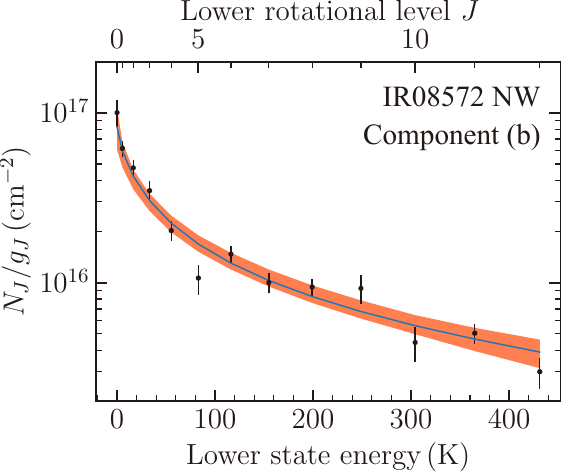}
        &
        \includegraphics[width=0.31\linewidth]{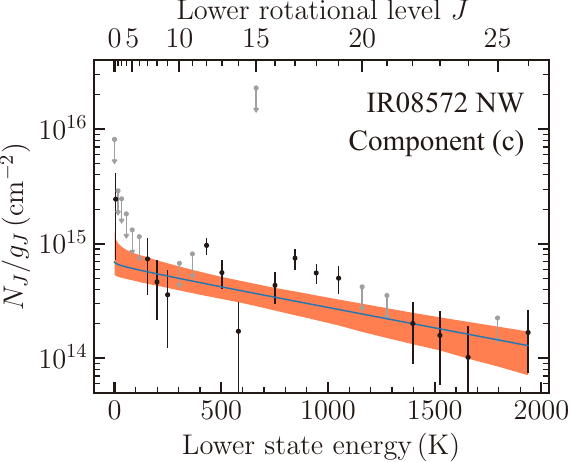}\\[2em]
        \includegraphics[width=0.3\linewidth]{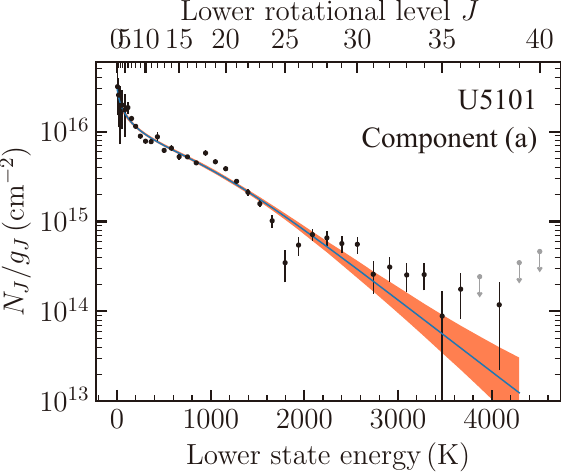}
        &
        \includegraphics[width=0.3\linewidth]{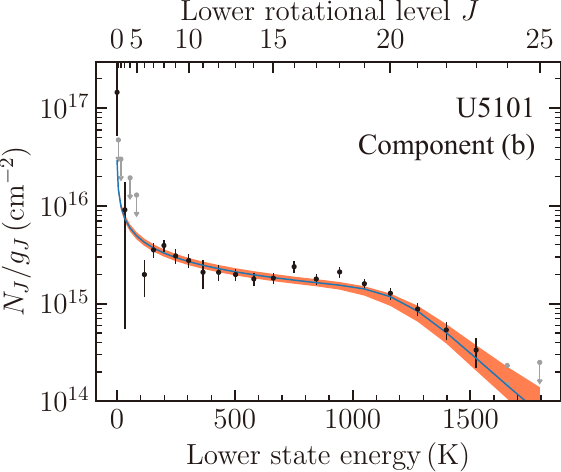}
        &
        {}\\[2em]
        \includegraphics[width=0.3\linewidth]{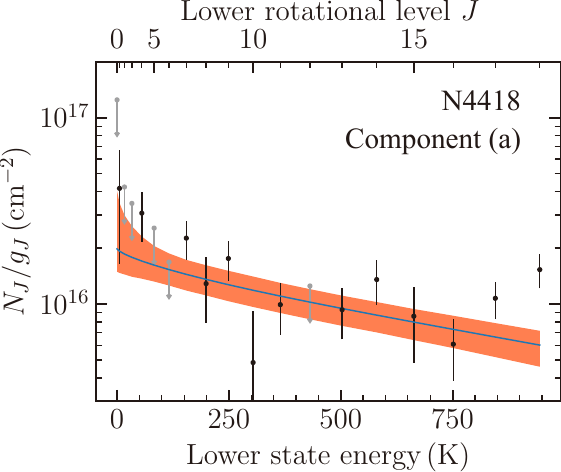}
        &
        \includegraphics[width=0.3\linewidth]{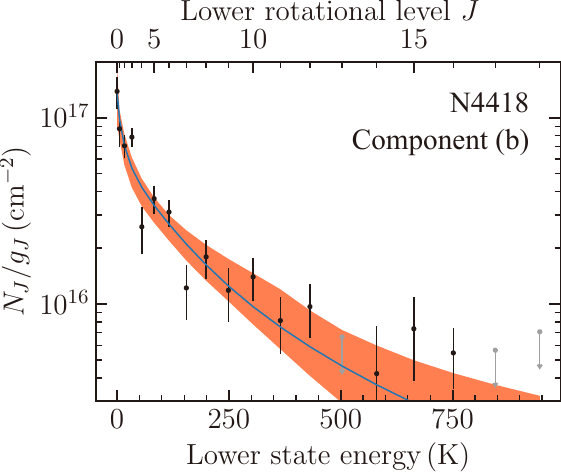}
        &
        {}\\
    \end{tabular}
    \caption{
        Population diagrams of each velocity component inside the sphere of influence mentioned in Section~\ref{subsubsec:est_rot_radius}.
        The column densities, whose upper limits are set, are colored in gray.
        The blue solid lines denote the best-fit RADEX models with the median value of the posterior distribution of each parameter, and the orange areas illustrate 95\% confidence bands.
        We herein adopt the RADEX model because it is preferable to a simple two-temperature model to reproduce the level populations.
        See Section~\ref{subsubsec:innermost_exc} for the details.
    }\label{fig:levelpops_cmps_torus}
\end{figure*}

The black data points with error bars in Figure~\ref{fig:levelpops_cmps_torus} show the level populations of each velocity component in a ``population diagram,'' for which the abscissa is the lower-state ($v=0$) energy ($E_J$) and the ordinate is the column density of CO molecules in the rotational level $v=0,J$ divided by its statistical weight ($N_J/g_J$) in the log scale.

Because the excitation temperature between the rotational levels $J$ and $J'$ is related to the gradient of the population diagram by
\begin{equation}
    \frac{1}{T_{\mathrm{ex},JJ'}}\propto \frac{\ln\qty(N_{J'}/g_{J'})-\ln\qty(N_J/g_J)}{E_{J'}-E_J},
\end{equation}
a shallower population diagram represents a higher excitation temperature.
In addition, data points in the population diagram align on a line if the level population is in local thermodynamic equilibrium (LTE).
\par
We modeled the observed level populations with the RADEX v08sep2017 code \citep{vanderTak2007}, which is a non-LTE model; i.e., it models the level populations of molecules without assuming LTE.
We used a non-LTE model herein because some population diagrams in Figure~\ref{fig:levelpops_cmps_torus} have curvatures even at high rotational levels, indicating that the level populations cannot be reproduced by isothermal LTE clumps.
The free parameters to model each velocity component are as follows: the kinetic temperature ($T_\mathrm{kin}$), the volume density of hydrogen molecules ($n_\mathrm{H_2}$), the brightness temperature of the FIR-to-(sub)millimeter background radiation field ($T_\mathrm{bg}$), and the column density of CO molecules ($N_\mathrm{CO}$).
\par
In this paper, we assume an escape probability based on the large-velocity-gradient (LVG) approximation \citep[e.g.,][]{Sobolev1960,Castor1970,Goldreich1974} because the velocity dispersion in the torus ($\sigma_V\sim 40\text{--}180\,\kmps$) is much larger than the thermal velocity of the CO molecules ($V_\mathrm{th}\lesssim 1\,\kmps$), for which the gas temperature is below the dust-sublimation temperature $\sim 1500\,\mathrm{K}$.
Note that RADEX only considers the transitions between the energy levels with $v=0,\ J\le 40$, where transitions occur in the FIR-to-(sub)millimeter wavelength range.
As mentioned in Section~\ref{subsec:vel_decomp}, the $v=1\to 0$ transitions of CO molecules are expected to be negligible compared to the $v=0\to 1$ transitions, and CO molecules at $v=0$ are dominant inside the torus; thus, this limitation of RADEX does not seem to change the results greatly.
\par
In addition, we consider the effect of non-unity area-covering factor ($C_f$) of the absorbing clumps relative to the NIR source.
Because we assumed $C_f=1$ in the velocity-decomposition process in Section~\ref{subsec:vel_decomp}, owing to the limitations of the traditional $\chi^2$ fitting, the column density at each rotational level ($N_J$) in Figure~\ref{fig:levelpops_cmps_torus} is underestimated by the expression
\begin{equation}
    N_{J}=m(C_f, N^\mathrm{int}_J)N^{\mathrm{int}}_{J},\label{eq:scaled_cdens}
\end{equation}
where $N_{J}$ is the estimated column density at the rotational level of $J$, with $C_f=1$ assumed, as shown in Figure~\ref{fig:levelpops_cmps_torus}; $N^\mathrm{int}_{J}$ is the intrinsic CO column density at level $J$ for the absorbing clumps with a covering factor $C_f\ne 1$, and the RADEX model gives this value; and $m(C_f, N^\mathrm{int}_J)$ is a scaling factor that accounts for the effect of the covering factor.
We formulate the scaling factor $m(C_f, N^\mathrm{int}_J)$ by the ratio of apparent peak optical depths with the non-unity area-covering factor $C_f$ to intrinsic ones.
In Section~\ref{subsec:vel_decomp}, we estimated the apparent peak optical depth from the continuum-normalized flux for each absorption line by $-\ln\qty(F_\lambda/F_\mathrm{c})$.
Because the normalized flux of CO absorption lines with the area-covering factor of the clumps $C_f$ are given as $F_\lambda/F_\mathrm{c}=C_f\exp\qty(-\tau^\mathrm{int}_\mathrm{p})+\qty(1-C_f)$, the scaling factor $m(C_f,N^\mathrm{int}_J)$, which is the ratio of the apparent optical depth to the intrinsic one, is given by the following equation:
\begin{equation}
    m(C_f,N^\mathrm{int}_J)\approx\frac{-\ln\qty[C_f\exp\qty(-\tau^\mathrm{int}_\mathrm{p})+\qty(1-C_f)]}{\tau^\mathrm{int}_\mathrm{p}}\label{eq:cfsc_fac_opt}
\end{equation}
\begin{eqnarray}
\begin{aligned}
    &\approx-\qty(\frac{\pi e^2}{m_\mathrm{e}c}\frac{N^\mathrm{int}_{J}f_{JJ'}\lambda_{JJ'}}{\sqrt{2\pi}\sigma_V})^{-1}\\
    &\quad\times\ln\bigg[C_f\exp\qty(-\frac{\pi e^2}{m_\mathrm{e}c}\frac{N^\mathrm{int}_{J}f_{JJ'}\lambda_{JJ'}}{\sqrt{2\pi}\sigma_V})\\
    &\qquad\qquad+(1-C_f)\bigg],
\end{aligned}\label{eq:cfsc_fac}
\end{eqnarray}
where $\tau^\mathrm{int}_\mathrm{p}$ is the intrinsic peak optical depth of the CO transition.
The scaling factor $m(C_f,N^\mathrm{int}_J)$ satisfies $0<m(C_f,N_J^\mathrm{int})<C_f\le 1$.
In addition, it becomes $m(C_f,N_J^\mathrm{int})\approx C_f$ when the absorption line is optically thin ($\tau^\mathrm{int}_\mathrm{p}\ll 1$), while $m(C_f,N_J^\mathrm{int})\ll 1$ when it is optically thick ($\tau^\mathrm{int}_\mathrm{p}\gg 1$).
This effect is worth considering because it produces curvature in the population diagram at low $J$ and makes the population diagram shallower at middle $J$, resulting in a higher apparent excitation temperature.
Although the formulation assumed in Equation~(\ref{eq:cfsc_fac}) may not be applicable when CO absorption lines are heavily saturated and the apparent line-profile differs from the intrinsic one significantly, we adopt this formulation because line profiles do not appear to vary greatly between optically thin ($J\gtrsim 15$) and thick ($J\sim 8$) transitions as determined by the fitting discussed in Section~\ref{subsec:vel_decomp}.
We therefore model the level populations using the RADEX model corrected for the covering factor, as shown in Equation~(\ref{eq:cfsc_fac}).
\par
We use the Markov chain Monte Carlo (MCMC) method to fit the RADEX model to the observed level populations because the free parameters mentioned above have skewed probability distributions and can sometimes be only upper or lower limited.
Box functions were used for the prior distributions, and 100 MCMC chains, each with a length of 25{,}000 trials and a burn-in length of 1000, were generated.\footnote{
    The length of each chain is determined so that it is more than 50 times longer than the lengths of the autocorrelating steps ($<500$).
}
See Appendix~\ref{sec:mcmc_fit} for more details, such as the boundary configurations of prior distributions.

\subsubsection{Excitation of the Innermost Components}\label{subsubsec:innermost_exc}
In this section, we discuss the physical properties of the innermost component (a) (Figure~\ref{fig:torus_geom}).
The left column of Figure~\ref{fig:levelpops_cmps_torus} shows the population diagram for component (a) of each target.
The innermost components (a), which exhibit shallower population diagrams than do the outer components (b), indicate higher excitation temperatures than the outer velocity components.
\par
We fitted the RADEX model to the level populations as explained in Section~\ref{subsubsec:radex_model}.
The blue solid line and the orange area in each panel of Figure~\ref{fig:levelpops_cmps_torus} show the model with the median parameter values and the 95\% confidence band, respectively.
In addition, Table~\ref{tab:res_mcmc} summarizes the estimated parameters.

\begin{splitdeluxetable*}{cccccBcccccccc}
\tablecaption{Ratios of Radii, Dynamical and Physical Properties of Components Inside the Molecular Torus\label{tab:res_mcmc}}
\tablehead{
{Target}  &  {Component}  &  {$V_\mathrm{LOS}$}  &  {$\sigma_V$}  &  {$R_\mathrm{rot}/R_\mathrm{rot,a}$}  &  {Target}  &  {Component}  &  {$C_f$}  &  {$T_\mathrm{kin}$}  &  {$\log(n_\mathrm{H_2}/\pcc)$}  &  {$\log(N_\mathrm{CO}/\mathrm{cm^{-2}})$}  &  {$T_\mathrm{bg}$}  &  {BG}\\
{}  &  {}  &  {$(\kmps)$}  &  {$(\kmps)$}  &  {}  &  {}  &  {}  &  {}  &  {(K)}  &  {}  &  {}  &  {(K)}  &  {}}
\decimalcolnumbers
\startdata
{\irone{}}  &  {(a)}  &  {$-210\pm 10$}  &  {$186\pm 4$}  &  {1{${}^*$}}  &  {\irone{}}  &  {(a)}  &  {$0.37^{+0.09}_{-0.05}$}  &  {$610\pm 30$}  &  {[6.3, 10]}  &  {$19.0\pm 0.1$}  &  {[2.73, 1500]}  &  {No}\\
{}  &  {(b)}  &  {$-70\pm 1$}  &  {$51.9\pm 0.8$}  &  {$12.8\pm 0.7$}  &  {}  &  {(b)}  &  {$0.83\pm 0.02$}  &  {$142\pm 8$}  &  {$6.5\pm 0.2$}  &  {$18.70\pm 0.04$}  &  {$\ge 441$}  &  {Yes}\\
{\ufive{}}  &  {(a)}  &  {$-241\pm 4$}  &  {$130\pm 5$}  &  {1{${}^*$}}  &  {\ufive{}}  &  {(a)}  &  {[0.38, 0.52]}  &  {$535\pm 28$}  &  {[6, 10]}  &  {$19.25\pm 0.04$}  &  {[2.73, 1500]}  &  {No}\\
{}  &  {(b)}  &  {$+29\pm 6$}  &  {$86\pm 6$}  &  {$2.3\pm 0.4$}  &  {}  &  {(b)}  &  {$0.183\pm 0.007$}  &  {$554^{+245}_{-170}$}  &  {$5.1^{+0.5}_{-0.4}$}  &  {$\ge 19.5$}  &  {$\le 301$}  &  {No}\\
{\nfour{}}  &  {(a)}  &  {$+3\pm 10$}  &  {$142\pm 13$}  &  {1{${}^*$}}  &  {\nfour{}}  &  {(a)}  &  {$\ge 0.34$}  &  {$\ge 550$}  &  {[6, 10]}  &  {$19.0^{+0.3}_{-0.2}$}  &  {[2.73, 1500]}  &  {No}\\
{}  &  {(b)}  &  {$-13\pm 2$}  &  {$64\pm 3$}  &  {$5\pm 1$}  &  {}  &  {(b)}  &  {$0.87^{+0.07}_{-0.05}$}  &  {$65^{+45}_{-24}$}  &  {$5.9\pm 0.5$}  &  {$18.8^{+0.2}_{-0.1}$}  &  {$\ge 18$}  &  {Yes}\\
{\ireight{}}  &  {(a)}  &  {$-161\pm 5$}  &  {$173\pm 4$}  &  {1{${}^*$}}  &  {\ireight{}}  &  {(a)}  &  {$\ge 0.45$}  &  {$\ge 849$}  &  {$6.10^{+0.07}_{-0.09}$}  &  {$18.72^{+0.11}_{-0.06}$}  &  {$\le 448${${}^\dagger$}}  &  {No}\\
{}  &  {(b)}  &  {$-165\pm 2$}  &  {$80\pm 3$}  &  {$4.8\pm 0.4$}  &  {}  &  {(b)}  &  {[0.59, 1]}  &  {$29^{+13}_{-9}$}  &  {$5.6^{+0.3}_{-0.4}$}  &  {$18.6\pm 0.1$}  &  {$\ge 192$}  &  {Yes}\\
{}  &  {(c)}  &  {$+100$ (fix)}  &  {42 (fix)}  &  {$\sim 17$}  &  {}  &  {(c)}  &  {[0.15, 1]}  &  {[2.73, 100]}  &  {$\le 5.3$}  &  {$17.7^{+0.3}_{-0.2}$}  &  {$\ge 684$}  &  {Yes}\\
\enddata
\tablecomments{Columns (3) and (4): LOS velocity and velocity dispersion, which are estimated in Section~\ref{subsec:vel_decomp}. Negative LOS velocities represent outflowing motion, while positive ones represent inflowing motion.  Column (5): the ratios of rotating radius of the component to that of the innermost component (a) based on the assumption of $\sigma_V\propto R_\mathrm{rot}^{-0.5}$ (See Section~\ref{subsec:location}). ${}^*$The ratio of component (a) is unity by definition. Column (8): covering factor. Column (9): gas kinetic temperature. Column (10): $\mathrm{H_2}$ molecular density. Column (11): CO column density. Column (12): brightness temperature of the FIR-to-(sub)millimeter background radiation. ${}^{\dagger}$The brightness temperature is not well constrained because of few data points at $J\ge 24$. (See text for the details.) Column (13): whether the FIR-to-(sub)millimeter background radiation stronger than the CMB affects the level population or not. When the background stronger than CMB or $T_\mathrm{bg}>2.73\,\mathrm{K}$ is detected, ``Yes'' is denoted, and ``No'' is denoted otherwise. For unconstrained parameters, the ranges of prior distributions are shown.}
\end{splitdeluxetable*}

In all targets, component (a) is attributed to hot and dense clumps with high gas-kinetic temperatures $T_\mathrm{kin,a}\gtrsim 500\,\mathrm{K}$ and $\mathrm{H_2}$ volume densities $\log\qty(n_\mathrm{H_2}/\pcc)\gtrsim 6$.
Because background temperatures are not required, the level population can be reproduced without strong background radiation.
This indicates that component (a) is attributed to collisionally excited hot clumps.
This interpretation is supported by the JWST detection of a similar hot ($\sim 500\,\mathrm{K}$) outflowing component in the CO rovibrational absorption band observed toward the s2 core in the southwest nucleus of VV~114~E, which is likely to be an AGN \citep{Gonzalez-Alfonso2024}, as mentioned in Section~\ref{sec:intro}.
In addition, the clumps cover $\gtrsim 40\%$ of the area of the NIR continuum source, which is probably the dust-sublimation layer, in all targets.
\par
Based on the CO column density and the lower limit of the molecular-hydrogen density, we can impose an upper limit on the geometrical thickness of component (a) along the LOS considering the volume-filling factor of the clumps ($\phi_V$):
\begin{equation}
    d_\mathrm{los}\sim \frac{N_\mathrm{H_2}}{n_\mathrm{H_2}\phi_V}=\frac{[\mathrm{H_2}]}{[\mathrm{CO}]}\frac{N_\mathrm{CO}}{n_\mathrm{H_2}\phi_V},\label{eq:los_thickness}
\end{equation}
where we assume that the ratio of the abundances of the CO molecules to the $\mathrm{H_2}$ molecules is $[\mathrm{CO}]/[\mathrm{H_2}]\sim 10^{-4}$ \citep{Dickman1978}.
The geometrical thickness of component (a) thus becomes $d_\mathrm{los,a}\lesssim 0.5$, $\lesssim 2$, $\lesssim 3$, and $\lesssim 0.4\,\mathrm{pc}$ for \irone{}, \ufive{}, \nfour{}, and \ireight{}, respectively, if we assume the volume-filling factor to be $\phi_V\sim 0.03$ as a typical value \citep[e.g.,][]{Beckert2004,Vollmer2004,Honig2007}.
These estimated values of $d_\mathrm{los,a}$ are consistent with the size of the torus, which is expected to be a few parsecs.
Thus, we attribute component (a) to hot and dense clumps with $T_\mathrm{kin,a}\gtrsim 500\,\mathrm{K}$ and $\log\qty(n_\mathrm{H_2,a}/\pcc)\gtrsim 6$ in the innermost region of the molecular torus.
\par
We note that the temperature gradient in an LTE clump is another potential candidate for producing curvature in a population diagram besides the area-covering factor mentioned in Section~\ref{subsubsec:radex_model}.
However, the kinetic-temperature gradient between illuminated and shaded areas in an LTE clump can be excluded because the column density of each velocity component can be as large as $N_\mathrm{CO}\gtrsim 10^{18.5}\,\psqc$, or $\tau_V\gtrsim 30$, and $\gtrsim 90\%$ of the clump volume has a temperature similar to that of the shaded area in such a situation \citep{Nenkova2008a}.
Thus, we rule out the temperature gradient as a candidate for the origin of the curvature in a population diagram in this paper.
See the discussion in Section~5.2.2 of \citet{Onishi2021} for more details.

\subsubsection{Excitation of Outer Components}\label{subsubsec:outer_exc}
We next estimate the physical properties of the outer velocity components (b) and (c) (Figure~\ref{fig:torus_geom}).
The middle and right columns of Figure~\ref{fig:levelpops_cmps_torus} show the population diagrams for the velocity components (b) and (c), respectively, for all targets.
The outer components exhibit population diagrams with more curvatures than the innermost component (a), indicating that the area-covering factor and the non-LTE populations have greater effects.
\par
As in component (a), we fit the RADEX model to the level populations of the outer velocity components.
Figure~\ref{fig:levelpops_cmps_torus} shows models with median parameters and their 95\% confidence bands, and Table~\ref{tab:res_mcmc} summarizes the estimated parameters for the outer components.
\par
Component (b) of \irone{}, \nfour{}, and \ireight{} is attributable to cool and dense clumps with gas-kinetic temperatures $T_\mathrm{kin,b}\sim 30\text{--}140\,\mathrm{K}$ and $\mathrm{H_2}$ volume densities $\log(n_\mathrm{H_2,b}/\pcc)\sim 5.6\text{--}6.5$.
On the other hand, component (b) of \ufive{} has the high kinetic temperature $T_\mathrm{kin,b}=554^{+245}_{-170}\,\mathrm{K}$.
This is consistent with the smaller relative radius of rotation ($R_\mathrm{rot,b}/R_\mathrm{rot,a}\sim 2.3$) of the velocity component than is the case for component (b) in the other targets, as estimated in Section~\ref{subsec:location}.
However, this temperature is ill-constrained, with a larger uncertainty than the temperature of component (b) in the other targets.
Component (b) also covers a large fraction ($\gtrsim 60\%$) of the NIR source in targets other than \ufive{,} where foreground emission may make the apparent absorption line depths smaller than intrinsic ones, as mentioned in Section~\ref{subsec:u51}.
Component (c) of \ireight{} is attributable to moderately dense clumps with $\log\qty(n_\mathrm{H_2}/\pcc)\lesssim 5.3$, while the kinetic temperature is not constrained.
\par
The most remarkable feature of the outer velocity components is that they exhibit contributions from FIR-to-(sub)millimeter background radiation with brightness temperatures $T_\mathrm{bg}\ge 441$, $\ge 18$, and $\ge 192\,\mathrm{K}$ in \irone{}, \nfour{}, and \ireight{}, respectively, as shown in Columns~(12) and (13) of Table~\ref{tab:res_mcmc}\footnote{
    Even in component (b) of \ufive{}, where only an upper limit is imposed on the background radiation, a contribution from such radiation is not excluded.
}.
The brightness temperatures of the background radiation, which are comparable to or higher than the kinetic temperatures in \irone{} and \ireight{}, thus indicate that the strong FIR-to-(sub)millimeter background excites the CO molecules of the outer velocity components radiatively.
Although the lower limit of $T_\mathrm{bg}$ in \nfour{} is lower than those in \irone{} and \ireight{}, it does not exclude the existence of the strong FIR-to-(sub)millimeter background.
\citet{Gonzalez-Alfonso2024} also indicates that radiative excitation is necessary to reproduce the high rotational temperatures of CO and $\mathrm{H_2O}$ molecules in the s2 core of VV~114~E.
In Section~\ref{subsec:comp_models}, we provide a brief discussion of the candidates for the origin of the FIR-to-(sub)millimeter background.
\par
The LOS geometrical thicknesses of components (b) and (c) can also be estimated using Equation~(\ref{eq:los_thickness}).
The results for component (b) are $d_\mathrm{los,b}\sim 0.2$, $\gtrsim 0.9$, and $\sim 1\,\mathrm{pc}$ in \irone{,} \nfour{,} and \ireight{,} respectively, if the volume-filling factor $\phi_V\sim 0.03$ is assumed, as in Section~\ref{subsubsec:innermost_exc}.
On the other hand, component (b) in \ufive{} exhibits a large lower limit for the CO column density, and the geometrical thickness is estimated to be $\gtrsim 30\,\mathrm{pc}$ if the volume-filling factor $\phi_V\sim 0.03$ is assumed.
Because component (b) in \ufive{} exhibits the high kinetic temperature $T_\mathrm{kin,b}\sim 400\text{--}800\,\mathrm{K}$ and is likely to be located inside the torus, this large estimated thickness is possibly be due to a volume-filling factor that may be larger than the assumed value of 0.03.
Although the estimated thickness is large, it is nevertheless consistent with the radius of the sphere of influence in Table~\ref{tab:radii_infl}.
The estimated thickness of component (c) in \ireight{} is $d_\mathrm{los,c}\gtrsim 0.3\,\mathrm{pc}$ with the assumption $\phi_V\sim 0.03$.
The LOS geometrical thicknesses of components (b) and (c) are thus consistent with the size of the molecular torus.
We therefore attribute components (b) and (c) to cold and (moderately) dense clumps with $T_\mathrm{kin,bc}\sim 30\text{--}140\,\mathrm{K}$ and $\log\qty(n_\mathrm{H_2,bc}/\pcc)\lesssim 6.5$, which are radiatively excited by FIR-to-(sub)millimeter background radiation in the outer region of the molecular torus.

\subsection{Comparison with Theoretical Models}\label{subsec:comp_models}
This section discusses whether our interpretation of the estimated locations and excitation mechanisms of the velocity components inside the torus is consistent with torus models.

\subsubsection{Kinematics}\label{subsubsec:kin_comp_models}
In this section, we compare the kinematics of the observed velocity components with the radiation-driven fountain model \citep{Wada2012a,Wada2016}.
Based on three-dimensional hydrodynamic simulations, the model proposes that the outflowing and inflowing gases are driven by radiation from the central accretion disk and the gravity of the central black hole to form the molecular torus.

\paragraph{Outflowing Components}
The radiation-driven fountain model indicates that the outflowing gas with a velocity $V_\mathrm{outflow}\lesssim 500\,\mathrm{km\,s^{-1}}$ is naturally reproduced around the boundary between the molecular torus and the ionizing cone \citep{Wada2016}.
In addition, both RHD and MHD models that consider infrared radiation pressure exhibit similar maximum velocities and regions \citep{Dorodnitsyn2016a,Chan2017}.
\par
Here, we compare the outflowing velocity in the models with the observed LOS velocities by assuming a value for the angle between the direction of the outflow and the LOS.
For \ireight{,} \ufive{,} and \nfour{,} we estimate the inclination angles to be $\Theta_\mathrm{los}\gtrsim 60^\circ$, where $\Theta_\mathrm{los}=90^\circ$ is the edge-on angle\footnote{
    Refer to Figure~11 of \citet{Onishi2021} for angle notations.
} \citep{Vega2008,Efstathiou2014,Martinez-Paredes2015,Yamada2021,Sakamoto2021a}.
In addition, we estimate the half-opening angles of the tori measured from the pole to be $\Theta_\mathrm{h}\lesssim 40^\circ$ for \ireight{} and \ufive{} based on model fitting to the infrared SEDs \citep{Vega2008,Efstathiou2014,Martinez-Paredes2015}.
\par
Thus, if we assume that the outflowing gas is distributed mainly in the high-latitude regions, as suggested in the radiation-driven fountain model, and if the outflowing motion is parallel to the boundary between the molecular torus and the ionizing cone, then the LOS velocity of $V_\mathrm{outflow}\cos(\Theta_\mathrm{los}-\Theta_\mathrm{h})\lesssim 450\,\mathrm{km\,s^{-1}}$ can be reproduced naturally.
If we assume that \irone{} has similar values of $\Theta_\mathrm{los}$ and $\Theta_\mathrm{h}$, the observed LOS velocities, $|V_\mathrm{los}|\sim 70\text{--}240\,\mathrm{km\,s^{-1}}$, of the outflowing components in \ireight{,} \irone{,} and \ufive{} are reasonable.
In addition, if we assume that outflowing gas in \nfour{} is perpendicular to the LOS; i.e., $\Theta_\mathrm{los}-\Theta_\mathrm{h}\sim 90^\circ$, then the observed LOS velocities $|V_\mathrm{los}|\sim 0\,\mathrm{km\,s^{-1}}$ are also reasonable.
The small value of $\Theta_\mathrm{h}$ in \nfour{} is supported by the observations that indicate that the extinction is larger than that in the other targets, as indicated by the deep silicate absorption and the non-detection of hard X-ray photons (Section~\ref{subsec:n44}).

\paragraph{Inflowing Components}
In contrast, the inflowing gas is predicted to pass through the dense and geometrically thin equatorial disk of the torus \citep{Wada2016,Izumi2018} in the radiation-driven fountain model.
However, gas near the equatorial plane is difficult to observed in CO absorption because the column density of the gas is too high for the infrared continuum source \citep{Wada2007}.
Actually, a radiative-transfer simulation based on the radiation-driven fountain model indicates that CO rovibrational absorption lines are not observed when $\Theta_\mathrm{los}\gtrsim 80^\circ$, owing to obscuration by the equatorial disk \citep{Matsumoto2022}.
Thus, the observed inflowing components in \ireight{} and \ufive{} are not attributable to inflowing gas in the equatorial disk.
\par
Instead, the inflowing components may be some other inflowing gas apart from the equatorial disk, such as a failed outflow.\footnote{
    Although this discrepancy may be caused by differences of the inner structures of the molecular tori of \ireight{} and \ufive{} from that of the Circinus galaxy, we herein assume that the structures are similar.
}
This scenario is also consistent with the LOS angles of $\Theta_\mathrm{los}\sim 77^\circ$ \citep{Vega2008} and $\sim 66^\circ$ \citep{Yamada2021}, which are $\lesssim 80^\circ$, in \ireight{} and \ufive{,} respectively.
\par
For these reasons, the LOS velocity and the location of the outflowing components are consistent with the radiation-driven fountain model.
On the other hand, the features of the inflowing components are not consistent with inflowing gas passing through the equatorial disk, as suggested in the model, but instead are probably inflowing gas separate from the equatorial disk.

\subsubsection{Excitation}\label{subsubsec:exc_comp_models}
In this section, by comparing the physical properties of some simple models to those derived in this paper, we discuss the sources of the excitation for the hot, dense clumps [component (a)] and for the cold clumps excited by the FIR-to-(sub)millimeter background [components (b) and (c)].
We also consider models other than the radiation-driven fountain model, because it is based on Seyfert galaxies with column densities of gas that are smaller than those derived for the U/LIRGs in this paper.
This difference is critical in discussing the excitation states of the CO molecules in the tori.

\paragraph{The Innermost Component}
We identify the innermost components (a) in the target U/LIRGs as collisionally excited hot clumps, as mentioned in Section~\ref{subsubsec:innermost_exc}.
This component is well thermalized, with a high kinetic temperature $\gtrsim 550\,\mathrm{K}$ and the large CO column density $N_\mathrm{CO}\gtrsim 10^{18.7}\,\psqc$, or $N_\mathrm{H_2}\gtrsim 10^{22.7}\,\psqc$, if the abundance ratio of CO to $\mathrm{H_2}$ is assumed to be $\mathrm{[CO]/[H_2]}\sim 10^{-4}$ \citep{Dickman1978}.
\par
\citet{Baba2018} consider the three candidates for the mechanisms responsible for exciting the molecular CO gas with a large column density and high kinetic temperature: the photodissociation regions (PDRs) formed by central UV radiation \citep[e.g.,][]{Tielens1985,Hollenbach1999,Meijerink2005,Wolfire2022}, mechanical shocks \citep[e.g.,][]{McKee1984,Hollenbach1989,Neufeld1989}, and X-ray-dominated regions (XDRs) formed by central X-ray radiation \citep[e.g.,][]{Maloney1996,Meijerink2005,Wolfire2022}.
The PDR and shock models, which were investigated by \citet{Meijerink2005} and \citet{McKee1984}, respectively, can reproduce a gas temperature $\gtrsim 10^2\,\mathrm{K}$ only when the column density of the gas is $N_\mathrm{H}\lesssim 10^{21}\,\psqc$.
Thus, central UV radiation and mechanical shocks are not suitable excitation mechanisms for the targets in this paper.
On the other hand, XDRs can achieve a high kinetic temperature $\gtrsim 550\,\mathrm{K}$ even when the column density of the gas is $N_\mathrm{H}\gtrsim 10^{23}\,\psqc$, because X-ray photons can penetrate deeper into the gas than UV photons\citep{Meijerink2005}.
\citet{Onishi2021} also found that the molecular CO gas attributed to the innermost component of \ireight{} is thermalized at a high kinetic temperature.\footnote{
    See Section~5.2 of \citet{Baba2018} and Section~5.2.3 of \citet{Onishi2021} for the details.
}
Accordingly, we find that the existence of a thermalized innermost component excited by XDRs is also applicable to targets other than \ireight{}.

\paragraph{The Outer Components}
We identify the outer components (b) and (c) in \irone{,} \nfour{,} and \ireight{} with radiationally excited cold clumps, as mentioned in Section~\ref{subsubsec:outer_exc}.
These components have low kinetic temperatures, $\sim 30\text{--}140\,\mathrm{K}$, and the high brightness temperature of the FIR-to-(sub)millimeter background is greater than $T_\mathrm{bg}\sim 20\text{--}700\,\mathrm{K}$.
The radius of rotation of the outer component is $\sim 1\text{--}5\,\mathrm{pc}$ (See Section~\ref{subsubsec:est_rot_radius}), if the radius of rotation of component (a) is assumed to be similar to the dust-sublimation radius.
\par
We herein discuss the origin of the high background radiation in the large regions with radii $\sim 1\text{--}5\,\mathrm{pc}$ in \irone{,} \nfour{,} and \ireight{.}
To keep the background radiation high even in the outer regions, the dust temperature must be high in such regions.
However, in Seyfert-galaxy models \citep[e.g.,][]{Matsumoto2022}, the dust temperature decreases steeply as the radius of the distribution increases, and a high level of background radiation cannot be reproduced because the exciting photons escape from the torus.
To resolve this problem, \citet{Gonzalez-Alfonso2019} proposed the ``greenhouse effect,'' which can heat the dust more efficiently in a large region by trapping FIR and MIR photons within a spherical gas shell around the AGN core, assuming the existence of such a gas-rich environment like U/LIRGs.
For \nfour{}, \citet{Ohyama2023} also argued that the greenhouse effect is likely to excite CO molecules efficiently in extended regions.
\par
In this paper, we estimate the dust temperature achievable in the outer regions given the infrared luminosities and molecular column densities of the targets.
Table~3 and Equation~(15) of \citet{Gonzalez-Alfonso2019} give the radial profile of the dust temperature in the greenhouse-effect model for some values of $L_\mathrm{IR}/\pi R_\mathrm{out}^2$ and $N_\mathrm{H_2}$, where $R_\mathrm{out}$ is the radius of the outer boundary of the torus.
\par
We thus have to obtain values of $R_\mathrm{out}$ for the targets \irone{,} \nfour{,} and \ireight{.}
For \nfour{}, \citet{Gonzalez-Alfonso2019} gives the radius of the outer boundary as $R_\mathrm{out}\sim 10.8\text{--}12.5\,\mathrm{pc}$\footnote{
    Values under the assumed cosmology in this paper.
} based on the previous submillimeter interferometry \citep{Sakamoto2013a} and FIR spectroscopy \citep{Gonzalez-Alfonso2012}.
For \ireight{}, \citet{Vega2008} estimates the radius of the outer boundary as $R_\mathrm{out}\sim 62.5\,\mathrm{pc}$ based on model fitting to the infrared SED.
For \irone{,} no previous studies have estimated the radius of the outer boundary.
We therefore scale the value of $R_\mathrm{out}$ from that of \ireight{} according to the dust-sublimation radius and obtain $R_\mathrm{out}\sim 18.8\,\mathrm{pc}$.
We adopt a configuration with $L_\mathrm{IR}/\pi R_\mathrm{out}^2\sim 10^8L_\odot\,\mathrm{pc}^{-2}$ and $N_\mathrm{H_2}\sim 10^{23}\,\psqc$ for the greenhouse-effect model because these values are the most similar to those estimated above.
As a result, the greenhouse-effect model gives the dust temperatures at the locations of the outermost components in \irone{,} \nfour{,} and \ireight{} as $T_\mathrm{dust}\sim 400$, $500$, and $400\,\mathrm{K}$, respectively.
These values do not strongly contradict the brightness temperature of the background radiation ($T_\mathrm{bg}$) for component (b) of \irone{,} \nfour{,} and \ireight{} in Table~\ref{tab:res_mcmc}.
Thus, radiation from dust is acceptable for the origin of the FIR-to-(sub)millimeter background radiation observed in component (b).
\par
In fact, a recent JWST observation toward the dust-embedded AGN II~Zw96-D1 indicates the existence of hot dust with $T_\mathrm{dust}\sim 400\,\mathrm{K}$ extended over a torus-sized region ($\sim 3\,\mathrm{pc}$) \citep{Garcia-Bernete2024}.
However, component (c) cannot be reproduced by this simple model, and more-realistic numerical models for U/LIRGs may be necessary to reproduce this component.
In addition, we note that the influence of vibrational transitions becomes greater in highly excited components such as component (c), and consideration of these transitions may be necessary, although the RADEX model used in this paper does not consider them.

\section{Conclusion}\label{sec:conclusion}
In this paper, we have decomposed the velocity components in CO rovibrational absorption lines ($v=0\to 1,\ \Delta{J}=\pm 1,\ \lambda\sim 4.67\,\micron$) in two AGNs (\irone{}, \ireight{}) and two AGN-starburst composites (\ufive{}, \nfour{}) using high-resolution ($R\sim 5000\text{--}10{,}000$) spectroscopy to probe the spatial distributions and determine the dynamical and physical properties of clumps inside the molecular torus.
Our findings in this paper are summarized as follows:

\begin{enumerate}
\item
We have decomposed discrete velocity components with multiple LOS velocities, velocity dispersions, and excitation states in each CO rovibrational absorption transition for all targets.
From the velocity decompositions, we have detected two or more velocity components, i.e., (a) and (b) in \irone{,} \ufive{,} and \nfour{,} and (a)--(c) in \ireight{,} associated with the molecular torus.

\begin{enumerate}[label=\roman*.]
\item
Component (a) is outflowing with the high LOS velocity $|V_\mathrm{LOS}|\sim 160\text{--}240\,\kmps$, and it has the largest velocity dispersion $\sigma_V\sim 130\text{--}180\,\kmps$ in all objects other than \nfour{}.
In \nfour{,} component (a) also has the largest velocity dispersion, $\sigma_V\sim 140\,\kmps{}$, but it has a systemic LOS velocity.
In addition, component (a) is detected up to high rotational levels with $J\sim 20\text{--}40$ in all targets.
\item
Component (b) is also outflowing with the high LOS velocity $|V_\mathrm{LOS}|\gtrsim 80\,\kmps$ in \irone{} and \ireight{}.
On the other hand, component (b) is systemic or inflowing with the low LOS velocity $|V_\mathrm{LOS}|\lesssim 30\,\kmps$ in \ufive{} and \nfour{}.
Moreover, this component has the velocity dispersion $\sigma_V\sim 50\text{--}80\,\kmps$, which is smaller than that of component (a) in all objects.
Although these components are also detected up to high rotational levels with $J\sim 10\text{--}30$, the highest rotational level is lower than that of component (a) in each target.
\item
Component (c) of \ireight{} is inflowing with the LOS velocity $V_\mathrm{LOS}\sim 100\,\kmps$ and the velocity dispersion $\sigma_V\sim 40\,\kmps$, and it is highly excited up to $J\sim 26$.
\end{enumerate}

\item
The detection of multiple distinct velocity components in the torus indicates that the torus medium is a clumpy medium and that the clumps form multiple clusters in separated spatial regions in the torus of each object.
\item Assuming that the radius of rotation of each velocity component is related to its velocity dispersion as $R_\mathrm{rot}\propto \sigma_V^{-2}$, we estimate the ratio of the radius of rotation for each velocity component to that of component (a), which has the largest velocity dispersion and is probably located in the innermost region of the torus.
As a result, component (b) in \ireight{} and in \irone{} is outflowing in the outer region of the torus, with the ratio $R_\mathrm{rot,b}/R_\mathrm{rot,a}\sim 5\text{--}13$.
On the other hand, component (b) in \ufive{} and \nfour{} is systemic or inflowing in more compact regions with the ratios with $R_\mathrm{rot,b}/R_\mathrm{rot,a}\sim 2\text{--}5$.
In addition, component (c) in \ireight{} is inflowing in the outermost region with $R_\mathrm{rot,c}/R_\mathrm{rot,a}\sim 17$.
\item
These results indicate the dynamical structure of \ireight{} and \irone{}, where clumps outflow in inner regions of the torus (outflow-dominant regions), with $R_\mathrm{rot}/R_\mathrm{rot,a}\sim 5\text{--}13$ and inflow in outer regions with $R_\mathrm{rot}/R_\mathrm{rot,a}\sim 17$ (Figure~\ref{fig:torus_geom}).
On the other hand, the radius of the outflow-dominant region is more compact in \ufive{} with $R_\mathrm{rot}/R_\mathrm{rot,a}\sim 2$ and is non-existent in \nfour{}, where outflowing clumps are not detected even in the vicinity of the AGN core.
Accordingly, we conclude that the molecular torus has a dynamic inner structure with outflows and inflows rather than a static structure.
\item
The radiation-driven fountain model \citep{Wada2016} can naturally explain both LOS velocities $|V_\mathrm{LOS}|\lesssim 240\,\kmps$ and the locations of the innermost component (a).
On the other hand, another model may be necessary to describe the inflow, apart from the equatorial disk, in order to reproduce the inflowing components in \ireight{} and \ufive{}.
\item
We have estimated the physical properties of each velocity component based on the level populations and the non-LTE model, RADEX:

\begin{enumerate}[label=\roman*.]
\item
The innermost component (a), which is detected in all objects, is attributable to hot ($T_\mathrm{kin}\gtrsim 550$) and dense ($\log\qty(n_\mathrm{H_2}/\pcc)\gtrsim 6$) clumps.
In addition, this component covers more than $\sim 30\%$ of the NIR continuum source, which is expected to be the dust sublimation layer with a radius of $\lesssim 1\,\mathrm{pc}$, and has the large CO column density $N_\mathrm{CO}\sim 10^{18.7\text{--}19.3}\,\mathrm{cm^{-2}}$.
\item
On the other hand, the outer component (b) is attributable to colder ($T_\mathrm{kin}\lesssim 30\text{--}140\,\mathrm{K}$) and less dense ($\log\qty(n_\mathrm{H_2}/\pcc)\sim 5.6\text{--}6.5$) clumps than those of component (a).
Component (b) also covers the NIR continuum source, with a large covering factor $\gtrsim 60\%$.
In addition, component (b) is likely illuminated by an FIR-to-(sub)millimeter background higher than $T_\mathrm{bg}\sim 20\text{--}400\,\mathrm{K}$.
\end{enumerate}

In conclusion, we have observationally found that the inner structure of the molecular torus in each of the four targets we have studied has a dynamic structure containing outflowing and inflowing clumps.
We have also find that the clumps inside the torus are collisionally excited in the innermost regions, whereas they are radiationally excited by the FIR-to-(sub)millimeter background in the outer regions.
\end{enumerate}

\begin{acknowledgments}
    We thank all the operating staff in Subaru Telescope and NAOJ for their great help in our observations. In particular, we are grateful to Dr.~Tae-Soo Pyo, Dr.~Takagi, and Dr.~Mieda for their support in the IRCS observations in 2010 and 2019. In addition, we thank Mr.~Doi for his fruitful discussions. This work was supported by JSPS KAKENHI grant Nos. JP19J21010 (S.O.), JP21H04496, JP23H05441 (T.N.), and JP19J00892 (S.B.). K.M. is a Ph.D. fellow of the Flemish Fund for Scientific Research (FWO-Vlaanderen) and acknowledges the financial support provided through Grant number 1169822N.
\end{acknowledgments}

%

\facilities{Subaru (IRCS), Shane (The Kast Double Spectrograph)}


\software{IRAF v2.16.1 \citep{Tody1986,Tody1993}, PyRAF v2.1.15 \citep{Pyraf}, Molecfit v1.5.9 \citep{Kausch2015,Smette2015}, TIPS \citep{Gamache2017}, RADEX v08sep2017 \citep{vanderTak2007}, Numpy v1.18.5 \citep{Harris2020}, Matplotlib v3.2.2 \citep{Hunter2007a}, Scipy v1.4.1 \citep{Virtanen2020}, Astropy v4.0.1 \citep{Astropy2013,Astropy2018}, Lmfit v1.0.0 \citep{Lmfit}, Emcee v3.0.2 \citep{Foreman-Mackey2013}, Pandas v1.0.3 \citep{Pandas}, Jupyter v1.0.0 \citep{Kluyver2016}, ASTEVAL v0.9.18 \citep{Asteval}}

\appendix

\section{Redshift of IRAS~01250+2832}\label{app:rs_ir01}
Because the precise estimate of the redshift of \irone{} has not been available, it is necessary to determine the redshift of \irone{} and its associated uncertainty  to determine the velocity centroid of each velocity component in the CO rovibrational absorption lines.
The first estimate of the redshift by \citet{Oyabu2011} was rough, and its uncertainty was not evaluated.
Thus, we have reanalyzed the optical spectrum of \irone{} derived by \citet{Oyabu2011} and have determined the redshift by fitting some Fraunhofer lines.

\subsection{Observational Data}
\begin{figure}
    \centering
    \includegraphics[width=\linewidth]{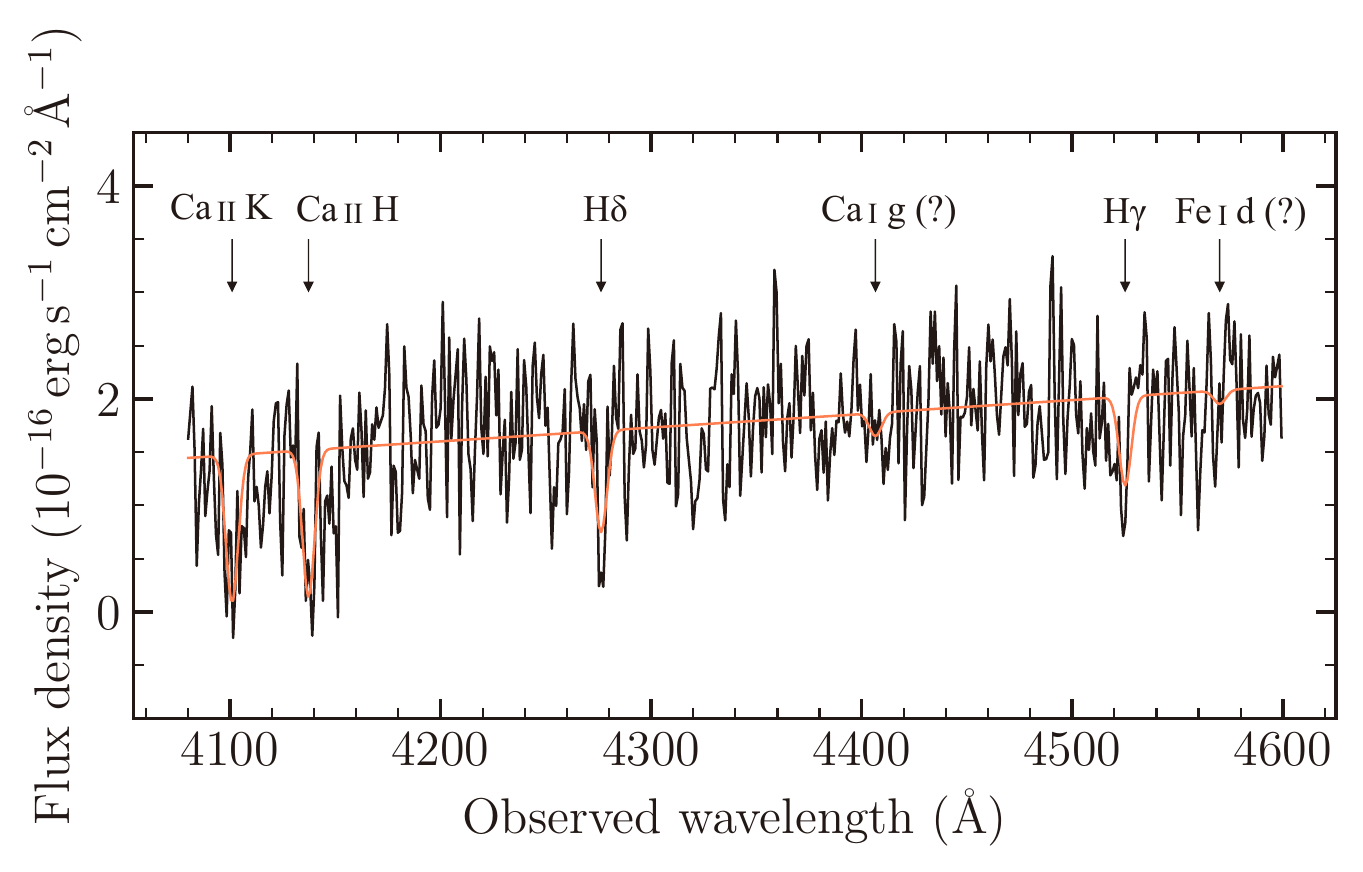}
    \caption{
        Optical spectrum of \irone{}.
        The spectrum is colored black, and the best-fit model is colored orange.
        Lines with less than 3$\sigma$ detections are denoted by question marks.
    }\label{fig:ir01_reds}
\end{figure}

In August 2009, \citet{Oyabu2011} observed the optical spectrum of \irone{} with the Kast Double Spectrograph on the Shane $3\,\mathrm{m}$ telescope at Lick Observatory using the 600/4310 grism and the 600/7500 gratings.
This instrument splits the incident beam into blue and red beams with a beamsplitter and simultaneously provides grism and grating spectroscopy for the blue and red beams, respectively.
In this work, we used only a part of the blue spectrum, in the wavelength range $\sim 4100\text{--}4600\,\text{\AA}$, to determine the redshift of \irone{} because the continuum had no severe curvature, and the absorption lines were well detected in this range.
We reduced the spectrum in the standard manner as done in \citet{Oyabu2011}, with the standard star Feige~15 for sensitivity correction and flux calibration.
We performed wavelength calibration using the spectrum from a He-Hg-Cd lamp, and the achieved systematic error of the calibration was $\sim 0.1\,\text{\AA}$.
We then corrected for Earth's motion and converted the wavelength to LSR values using the \verb|rvcorrect| task of IRAF.
The derived spectrum is shown in Figure~\ref{fig:ir01_reds}.

\subsection{Methods and Results}
We performed line fitting for the derived spectrum to determine the redshift, using equal weights for all data points.
We considered the lines $\mathrm{Ca_{II}\ K,H}$ ($3933.66$, $3968.47\,\text{\AA}$), $\mathrm{H\delta}$ ($4101.734\,\text{\AA}$), $\mathrm{Ca_I\ g}$ ($4226.73\,\text{\AA}$), $\mathrm{H\gamma}$ ($4340.472\,\text{\AA}$), and $\mathrm{Fe_I\ d}$ ($4383.5447\,\text{\AA}$) and fitted them with Gaussian profiles having a common wavelength width and redshift and with the individual absorption depths of the lines as free parameters.
We extracted the rest wavelength of each line from the NIST Atomic Spectra Database \citep{NISTASD}.
In addition, we assumed the continuum to be linear, with its slope and y-intercept as free parameters.
\par
The resulting redshift we derived for \irone{} in the LSR was
\begin{equation}
    z=0.04254\pm 0.00010\,\text{(stat.)}\pm 0.00002\,\text{(sys.)},
\end{equation}
where the statistical errors originate from the fitting errors and the systematic errors are due to wavelength-calibration uncertainties.
The derived FWHMs of the lines were $7\pm 1\,\text{\AA}$.
The best-fit model is shown in Figure~\ref{fig:ir01_reds}.
Hence, we use this redshift for \irone{} in this paper.

\section{Number of Velocity Components in Each Target}\label{sec:aic}

For robust fitting of gaseous CO rovibrational lines with multiple Gaussians, we determine the number of Gaussian velocity components based on the Akaike information criterion (AIC).
This quantity is defined as $\mathrm{AIC}\equiv \chi^2_\mathrm{min}+2N_\mathrm{vary}$, where $\chi^2_\mathrm{min}$ is the best-fit chi-squared value, and $N_\mathrm{vary}$ is the number of free parameters \citep{Akaike1974}.
A more favorable model has a smaller AIC value; thus, the model with the minimum AIC value is the most favorable.
In addition, the difference between the AIC and its minimum ($\Delta\mathrm{AIC}\equiv \mathrm{AIC}-\mathrm{AIC_{min}}$) gives the probability that a model with larger AIC is more likely than the one with the minimum AIC as $\exp(-\Delta\mathrm{AIC}/2)$.
Hence, $\Delta\mathrm{AIC}>14$ means that a model with a larger AIC is rejected with a significance level of $>99.9\%$.
\par
\begin{figure}
    \centering
    \begin{tabular}{cc}
        \includegraphics[width=0.45\linewidth]{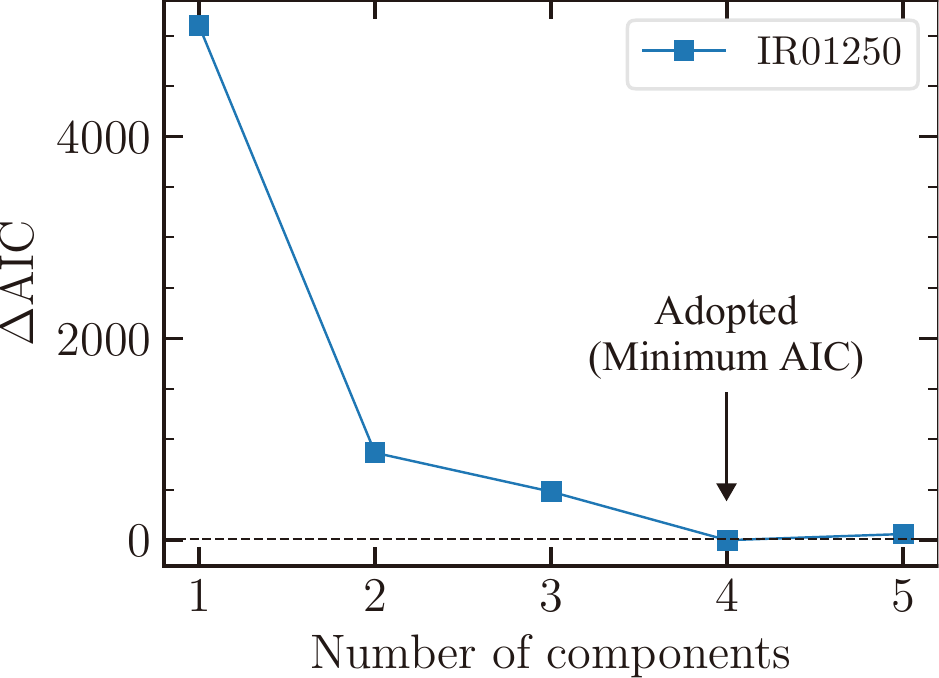}
        &\includegraphics[width=0.45\linewidth]{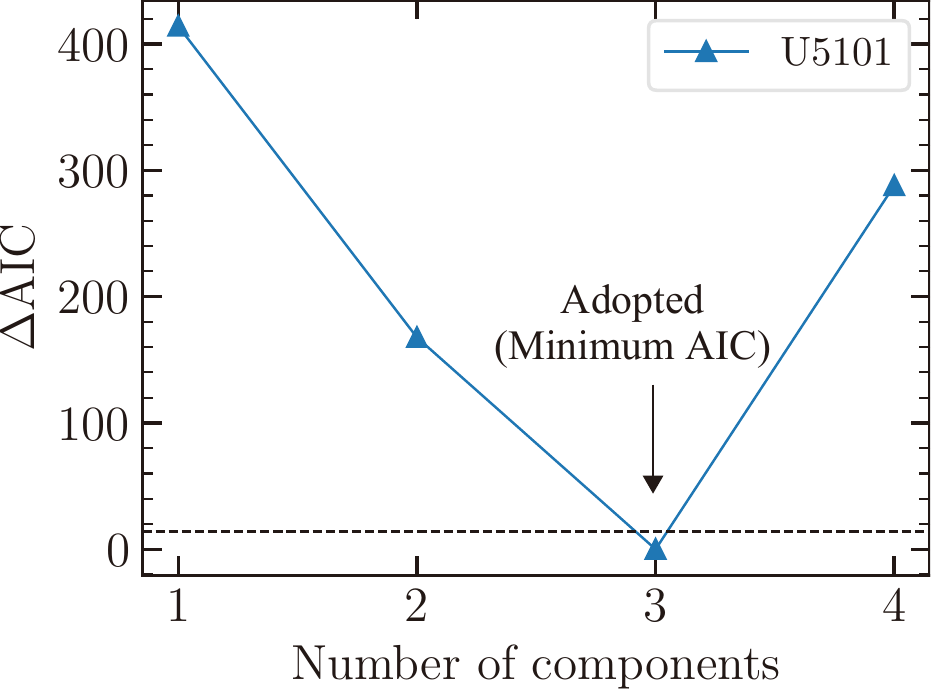}\\
        \includegraphics[width=0.45\linewidth]{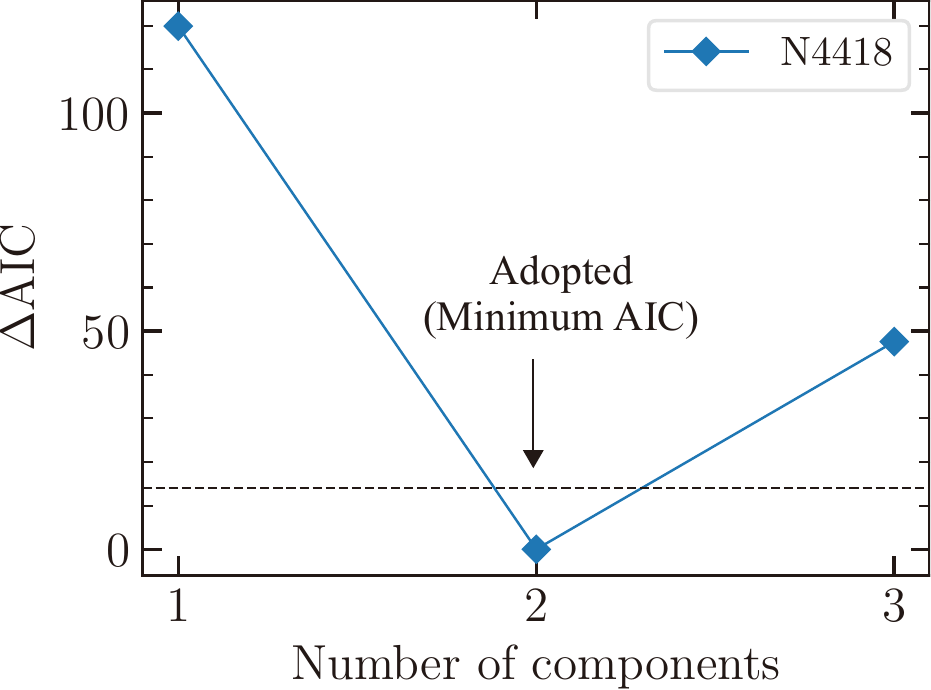}
        &\includegraphics[width=0.45\linewidth]{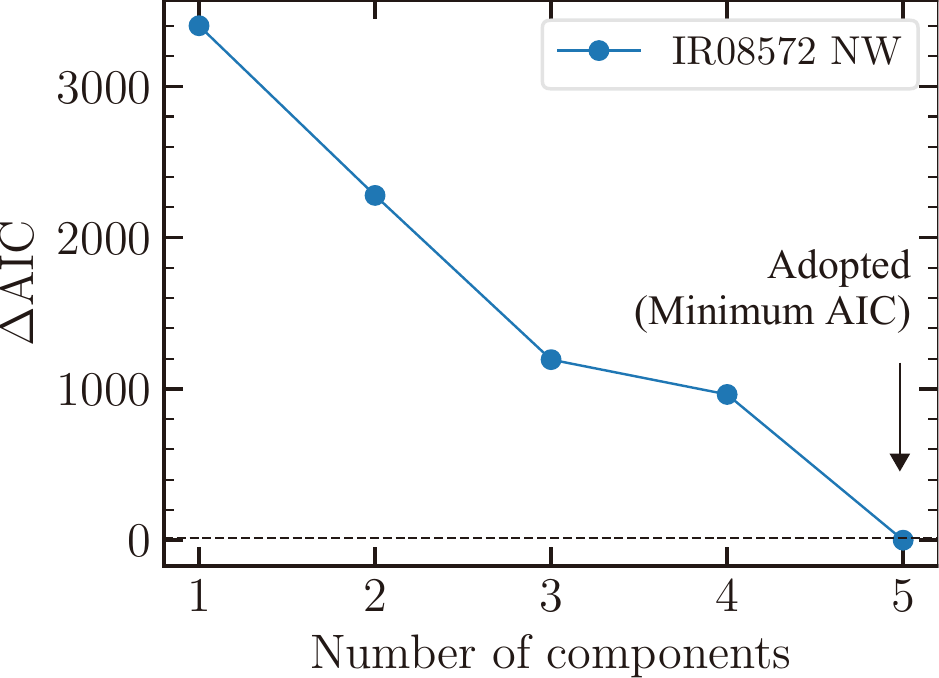}
    \end{tabular}
    \caption{
        Relationship between the number of velocity components considered and the difference between the AIC and its minimum ($\Delta\mathrm{AIC}$) for \irone{,} \ufive{,} \nfour{,} and \ireight{}.
        The name of the target in each panel is shown in the legend.
        The dashed horizontal line denotes the 0.1\% rejection threshold ($\Delta\mathrm{AIC}$=14).
        Note that for \irone{} the model with five velocity components gives $\Delta\mathrm{AIC}=61$, which exceeds the rejection threshold.
        The adopted number of velocity components is denoted by a black arrow.
    }\label{fig:AIC_check_ir01}
\end{figure}

We tried from one to five Gaussian velocity components as long as the following criteria were satisfied:
\begin{enumerate}[label=\arabic*.]
    \item The AIC is decreased by adding another velocity component.
    \item The velocity centroid of each velocity component settles within the range $(V_\mathrm{0}/\kmps)\in[-500, 200]$, and the velocity dispersion settles within the range $(\sigma_V/\kmps)\in[0, 300]$, so that they do not become degenerate with those of the velocity components from an adjacent transition.
\end{enumerate}
Figure~\ref{fig:AIC_check_ir01} shows the number of velocity components tried and the $\Delta\mathrm{AIC}$ values for \irone{,} \ufive{,} \nfour{,} and \ireight{}.

\section{Decomposed Spectra of \ufive{,} \nfour{,} and \ireight{}}\label{sec:decomp_spec_others}

\begin{sidewaysfigure*}
    \centering
    \includegraphics[width=0.9\linewidth]{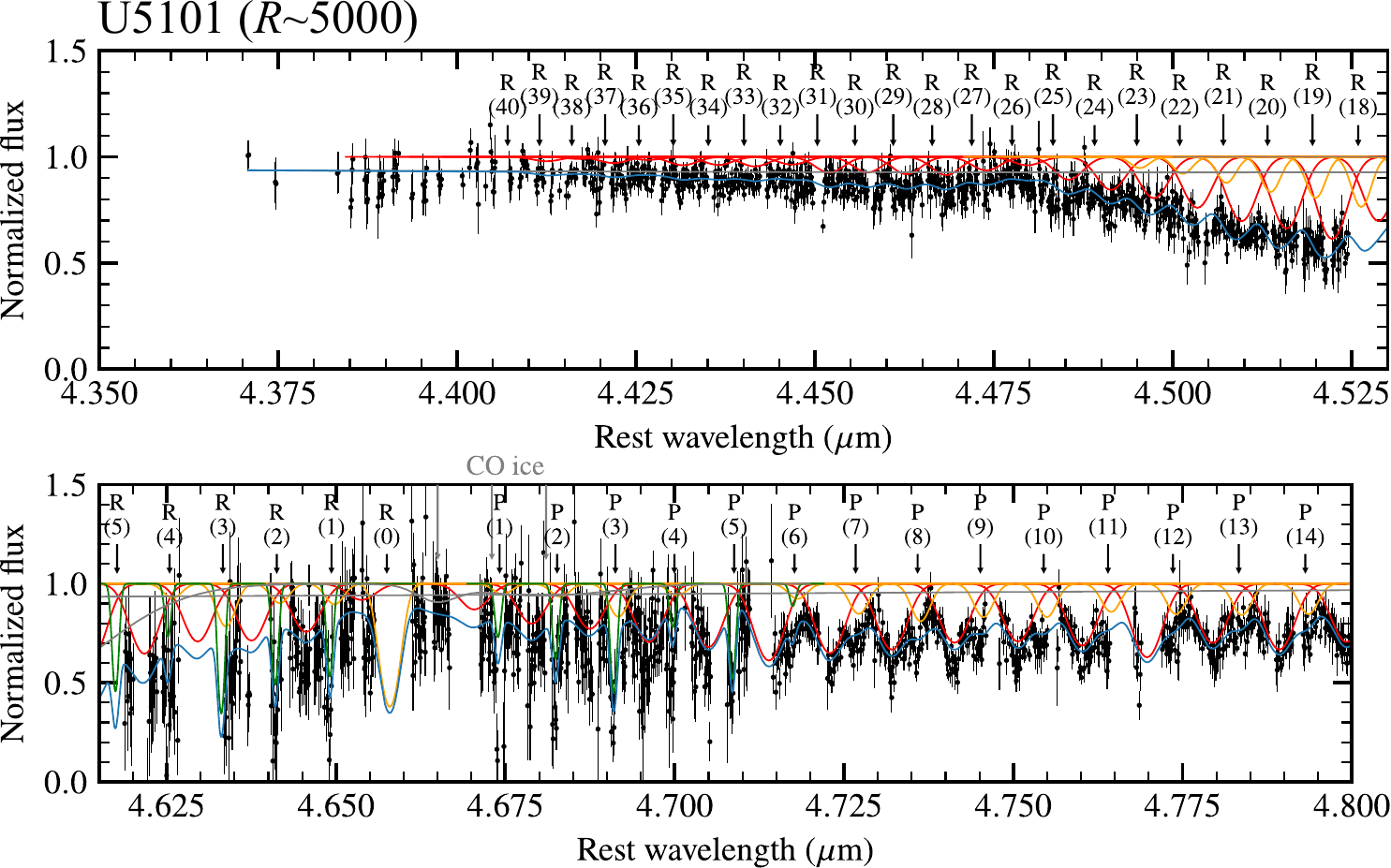}
    \caption{
        Best-fit model of the gaseous CO rovibrational absorption lines in \ufive{}.
        The abscissa is the rest wavelength.
        The ordinate is the normalized flux.
        Components (a)--(c) and their sum are denoted with red, orange, green, and sky-blue lines, respectively.
        The black arrows denote the rest wavelength of each CO transition.
        The $\mathrm{H_I\ Pf\beta}$ and the $\mathrm{H_2}\ v=0\to 0\ S(9)$ emission line are colored in olive-green.
        Ice features of $\mathrm{H_2O}$, $\mathrm{OCN^{-}}$, CO, and OCS are denoted with solid gray lines.
    }\label{fig:fitres_colines_u51}
\end{sidewaysfigure*}

As explained in Section~\ref{subsec:vel_decomp}, we decompose the velocity components in each CO ro-vibrational transition.
Figures~\ref{fig:fitres_colines_u51}--\ref{fig:fitres_colines_ir08} show the velocity-decomposed spectra of \ufive{,} \nfour{,} and \ireight{}.
(The spectrum of \irone{} is shown in Figure~\ref{fig:fitres_colines_ir01}.)
Note that the results of our velocity decomposition for \ireight{} are slightly different from those in \citet{Onishi2021}.
This occurred because we consider the emission lines of $\mathrm{H_I\ Pf\beta}$ and $\mathrm{H_2}\ v=0\to 0\ S(9)$ in the model fitting in this paper, as well as the gaseous CO absorption lines, whereas \citet{Onishi2021} excluded the emission lines in the continuum-normalization step before the fitting.

\onecolumngrid{}

\addtocounter{figure}{-1}
\begin{sidewaysfigure*}
    \centering
    \includegraphics[width=0.9\linewidth]{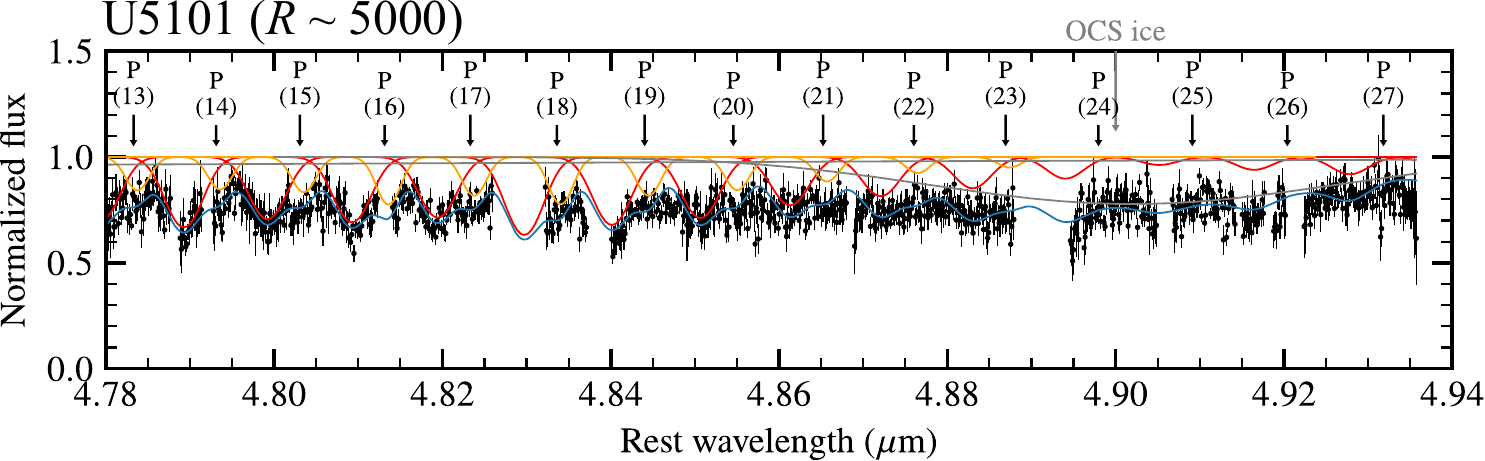}
    \caption{
        (Continued) Best-fit model of the gaseous CO rovibrational absorption lines in \ufive{}. The abscissa is the rest wavelength.
        The ordinate is the normalized flux.
        Components (a)--(c) and their sum are denoted with red, orange, green, and sky-blue lines, respectively.
        The black arrows denote the rest wavelength of each CO transition.
        The $\mathrm{H_I\ Pf\beta}$ and the $\mathrm{H_2}\ v=0\to 0\ S(9)$ emission line are colored in olive-green.
        Ice features of $\mathrm{H_2O}$, $\mathrm{OCN^{-}}$, CO, and OCS are denoted with solid gray lines.
    }
\end{sidewaysfigure*}

\begin{sidewaysfigure*}
    \centering
    \includegraphics[width=0.9\linewidth]{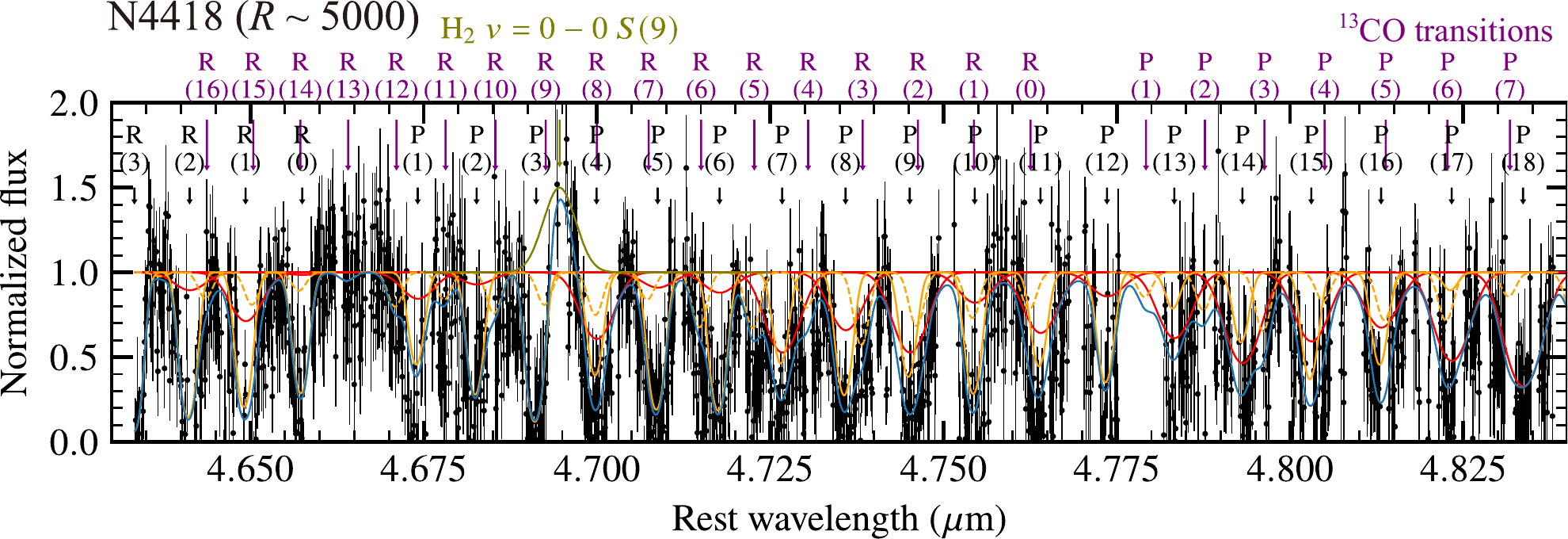}
    \caption{
        Best-fit model of the gaseous CO rovibrational absorption lines in \nfour{}.
        The abscissa is the rest wavelength.
        The ordinate is the normalized flux.
        Components (a) and (b) and their sum are denoted with red, orange, and sky-blue lines, respectively.
        In addition, component (b) of $\mathrm{{}^{13}CO}$ absorption lines are shown with orange dashed lines.
        The black arrows denote the rest wavelength of each CO transition.
        The $\mathrm{H_I\ Pf\beta}$ and the $\mathrm{H_2}\ v=0\to 0\ S(9)$ emission line are colored olive-green.
    }\label{fig:fitres_colines_n44}
\end{sidewaysfigure*}

\begin{sidewaysfigure*}
    \centering
    \includegraphics[width=0.9\linewidth]{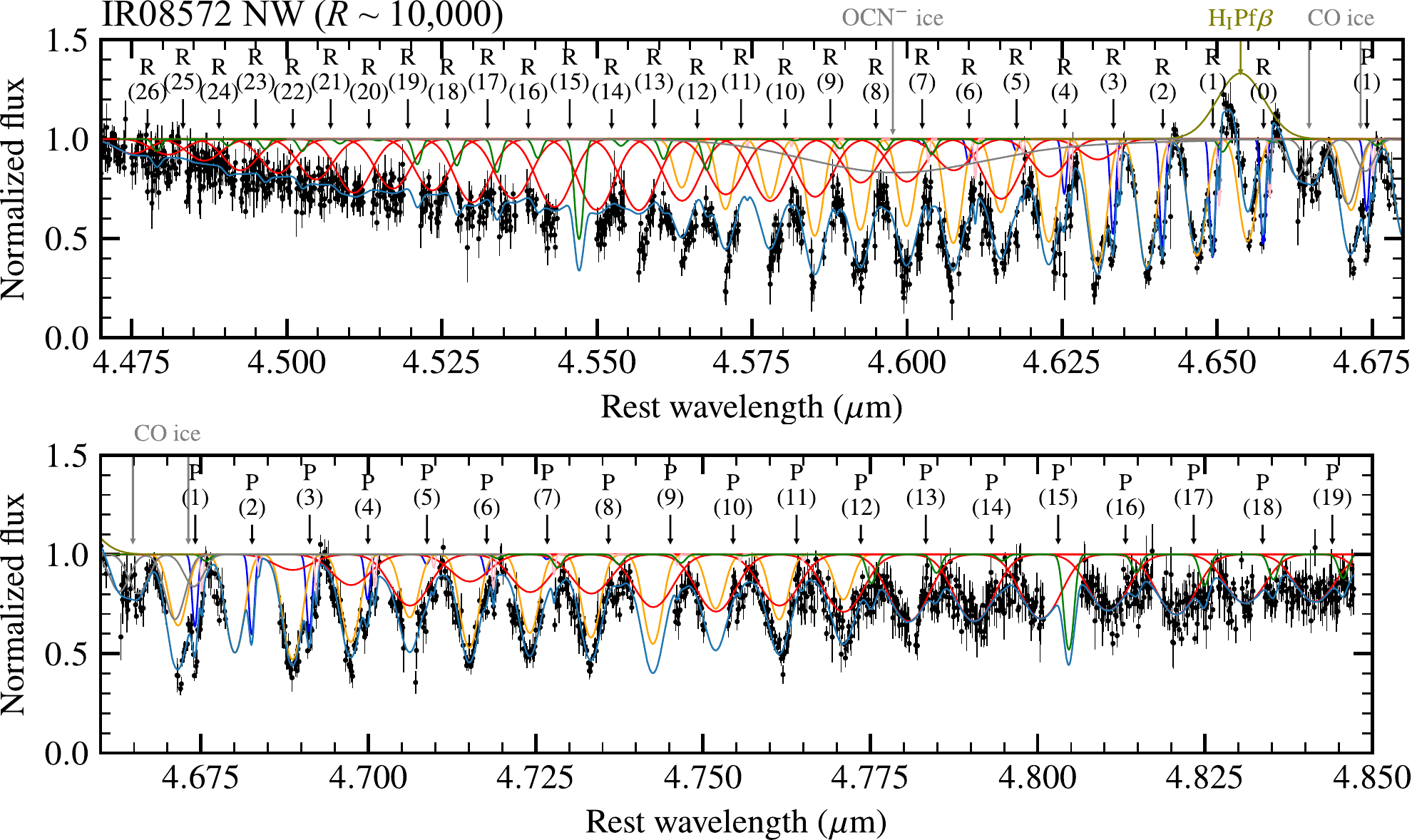}
    \caption{
        Best-fit model of the gaseous CO rovibrational absorption lines in \ireight{}.
        The abscissa is the rest wavelength.
        The ordinate is the normalized flux.
        Components (a)--(e) and their sum are denoted in red, orange, green, dark blue, pink, and sky-blue lines, respectively.
        Black arrows denote the rest wavelength of each CO transition.
        The emission lines of $\mathrm{H_I\ Pf\beta}$ and $\mathrm{H_2}\ v=0\to 0\ S(9)$ are colored olive-green.
        Ice features of $\mathrm{H_2O}$, $\mathrm{OCN^{-}}$, CO, and OCS are denoted with solid gray lines.
        Note that the results of our velocity decomposition for \ireight{} are slightly different from those in \citet{Onishi2021}.
        This occurred because we consider the emission lines of $\mathrm{H_I\ Pf\beta}$ and $\mathrm{H_2}\ v=0\to 0\ S(9)$ in the model fitting in this paper, as well as the gaseous CO absorption lines, whereas \citet{Onishi2021} excluded the emission lines in the continuum-normalization step before the fitting.
    }\label{fig:fitres_colines_ir08}
\end{sidewaysfigure*}

\twocolumngrid{}

\clearpage

\section{Model Fitting to the Population Diagram with MCMC}\label{sec:mcmc_fit}
We fitted the RADEX model for the CO level populations to the observed population diagrams using the MCMC method with Emcee v3.0.2 \citep{Foreman-Mackey2013} and Lmfit v1.0.0 \citep{Lmfit} packages.
This approach enables us to investigate reasonable solutions for which the distributions in parameter space are far from normal ones, as described in Section~\ref{subsubsec:radex_model}.
\par
The free parameters we consider are $C_f$, $T_\mathrm{kin}$, $n_\mathrm{H_2}$, $T_\mathrm{bg}$, and $N_\mathrm{CO}$.
The prior distribution of each parameter is a box function.
We assume the following conditions unless otherwise stated:
\begin{enumerate}[label=(\arabic*)]
    \item
    The covering factor is lower than unity and higher than zero; i.e., $C_f\in [0,1]$.
    For velocity components where the lower limits of $C_f$ are not constrained, we determine the lower boundaries of the prior distributions on the basis of the peak depths of the absorption lines.
    \item
    The brightness temperature of the background radiation $T_\mathrm{bg}$ is lower than the dust sublimation temperature of $1500\,\mathrm{K}$ and higher than the cosmic microwave background (CMB) temperature of $2.73\,\mathrm{K}$; i.e., $(T_\mathrm{bg}/\mathrm{K})\in [2.73, 1500]$.
    The kinetic temperature also is higher than $2.73\,\mathrm{K}$, but it is lower than $1000\,\mathrm{K}$ so that it is observed in absorption against the NIR source, which is the dust sublimation layer; i.e., $(T_\mathrm{kin}/\mathrm{K})\in [2.73, 1000]$.
    \item
    The density ($n_\mathrm{H_2}$) of molecular hydrogen is smaller than the typical value in a maser disk ($10^{10}\,\mathrm{cm^{-3}}$), which is expected to be located in the dense inner region of the torus \citep{Taniguchi1998}.
    The prior distribution of the molecular-hydrogen density is thus a box function in the range of $\log\qty(n_\mathrm{H_2}/\mathrm{cm^{-3}})\in [3, 10]$.
    For the innermost components (a), the lower boundary of the $n_\mathrm{H_2}$ prior distribution is determined so that it is larger than that of the outer components because an MHD torus model \citep{Chan2017} predicts that the volume density of the gas decreases as the distance from the central black hole increases.
    \item
    The prior distribution for the CO column density is a box function in the range of $\log\qty(N_\mathrm{CO}/\mathrm{cm^{-2}})\in [15, 20]$.
    The upper limit corresponds to the hydrogen column density $2N_\mathrm{H_2}\sim 10^{24}\,\mathrm{cm^{-2}}$ or the visual extinction $A_V\sim 10^3$ with the abundance ratio $\mathrm{[CO]/[H_2]}=10^{-4}$ \citep{Dickman1978} and the ratio between the visual extinction and the hydrogen column density $N_\mathrm{H}/A_V=1.8\times 10^{21}\psqc$ \citep{Bohlin1978a} assumed.
    The equivalent $M$-band extinction is $A_M\sim 10$ if the Milky Way dust model is assumed \citep{Draine2003}.
    Above this limit, the NIR continuum is heavily extinguished and the CO absorption should be hard to be observed.
\end{enumerate}
\par
The posterior distribution $p(\vb*{w}|\vb*{X})$ is
\begin{eqnarray}
    p(\vb*{w}|\vb*{X})&\propto&p(\vb*{X}|\vb*{w})p(\vb*{w}),\\
    p(\vb*{X}|\vb*{w})&\propto&\exp\qty[-\frac{1}{2}\sum_{J}\qty(\frac{X^J-f(J;\vb*{w})}{\delta{X^J}})^2]
\end{eqnarray}
where $\vb*{w}$ is a vector consisting of the five free parameters mentioned above and $p(\vb*{w})$ is their prior distributions; $\vb*{X}$ is a vector containing the observed column densities of the excited CO molecules in each rotational level divided by their statistical weights ($N_J/{2J+1}$) and $\vb*{\delta X}$ is a vector containing the corresponding errors ($\delta N_J/{2J+1}$); and $f(J;\vb*{w})$ is the RADEX model function for $N_J/{2J+1}$, given the physical parameters of $\vb*{w}$.
Here, 100 chains, each of which is 25{,}000 trials long, are generated, each with a burn-in length of 1000.
We can thus generate chains for which the length is at least 50 times longer than the number of integrated autocorrelation steps for each free parameter ($<500$).


\bibliographystyle{aasjournal}
\bibliography{library_co_paper2024}



\end{document}